\documentclass{vldb}
\usepackage{graphicx}
\usepackage{textcomp}
\usepackage{xcolor}
\usepackage[utf8]{inputenc}
\usepackage{amsmath}
\usepackage{booktabs}
\usepackage{mathtools}
\usepackage[ruled, vlined, linesnumbered]{algorithm2e}
\usepackage{verbatim}
\usepackage{subfig}
\usepackage[font=bf]{caption}
\usepackage{multirow,multicol,enumitem,epsfig,url,array,makecell,balance,tabularx,cite}

\vldbTitle{Homogeneous Network Embedding for Massive Graphs via Reweighted Personalized PageRank}
\vldbAuthors{Renchi Yang, Jieming Shi, Xiaokui Xiao, Yin Yang, and Sourav S. Bhowmick}
\vldbDOI{https://doi.org/10.14778/3377369.3377376}
\vldbVolume{13}
\vldbNumber{5}
\vldbYear{2020}

\newcommand{\jieming}[1]{{\color{blue}{\bf [Jieming: #1]}}}

\makeatletter
\newcommand*\bigcdot{\mathpalette\bigcdot@{.5}}
\newcommand*\bigcdot@[2]{\mathbin{\vcenter{\hbox{\scalebox{#2}{$\m@th#1\bullet$}}}}}
\makeatother

\def\done{\hspace*{\fill} {$\square$}}
\def\header{\vspace{1mm} \noindent}

\newcommand{\PM}{\mathbf{P}}
\newcommand{\AM}{\mathbf{A}}
\newcommand{\DM}{\mathbf{D}}
\newcommand{\IM}{\mathbf{I}}

\newcommand{\UM}{\mathbf{U}}
\newcommand{\VM}{\mathbf{V}}

\newcommand{\XM}{\mathbf{X}}
\newcommand{\YM}{\mathbf{Y}}
\newcommand{\PiM}{\boldsymbol{\Pi}}
\newcommand{\PiMp}{\boldsymbol{\Pi^\prime}}
\newcommand{\kp}{k^\prime}

\newcommand{\app}{$\mathsf{APP}$\xspace}
\newcommand{\arope}{$\mathsf{AROPE}$\xspace}
\newcommand{\randne}{$\mathsf{RandNE}$\xspace}
\newcommand{\versealg}{$\mathsf{VERSE}$\xspace}
\newcommand{\rare}{$\mathsf{RaRE}$\xspace}
\newcommand{\ga}{$\mathsf{GA}$\xspace}
\newcommand{\gae}{$\mathsf{GAE}$\xspace}
\newcommand{\nethiex}{$\mathsf{NetHiex}$\xspace}
\newcommand{\nrp}{$\mathsf{NRP}$\xspace}
\newcommand{\nrpb}{NRP\xspace}

\newcommand{\ane}{$\mathsf{ANE}$\xspace}
\newcommand{\linealg}{$\mathsf{LINE}$\xspace}
\newcommand{\deepwalk}{$\mathsf{DeepWalk}$\xspace}
\newcommand{\nodetovec}{$\mathsf{node2vec}$\xspace}
\newcommand{\structovec}{$\mathsf{struc2vec}$\xspace}
\newcommand{\walklets}{$\mathsf{Walklets}$\xspace}
\newcommand{\harp}{$\mathsf{HARP}$\xspace}
\newcommand{\sdne}{$\mathsf{SDNE}$\xspace}
\newcommand{\dngr}{$\mathsf{DNGR}$\xspace}
\newcommand{\graphgan}{$\mathsf{GraphGAN}$\xspace}
\newcommand{\prune}{$\mathsf{PRUNE}$\xspace}
\newcommand{\graphwave}{$\mathsf{GraphWave}$\xspace}
\newcommand{\drne}{$\mathsf{DRNE}$\xspace}
\newcommand{\seedne}{$\mathsf{SeedNE}$\xspace}
\newcommand{\nodehash}{$\mathsf{node2hash}$\xspace}
\newcommand{\dvne}{$\mathsf{DVNE}$\xspace}
\newcommand{\netra}{$\mathsf{NetRA}$\xspace}
\newcommand{\approxppr}{$\mathsf{ApproxPPR}$\xspace}
\newcommand{\baseline}{$\mathsf{ApproxPPR}$\xspace}
\newcommand{\netmf}{$\mathsf{NetMF}$\xspace}
\newcommand{\netsmf}{$\mathsf{NetSMF}$\xspace}
\newcommand{\pbg}{$\mathsf{PBG}$\xspace}
\newcommand{\graphy}{$\mathsf{Graphy}$\xspace}
\newcommand{\gra}{$\mathsf{GRA}$\xspace}
\newcommand{\rsvd}{$\mathsf{BKSVD}$\xspace}
\newcommand{\strap}{$\mathsf{STRAP}$\xspace}

\newcommand{\topppr}{$\mathsf{TopPPR}$\xspace}

\newcommand{\prone}{$\mathsf{ProNE}$\xspace}
\newcommand{\graphcsc}{$\mathsf{GraphCSC\text{-}M}$\xspace}
\newcommand{\dwns}{$\mathsf{DWNS}$\xspace}

\newcommand{\wra}{\overrightarrow{w}}
\newcommand{\wla}{\overleftarrow{w}}
\newcommand{\wrab}{\overrightarrow{\mathbf{w}}}
\newcommand{\wlab}{\overleftarrow{\mathbf{w}}}

\newtheorem{example}{Example}

\newtheorem{theorem}{Theorem}
\newtheorem{lemma}{Lemma}

\newcommand{\proofref}{Appendix A\xspace}

\newcommand{\fwdref}{Appendix B\xspace}

\def\BibTeX{{\rm B\kern-.05em{\sc i\kern-.025em b}\kern-.08em
    T\kern-.1667em\lower.7ex\hbox{E}\kern-.125emX}}

\begin{document}

\title{Homogeneous Network Embedding for Massive Graphs via Reweighted Personalized PageRank}
\subtitle{[Technical Report]}

\numberofauthors{1}
\author{
\alignauthor
Renchi Yang$^\ast$, Jieming Shi$^\dagger$, Xiaokui Xiao$^\dagger$, Yin Yang$^\S$, Sourav S. Bhowmick$^\ast$\\
\affaddr{$^\ast$School of Computer Science and Engineering, Nanyang Technological University, Singapore}\\
\affaddr{$^\dagger$School of Computing, National University of Singapore, Singapore}\\
\affaddr{$^\S$College of Science and Engineering, Hamad Bin Khalifa University, Qatar}\\
\email{$^\ast$\{yang0461,assourav\}@ntu.edu.sg,$^\dagger$\{shijm,xkxiao\}@nus.edu.sg,$^\S$yyang@hbku.edu.qa}
}

\maketitle

\begin{abstract}
Given an input graph $G$ and a node $v \in G$, homogeneous network embedding (HNE) maps the graph structure in the vicinity of $v$ to a compact, fixed-dimensional feature vector. 
This paper focuses on HNE for massive graphs, \textit{e.g.}, with billions of edges. On this scale, most existing approaches fail, as they incur either prohibitively high costs, or severely compromised result utility. 

Our proposed solution, called Node-Reweighted PageRank (\nrp), is based on a classic idea of deriving embedding vectors from pairwise personalized PageRank (PPR) values. Our contributions are twofold: first, we design a simple and efficient baseline HNE method based on PPR that is capable of handling billion-edge graphs on commodity hardware; second and more importantly, we identify an inherent drawback of vanilla PPR, and address it in our main proposal \nrp. Specifically, PPR was designed for a very different purpose, {\it i.e.}, ranking nodes in $G$ based on their \textit{relative} importance from a source node's perspective. In contrast, HNE aims to build node embeddings considering the \textit{whole} graph. Consequently, node embeddings derived directly from PPR are of suboptimal utility.

The proposed \nrp approach overcomes the above deficiency through an effective and efficient node reweighting algorithm, which augments PPR values with node degree information, and iteratively adjusts embedding vectors accordingly. Overall, \nrp takes $O(m\log n)$ time and $O(m)$ space to compute all node embeddings for a graph with $m$ edges and $n$ nodes.
Our extensive experiments that compare \nrp against $18$ existing solutions over 7 real graphs demonstrate that \nrp achieves higher result utility than all the solutions for link prediction, graph reconstruction and node classification, while being up to orders of magnitude faster. In particular, on a billion-edge Twitter graph, \nrp terminates within 4 hours, using a single CPU core.
\end{abstract}


\thispagestyle{plain}
\pagestyle{plain}
\section{Introduction}\label{sec:intro}

Given a graph $G=(V,E)$ with $n$ nodes, a {\it network embedding} maps each node $v \in G$ to a compact feature vector in $\mathbb{R}^k$ ($k \ll n$), such that the embedding vector captures the graph features surrounding $v$. These embedding vectors are then used as inputs in downstream machine learning operations \cite{AlibabaGE18,FaloutsosTSDMZ18,WuFCABW18}. A {\it homogeneous network embedding} (\textit{HNE}) is a type of network embedding that reflects the topology of $G$ rather than labels associated with nodes or edges. HNE methods have been commonly applied to various graph mining tasks based on neighboring associated similarities, including node classification~\cite{rcos13}, link prediction~\cite{bljl11}, and graph reconstruction \cite{HOPE16}. This paper focuses on HNE computation on massive graphs, \textit{e.g.}, social networks involving billions of connections. Clearly, an effective solution for such a setting must be highly scalable and efficient, while obtaining high result utility.

HNE is a well studied problem in the data mining literature, and there are a plethora of solutions. However, most existing solutions fail to compute effective embeddings for large-scale graphs. For example, as we review in Section \ref{sec:rw}, a common approach is to learn node embeddings from random walk simulations, \textit{e.g.}, in \cite{deepwalk14,node2vec16}. However, the number of possible random walks grows exponentially with the length of the walk; thus, for longer walks on a large graph, it is infeasible for the training process to cover even a considerable portion of the random walk space. Another popular methodology is to construct node embeddings by factorizing a proximity matrix, \textit{e.g.}, in \cite{AROPE18}. The effectiveness of such methods depends on the proximity measure between node pairs. As explained in Section \ref{sec:rw}, capturing multi-hop topological information generally requires a sophisticated proximity measure; on the other hand, the computation, storage and factorization of such a proximity matrix often incur prohibitively high costs on large graphs.

\begin{figure*}[!t]
\centering
\hspace{-9mm}
\begin{minipage}{0.3\textwidth}
\centering
\begin{small}
\vspace{3mm}
\includegraphics[width=0.9\columnwidth]{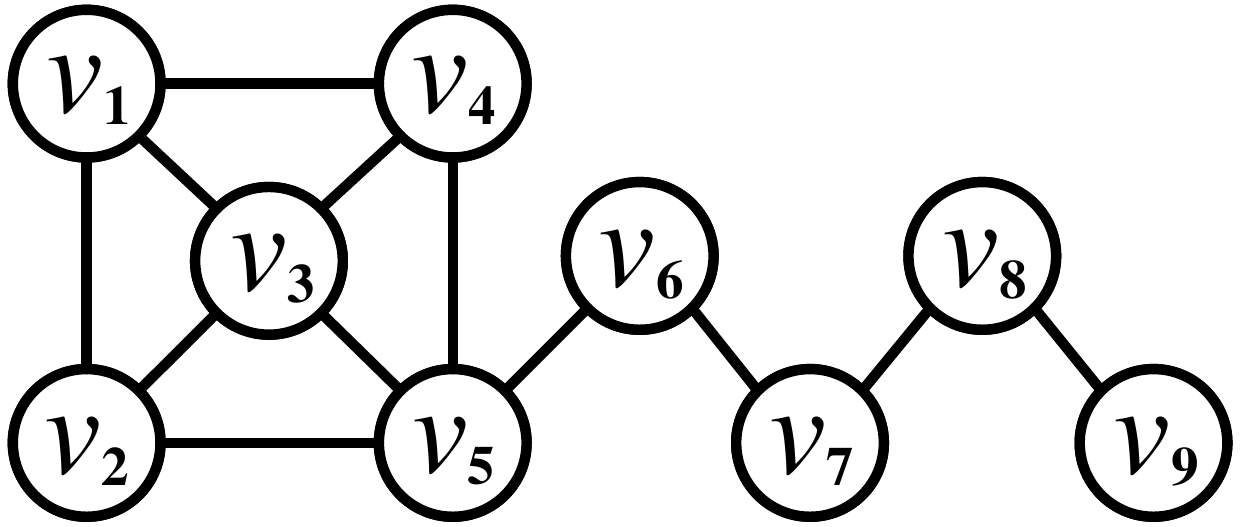}
\vspace{0mm}
\caption{An example graph $G$.}\label{fig:over-ppr}
\end{small}
\end{minipage}
\hspace{4mm}
\begin{minipage}{0.6\textwidth}
\centering
\captionsetup{type=table}
\renewcommand{\arraystretch}{1.2}
\begin{small}
\caption{PPR for $\boldsymbol{v_2}$ and $\boldsymbol{v_9}$ in Fig.~\ref{fig:over-ppr} ($\alpha = 0.15$).} \label{tbl:over-ppr}
\vspace{-2mm}
\begin{tabular}{|c|c|c|c|c|c|c|c|c|c|}
\hline
$\boldsymbol{v_i}$ & $v_1$ & $v_2$ & $v_3$ & $v_4$ & $v_5$ & $v_6$ & $v_7$ & $v_8$ & $v_9$ \\
\hline
$\boldsymbol{\pi(v_2, v_i)}$  & 0.15  &  0.269  &  0.188  &  \bf{0.118}  &  0.17 &   0.048  &  0.029  &  0.019  &  0.008  \\
\hline
$\boldsymbol{\pi(v_4, v_i)}$  & 0.15  &  {\bf 0.118}  &  0.188  &  0.269  &  0.17 &   0.048 &  0.029  &  0.019  &  0.008  \\
\hline
$\boldsymbol{\pi(v_7, v_i)}$  & 0.036  &  0.043  &  0.056  &  0.043  & 0.093  &  0.137  &  0.29  &  0.187  &  {\bf 0.12}  \\
\hline
$\boldsymbol{\pi(v_9, v_i)}$  &  0.02  &  0.024  &  0.031  &  0.024  &  0.056 &   0.083  &  \bf{0.168}  &  0.311  &  0.282   \\
\hline
\end{tabular}
\end{small}
\end{minipage}
\vspace{-4mm}
\end{figure*}

This paper revisits an attractive idea: constructing HNEs by taking advantage of \textit{personalized PageRank} (\textit{PPR}) \cite{PPR03}. Specifically, given a pair of nodes $u, v \in G$, the PPR value $\pi(u, v)$ of $v$ with respect to $u$ is the probability that a random walk from $u$ terminates at $v$. PPR values can be viewed as a concise summary of an infinite number of random walk simulations, which, intuitively, should be helpful in building node embeddings. Realizing the full potential of PPR for scalable HNE computation, however, is challenging. One major hurdle is cost: materializing the PPR between each pair of nodes clearly takes $O(n^2)$ space for $n$ nodes (\textit{e.g.}, in \cite{yin2019scalable}), and evaluating even a single PPR value can involve numerous random walk simulations (\textit{e.g.}, in \cite{APP17,verse18}).

Further, we point out that even without considering computational costs, it is still tricky to properly derive HNEs from PPR values. The main issue is that PPR was designed to serve a very different purpose, \textit{i.e.}, ranking nodes in $G$ based on their relative importance from a source node's perspective. In other words, PPR is essentially a \textit{local} measure. On the other hand, HNE aims to summarize nodes from the view of the whole graph. To illustrate this critical difference, consider the example in Fig.~\ref{fig:over-ppr} with nodes $v_1$-$v_9$. Observe that between the node pair $v_2$ and $v_4$, there are three different nodes connecting them, {\it i.e.}, $v_1$, $v_3$ and $v_5$. In contrast, there is only one common neighbor between $v_9$ and $v_7$. Intuitively, if we were to predict a new edge in the graph, it is more likely to be ($v_2$, $v_4$) than ($v_9$, $v_7$). For instance, in a social network, the more mutual friends two users have, the more likely they know each other \cite{brzozowski2011should}. However, as shown in Table~\ref{tbl:over-ppr}, in terms of PPR values, we have $\pi(v_9, v_7)=0.168 > \pi(v_2, v_4)=0.118$, which tends to predict $(v_9,v_7)$ over $(v_2,v_4)$ and contradicts with the above intuition. This shows that PPR by itself is \textit{not} an ideal proximity measure, at least for the task of link prediction. This problem is evident in PPR-based HNE methods, \textit{e.g.}, in \cite{verse18, APP17}, and a similar issue limits the effectiveness of a recent proposal \cite{yin2019scalable}, as explained in Section \ref{sec:rw}.


This paper addresses both the scalability and result utility issues of applying PPR to HNE computation with a novel solution called \textit{Node-Reweighted PageRank} (\nrp). Specifically, we first present a simple and effective baseline approach that overcomes the efficiency issue of computing node embeddings using PPR values. The main proposal \nrp then extends this baseline by addressing the above-mentioned deficiency of conventional PPR. Specifically, \nrp augments PPR values with additional reweighting steps, which calibrate the embedding of each node to its in- and out- degrees. Intuitively, when a node has a large number of neighbors ({\it e.g.}, $v_2$ in Fig. \ref{fig:over-ppr}), its embedding vector should be weighted up in accordance with its degree, so as to reflect the importance of the node from the perspective of the whole graph. In \nrp, node reweighting is performed using an effective and scalable algorithm that iteratively adjusts node embeddings, whose cost is small compared to PPR computations. 
Overall, \nrp takes $O\left( k(m+kn)\log n\right)$ time and $O(m+nk)$ space to construct length-$k$ embeddings of all nodes in a graph with $n$ nodes and $m$ edges. In the common case that $k$ is small and the graph is sparse, the above complexities reduce to $O(m \log n)$ time and $O(m)$ space.

We have conducted extensive experiments on 7 popular real datasets, and compared \nrp against 18 existing HNE solutions on three tasks: link prediction, graph reconstruction and node classification. In all settings, \nrp achieves the best result utility. Meanwhile, with a few exceptions, \nrp is often orders of magnitude faster than its competitors.
In particular, on a Twitter graph with 1.2 billion edges, \nrp terminates within 4 hours on a single CPU core. 
\section{Related Work}\label{sec:rw}

Network embedding is a hot topic in graph mining, for which there exists a large body of literature as surveyed in \cite{CaiZC18,abs-1711-08752,abs-1801-05852}. Here we review the HNE methods that are most relevant to this work.

\header
{\bf Learning HNEs from random walks.} A classic methodology for HNE computation is to learn embeddings from random walk simulations. 
Earlier methods in this category include \deepwalk \cite{deepwalk14}, \linealg \cite{LINE15}, \nodetovec \cite{node2vec16} and \walklets \cite{walklets17}. 
The basic idea is to learn the embedding of a node $v$ by iteratively ``pulling'' the embeddings of positive context nodes ({\it i.e.}, those that are on the random walks originating from $v$) towards that of $v$, and ``pushing'' the embeddings of negative context nodes ({\it i.e.}, the nodes that are not connected to $v$) away from $v$. Subsequent proposals \cite{HARP18,ribeiro2017struc2vec} construct a multilayer graph over the original graph $G$, and then perform random walks on different layers to derive more effective embeddings. Instead of using a predefined sampling distribution, \seedne \cite{gao2018self} adaptively sample negative context nodes in terms of their informativeness.  \graphcsc \cite{chen2019exploiting} learns the embeddings of different centrality-based random walks, and combines these embeddings into one by weighted aggregation. Recent techniques \app \cite{APP17} and \versealg \cite{verse18} improve the quality of embeddings by refining the procedures for learning from PPR-based random walk samples. However, neither of them addresses the deficiency of conventional PPR as described in Section \ref{sec:intro}.

The main problem of random-walk-based HNE learning in general is their immense computational costs (proportional to the number of random walks), which can be prohibitive for large graphs. The high running time could be reduced with massively-parallel hardware, \textit{e.g.}, in \pbg \cite{lerer2019pytorch}, and/or with GPU systems, {\it e.g.}, in \graphy \cite{zhu2019graphy}. Nevertheless, they still incur a high financial cost for consuming large amounts of computational resources.

\header
{\bf Learning HNEs without random walks.} HNEs can also be learned directly from the graph structure using a deep neural network, without performing random walks. Training such a deep neural network, however, also incurs very high computational overhead, especially for large graphs \cite{verse18}. Notably, \sdne \cite{SDNE16} and \dngr \cite{DNGR16} employ multi-layer auto-encoders with a target proximity matrix to learn embeddings. \gae \cite{kipf2016variational} combines the graph convolutional network \cite{kipf2016semi} and auto-encoder models to learn embeddings. \prune \cite{lai2017preserving} utilizes a Siamese neural network to preserve both pointwise mutual information and global PageRank of nodes. \netra \cite{yu2018learning} and \drne \cite{tu2018deep} learn embeddings by feeding node sequences to a long short-term memory model  (LSTM). \dvne \cite{zhu2018deep} learns a Gaussian distribution in the Wasserstein space with the deep variational model as the latent representation of each node. \ga
\cite{abu2018watch} applies graph attention mechanism to a closed-form expectation of the limited random-walk co-occurrence matrices \cite{deepwalk14} to learn the embeddings.
\graphgan \cite{wang2018graph}, \ane \cite{dai2018adversarial} and \dwns \cite{dai2019adversarial} adopt the popular generative adversarial networks (GAN) to accurately model the node connectivity probability. As demonstrated in our experiments, none of these methods scale to large graphs.

\header
{\bf Constructing HNEs through matrix factorization.} 
Another popular methodology for HNE is through factorizing a \textit{proximity matrix} $\mathbf{M} \in \mathbb{R}^{n\times n}$, where $n$ is the number of nodes in the input graph $G$, and each entry $\mathbf{M}[i, j]$ signifies the \textit{proximity} between a pair of nodes $v_i, v_j \in G$. The main research question here is how to choose an appropriate $\mathbf{M}$ that (i) captures the graph topology well and (ii) is efficient to compute and factorize on large graphs. Specifically, to satisfy (i), each entry $\mathbf{M}[i,j] \in \mathbf{M}$ should accurately reflect the proximity between nodes $v_i, v_j$ via indirect connections, which can be long and complex paths. Meanwhile, to satisfy (ii) above, the computation / factorization of $\mathbf{M}$ should be done in memory. This means that $\mathbf{M}$ should either be sparse, or be efficiently factorized without materialization. In addition, note that for a directed graph $G$, the proximity is also directed, meaning that it is possible that $\mathbf{M}[i,j] \neq \mathbf{M}[j, i]$. Thus, methods that require $\mathbf{M}$ to be symmetric are limited to undirected graphs and cannot handle directed graphs.

Earlier factorization-based work, including \cite{tang2011leveraging,ahmed2013distributed,GraRep15,MNMF}, directly computes $\mathbf{M}$ before factorizing it to obtain node embeddings. 
For instance, spectral embedding \cite{tang2011leveraging} simply outputs the top $k$ eigenvectors of the Laplacian matrix of an undirected graph $G$ as node embeddings. This method has limited effectiveness \cite{deepwalk14,node2vec16}, as the Laplacian matrix only captures one-hop connectivity information for each node. To remedy this problem, one idea is to construct a higher-order proximity matrix $\mathbf{M}$ to capture multi-hop connectivity for each node \cite{GraRep15,MNMF,NEU17}. However, such a higher-order proximity matrix $\mathbf{M}$ is usually no longer sparse; consequently, materializing $\mathbf{M}$ becomes prohibitively expensive for large graphs due to the $O(n^2)$ space complexity for $n$ nodes.

Recent work \cite{HOPE16,AROPE18,RandNE18} constructs network embeddings without materializing $\mathbf{M}$, to avoid extreme space overhead. Many of these methods, however, rely on the assumption that $\mathbf{M}$ is symmetric; consequently, they are limited to undirected graphs as discussed above. For instance, \arope \cite{AROPE18} first applies an eigen-decomposition on the adjacency matrix $\AM$ of an undirected graph $G$, and then utilizes the decomposition results to derive each node's embedding to preserve proximity information, without explicitly constructing $\mathbf{M}$. Similar approaches have been adopted in \cite{HOPE16,RandNE18}. In particular, \randne \cite{RandNE18} uses a Gaussian random projection of $\mathbf{M}$ directly as node embeddings without factorization, in order to achieve high efficiency, at the cost of lower result utility.

The authors of \cite{NetMF18} prove that random-walk-based methods such as as \deepwalk, \linealg and \nodetovec essentially perform matrix factorizations. Thus, they propose \netmf, which factorizes a proximity matrix $\mathbf{M}$ that approximates the closed form representation of \deepwalk's implicit proximity matrix. However, \netmf requires materializing a dense $\mathbf{M}$, which is infeasible for large graphs. \netsmf \cite{qiu2019netsmf} improves the efficiency of \netmf by sparsifying $\mathbf{M}$ using the theory of spectral sparsification. However, \netsmf 
is still rather costly as it requires simulating a large number of random walks to construct $\mathbf{M}$. \prone \cite{zhang2019prone} learns embeddings via matrix factorization with the enhancement of spectral propagation. However, \prone is mainly designed for node classification, and its accuracy is less competitive for other tasks such as link prediction and graph reconstruction. \gra \cite{liu2019general} iteratively fine-tunes the proximity matrix to obtain enhanced result effectiveness, at the expense of high comptuational costs. 

\header
{\bf HNE via Approximate Personalized PageRank.} 
Although the idea of using PPR as the proximity measure to be preserved in the embeddings is often mentioned in random-walk-based solutions \cite{APP17,verse18}, these methods largely fail to scale due to numerous random walk simulations for the costly training process.
A recent work \strap \cite{yin2019scalable} obtains better scalability by computing and factorizing a PPR-based proximity matrix instead. Specifically, \strap builds node embeddings by factorizing the {\em transpose proximity matrix}, defined as $\mathbf{M}=\boldsymbol{\Pi}+\boldsymbol{\Pi}^{\top}$, where $\boldsymbol{\Pi}$ and $\boldsymbol{\Pi}^{\top}$ represent the approximate PPR matrices of the original graph $G$ and its transpose graph (\textit{i.e.}, obtained by reversing the direction of each edge in $G$), respectively. 

The space and time complexities of \strap are $O(\frac{n}{\delta})$ and $O(\frac{m}{\delta}+nk^2)$, respectively, where $\delta$ is the error threshold for PPR values. In the literature of approximate PPR processing (\textit{e.g.}, \cite{FORA17,TopPPR18,wang2019efficient,wang2019parallelizing,shi2019realtime}), $\delta$ is commonly set to $\frac{1}{n}$, which would lead to prohibitively high space (\textit{i.e.}, $O(n^2)$) and time (\textit{i.e.}, $O(mn + nk^2)$) costs in \strap. Instead, in \cite{yin2019scalable}, the authors fix $\delta$ to a constant $10^{-5}$ and only retain PPR values greater than $\frac{\delta}{2}$, which compromises result utility.
Even so, \strap is still far more expensive than the proposed solution \nrp, as shown in our experiments.

Further, as explained in Section \ref{sec:intro}, conventional PPR is not an ideal proximity measure for the purpose of HNE due to the former's relative nature; this problem propagates to \strap which uses the PPR-based transpose proximity measure, \textit{i.e.}, $\pi(u,v)+\pi(v,u)$ for each node pair $u, v \in G$. For instance, in the example of Table \ref{tbl:over-ppr}, we have $\pi(v_7,v_9)+\pi(v_9,v_7)=0.288 > \pi(v_2,v_4)+\pi(v_4,v_2)=0.236$, indicating that \strap also tends to predict $(v_9, v_7)$ over $(v_2, v_4)$ in link prediction, which is counter-intuitive as we have explained in Section \ref{sec:intro}.



\header
{\bf Other HNE methods.} There also exist several techniques that generate embeddings without random walks, neural networks or matrix factorization. In particular, \nethiex \cite{ma2018hierarchical} applies expectation maximization to learn embeddings that capture the neighborhood structure of each node, as well as the latent hierarchical taxonomy in the graph. \rare \cite{RaRE18} considers both the proximity and popularity of nodes, and derive embeddings by maximizing a posteriori estimation using stochastic gradient descent. \graphwave \cite{donnat2018} represents each node’s neighborhood via a low-dimensional embedding by leveraging heat wavelet diffusion patterns, so as to capture structural roles of nodes in networks. \nodehash \cite{wang2018feature} transplants the feature hashing technique for word embeddings to embed nodes in networks. A common problem with the above methods is that they do not aim to preserve proximity information between nodes; consequently, they are generally less effective for tasks such as link prediction and graph reconstruction, as demonstrated in our experiments in Section \ref{sec:exp}.

\section{Scalable PPR Computation and Factorization}\label{sec:over}
This section presents \baseline, a simple and effective baseline approach to HNE that obtains node embeddings through factorizing a conceptual approximate PPR proximity matrix. Unlike previous methods, \baseline scales to billion-edge graphs without seriously compromising result quality. \baseline forms the foundation of the our main proposal \nrp, presented in Section \ref{sec:algo}. In what follows, Section \ref{sec:over-basic} overviews \baseline and formally defines the main concepts. Section \ref{sec:algo-ppr} presents the main contribution in \baseline: a scalable approximate PPR factorization algorithm. Table \ref{tbl:notations} summarizes frequent notations used throughout the paper.

\begin{table}[!t]
\centering
\renewcommand{\arraystretch}{1.1}
\caption{Frequently used notations.} \label{tbl:notations}
\begin{small}
	\begin{tabular}{|p{0.5in}|p{2.45in}|}
		\hline
		{\bf Notation} &  {\bf Description}\\
		\hline
		$G$=$(V,E)$   & A graph $G$ with node set $V$ and edge set $E$\\
		\hline
		$n, m$   & The numbers of nodes and edges in $G$, respectively\\
		\hline
		$d_{in}(v_i)$   & The in-degree of node $v_i$ \\
		\hline
		$d_{out}(v_i)$   & The out-degree of node $v_i$ \\
		\hline
		$\AM, \DM, \PM$   & The adjacency, out-degree and transition matrices of $G$ \\
		\hline
		$\alpha$ & The random walk decay factor \\
		\hline
		$k$   & The dimensionality of the embedding vectors \\
		\hline
		$\XM$, $\YM$   & The forward and backward embeddings, respectively \\
		\hline
		$\wra_v$, $\wla_v$ & The forward and backward weights for $v$'s forward and backward embeddings, respectively\\
		\hline
	\end{tabular}
\end{small}
\vspace{-4mm}
\end{table}

\subsection{Overview} \label{sec:over-basic}
As mentioned in Section \ref{sec:intro}, given an input graph $G = (V, E)$, the goal of HNE is to construct a size-$k$ embedding for each node $v \in G$, where $k$ is a user-specified per-node space budget. The input graph $G$ can be either directed or undirected. For simplicity, in the following we assume that $G$ is directed; for an undirected graph, we simply replace each undirected edge ($u, v$) with two directed ones with opposing direction, \textit{i.e.}, ($u, v$) and ($v, u$). Note that the capability to handle directed graphs is an advantage of our methods, compared to existing solutions that are limited to undirected graphs, {\it e.g.}, \cite{AROPE18,RandNE18,qiu2019netsmf,ma2018hierarchical,zhang2019prone}.

In a directed graph, each node plays two roles: as the incoming end and outgoing end of edges, respectively. These two roles can have very different semantics. For instance, in a social graph, a user can deliberately choose to \textit{follow} users who he/she is interested in, and is \textit{followed} by users who are interested in him/her. The follower relationships and followee relationships of the same user should have different representations.
This motivates building two separate embedding vectors $\XM_v$ and $\YM_v$ for each node $v$, referred to as the {\it forward} and {\it backward} embeddings of $v$, respectively. In our solutions, we assign equal space budget (\textit{i.e.}, $\frac{k}{2}$) to $\XM_v$ and $\YM_v$.


One advantage of \baseline is that, it uses the PPR proximity matrix to do the factorization, without actually matrializing the matrix. Specifically, the definition of PPR is based on random walks, as follows. Suppose that we start a random walk from a source node $u$. At each step, we terminate the walk with probability $\alpha$, and continue the walk ({\it i.e.}, moving on to a random out-neighbor of $u$) with probability $1-\alpha$. Then, for each node $v \in G$, we define its PPR  $\pi(u, v)$ with respect to  source node $u$ as the probability that the walk originating from $u$ terminates at $v$. 

Formally, let $\PiM$ be an $n \times n$ matrix where $\PiM[i, j] = \pi(v_i, v_j)$ for the $i$-th node $v_i$ and $j$-th node $v_j$ in $G$, and $\PM$ be the probability transition matrix of $G$, \textit{i.e.}, $\PM[i, j] = \frac{1}{d_{out}(v_i)}$, where $v_j$ is an out-neighbor of $v_i$ and $d_{out}(v_i)$ denotes the out-degree of $v_i$. Then,
\begin{equation} \label{eq:over-ppr}
\textstyle \PiM = \sum_{i=0}^{\infty}{\alpha(1-\alpha)^i\cdot\PM^i}.
\end{equation}


\baseline directly uses $\PiM$ as the proximity matrix, \textit{i.e.}, $\mathbf{M}=\PiM$. The goal is then to factorize $\PiM$ into the forward and backward embeddings of nodes of the input graph $G$, such that for each pair of nodes $u$ and $v$, \textit{i.e.}:
\begin{equation} \label{eq:baseline-objective}
\XM_u\YM_v^{\top}\approx \pi(u, v)  
\end{equation}

\header
\textbf{Remark.} Note that directly computing $\PiM$ (and subsequently factorizing it into the node embeddings $\XM$ and $\YM$) is infeasible for a large graph. In particular, $\PiM$ is a dense matrix that requires $O(n^2)$ space for $n$ nodes, and Eq.~\eqref{eq:over-ppr} involves summing up an infinite series. To alleviate this problem, we could apply an approximate PPR algorithm \cite{HubPPR16,FORA17,TopPPR18} to compute the top-$L$ largest PPR values for each node in $G$, which reduces the space overhead to $O(nL)$. Unfortunately, even the state-of-the-art approximate top-$L$ PPR algorithm, {\it i.e.}, \topppr, is insufficient for our purpose. Specifically, \topppr takes $O\left(\frac{L^{\frac{1}{4}}n^{\frac{3}{4}}\log{n}}{\sqrt{gap_{\rho}}}\right)$ time to compute the approximate top-$L$ PPR values for each node, where $gap_{\rho} \leq 1$ is a parameter that quantifies the difference between the top-$L$ and non-top-$L$ PPR values \cite{TopPPR18}. To approximate the entire $\PiM$, we would need to invoke \topppr for every node, which incurs time super-quadratic to $n$. Empirically, Ref. \cite{TopPPR18} reports that running \topppr on a billion-edge Twitter graph (used in our experiments as well) takes about 15 seconds CPU time, for $L=500$. The same graph contains over 41 million nodes, meaning that running \topppr on each of them would cost over 19 years of CPU time, which is infeasible even for a powerful computing cluster. While it is theoretically possible to reduce computational costs by choosing a small $L$ and/or a large error threshold in \topppr, doing so would lead to numerous zeros in $\PiM$, which seriously degrades the result quality. We address this challenge in the next subsection with a simple and effective solution.

\subsection{PPR Approximation} \label{sec:algo-ppr}

Observe that our goal is to obtain the node embeddings $\XM$ and $\YM$, rather than the PPR matrix $\PiM$ itself. Thus, the main idea of \baseline is to integrate the computation and factorization of $\PiM$ in the same iterative algorithm. Specifically, according to Eq. \eqref{eq:over-ppr}, $\PiM$ can be viewed as the weighted sum of proximity values of different orders, \textit{i.e.}, one-hop proxmity, two-hop proximity, {\it etc}. Therefore, instead of first computing $\PiM$ and then factorizing this {\em dense} matrix into node embeddings, we can instead start by factorizing the {\em sparse} first-order proximity matrix ({\it i.e.}, $\PM$) into the initial embeddings $\XM$ and $\YM$, and then iteratively refine $\XM$ and $\YM$, thereby incorporating higher-order information into them.  This allows us to avoid the substantial space and computational overheads incurred for the construction and factorization of the $n\times n$ dense matrix $\PiM$.

First, we consider a {\it truncated} version of $\PiM$ as follows:
\begin{equation} \label{eq:algo-ppr}
\textstyle \PiMp =  \sum_{i=1}^{\ell_1}{\alpha(1-\alpha)^i\cdot\PM^{i}},
\end{equation}
where $\ell_1$ is a relative large constant ({\it e.g.}, $\ell_1 = 20$). In other words, we set
$$\textstyle \PiMp =  \PiM - \alpha \IM - \left(\sum_{i=\ell_1 + 1}^{+\infty}{\alpha(1-\alpha)^i\cdot\PM^{i}}\right),$$
where $\IM$ denotes an $n \times n$ identity matrix. The rationale is that when $i$ is sufficiently large, $\alpha(1-\alpha)^i$ is small, in which case $\sum_{i=\ell_1 + 1}^{+\infty}{\alpha(1-\alpha)^i\cdot\PM^{i}}$ becomes negligible. In addition, $\alpha \IM$ only affects the PPR $\pi(u, u)$ from each node $u$ to itself, which has no impact on our objective in Eq.~\eqref{eq:baseline-objective} since we only concern the PPR values between different nodes.

To decompose $\PiMp$, observe that
\begin{equation*} \label{eq:algo-pimp}
\textstyle \PiMp =  \left(\sum_{i=1}^{\ell_1}{\alpha(1-\alpha)^i\cdot\PM^{i-1}}\right)\DM^{-1} \AM,
\end{equation*}
where $\AM$ is the adjacency matrix of $G$, and $\DM$ is an $n \times n$ diagonal matrix where the $i$-th diagonal element is $d_{out}(v_i)$. Instead of applying exact singular value decomposition (SVD) that is very time consuming, we factorize $\AM$ using the \rsvd algorithm in \cite{musco2015randomized} for randomized SVD, obtaining two $n \times \kp$ matrices $\UM, \VM$ and a $\kp \times \kp$ diagonal matrix $\mathbf{\Sigma}$ given inputs $\AM$ and $\kp$, such that $\UM \mathbf{\Sigma} \VM^\top \approx \AM$.
In short, \rsvd reduces $\AM$ to a low-dimensional space by the Gaussian random projection and then performs SVD on the low-dimensional matrix. Given a relative error threshold $\epsilon$, \rsvd guarantees a $(1+\epsilon)$ error bound for spectral norm low-rank approximation, which is much tighter than the theoretical accuracy bounds provided by previous truncated SVD algorithms \cite{halko2011finding,sarlos2006improved,clarkson2017low}.

Given the output $\UM, \mathbf{\Sigma}, \VM$ from \rsvd, we set
$$\XM_1 = \DM^{-1} \UM \sqrt{\mathbf{\Sigma}} \;\textrm{ and }\; \YM = \VM \sqrt{\mathbf{\Sigma}}.$$
After that, we compute 
$${\XM_i = (1 - \alpha)\PM \XM_{i-1} + \XM_1} \ \textrm{for}\ i = 2, \ldots, \ell_1,$$
and set ${\XM = \alpha(1 - \alpha) \XM_{\ell_1}}$. This results in
\begin{align*}
&\textstyle \XM=\sum_{i=1}^{\ell_1}{\alpha(1-\alpha)^i\PM^{i-1}\XM_1}\ \textrm{and}\\
&\textstyle \XM\YM^{\top}=\sum_{i=1}^{\ell_1}{\alpha(1-\alpha)^{i}\PM^{i-1}\cdot \XM_1\YM^{\top}}
\end{align*}
Note that $\XM_1\YM^{\top}\approx\DM^{-1}\AM=\PM$. It can be verified that $\XM\YM^{\top}\approx \PiMp.$
Particularly, the following theorem establishes the accuracy guarantees of \approxppr. 
\begin{theorem}\label{thrm:ppr}
Given $\AM,\DM^{-1},\PM$, the dimensionality $\kp$, the random walk decay factor $\alpha$, the number of iterations $\ell_{1}$ and error threshold $\epsilon$ for \rsvd as inputs to Algorithm \ref{alg:ApproxPPR}, it returns embedding matrices $\XM$ and $\YM$ ($\XM, \YM \in \mathbb{R}^{n\times \kp}$ ) that satisfy, for every pair of nodes $(u,v)\in V\times V$ with $u\neq v$,
\begin{align*}
&\left|\PiM[u,v]-(\XM\YM^{\top})[u,v]\right|\\
&\le (1+\epsilon)\sigma_{\kp+1}(1-\alpha)(1-(1-\alpha)^{\ell_1})+(1-\alpha)^{\ell_1+1},
\end{align*}
and for every node $u\in V$,
\begin{align*}
&\sum_{v\in V}{\left|\PiM[u,v]-(\XM\YM^{\top})[u,v]\right|}\\
&\le \sqrt{n}(1+\epsilon)\sigma_{\kp+1}(1-\alpha)(1-(1-\alpha)^{\ell_1})+(1-\alpha)^{\ell_1+1},
\end{align*}
where $\sigma_{\kp+1}$ is the $(\kp+1)$-th largest singular value of $\AM$.
\begin{proof}
See \proofref for the proof.
\end{proof}
\end{theorem}
Theorem \ref{thrm:ppr} indicates that the PPR value between any pair of nodes preserved in the embedding vectors $\XM$ and $\YM$ has absolute error at most $(1+\epsilon)\sigma_{\kp+1}(1-\alpha)(1-(1-\alpha)^{\ell_1})+(1-\alpha)^{\ell_1+1}$ and average absolute error of $\frac{1}{\sqrt{n}}(1+\epsilon)\sigma_{\kp+1}(1-\alpha)(1-(1-\alpha)^{\ell_1})+\frac{1}{n}(1-\alpha)^{\ell_1+1}$. Observe that the accuracy of the preserved PPR is restricted by $\epsilon$ and $\sigma_{\kp+1}$, namely the accuracy of the low-rank approximation, {\it i.e.}, \rsvd. 

Finally, we use $\XM_v$ and $\YM_v$ as the initial forward and backward embeddings, respectively, for each node $v$. Algorithm~\ref{alg:ApproxPPR} summarizes the pseudo-code for this construction of $\XM$ and $\YM$. Next, we present a concrete example.


\begin{algorithm}[!t]
	\begin{small}
		\caption{\baseline}
		\label{alg:ApproxPPR}
		\KwIn{$\AM$, $\DM^{-1}$, $\PM$, $\alpha$, $\kp$, $\ell_1$, $\epsilon$.}
		\KwOut{$\XM$, $\YM$.}
		$[\UM,\mathbf{\Sigma},\VM] \gets \mathsf{BKSVD}(\AM, \kp, \epsilon)$\;
		$\XM_1 \gets \DM^{-1}\UM\sqrt{\mathbf{\Sigma}}, \quad \YM \gets \VM\sqrt{\mathbf{\Sigma}}$\;
		\For{$i \gets 2$ to $\ell_1$}{
			$\XM_i \gets (1 - \alpha)\PM \XM_{i-1} + \XM_1$\;
		}
		$\XM \gets \alpha(1-\alpha)\XM_{\ell_1}$\;
		\Return $\XM$, $\YM$\;
	\end{small}
\end{algorithm}

\begin{figure*}[!t]
\begin{equation*}
\footnotesize
\YM \! = \!
\begin{bmatrix}
\mathbf{Y}_{v_1}\\
\mathbf{Y}_{v_2}\\
\mathbf{Y}_{v_3}\\
\mathbf{Y}_{v_4}\\
\mathbf{Y}_{v_5}\\
\mathbf{Y}_{v_6}\\
\mathbf{Y}_{v_7}\\
\mathbf{Y}_{v_8}\\
\mathbf{Y}_{v_9}
\end{bmatrix}
\! = \!
\begin{bmatrix*}[r]
-0.652, & \!\!\!\!   0.243\\
-0.668, & \!\!\!\!  -0.359\\
-0.823, & \!\!\!\!  -0.142\\
-0.668, & \!\!\!\!  -0.359\\
-0.737, & \!\!\!\!  0.547\\
-0.314, & \!\!\!\!  -0.42\\
-0.105, & \!\!\!\!  0.633\\
-0.094, & \!\!\!\!  -0.225\\
-0.071, & \!\!\!\! 0.818  
\end{bmatrix*},\
\XM_1 \! = \!
\begin{bmatrix*}[r]
-0.217, & \!\!\!\!    -0.121\\
-0.223, & \!\!\!\!    0.091\\
-0.206, & \!\!\!\!    0.008\\
-0.223, & \!\!\!\!    0.091\\
-0.184, & \!\!\!\!    -0.13\\
-0.157, & \!\!\!\!    0.4\\
-0.083, & \!\!\!\!    -0.16\\
-0.047, & \!\!\!\!    0.481\\
-0.032, & \!\!\!\!    -0.034
\end{bmatrix*},
\cdots,
\XM \! = \!
\begin{bmatrix}
\mathbf{X}_{v_1}\\
\mathbf{X}_{v_2}\\
\mathbf{X}_{v_3}\\
\mathbf{X}_{v_4}\\
\mathbf{X}_{v_5}\\
\mathbf{X}_{v_6}\\
\mathbf{X}_{v_7}\\
\mathbf{X}_{v_8}\\
\mathbf{X}_{v_9}
\end{bmatrix}
\! = \!
\begin{bmatrix}
-0.182, & \!\!\!\!   -0.014\\
-0.18, & \!\!\!\!   0.004\\
-0.14, & \!\!\!\!   -0.002\\
-0.18, & \!\!\!\!   0.004\\
-0.13, & \!\!\!\!   -0.008\\
-0.182, & \!\!\!\!   0.075\\
-0.126, & \!\!\!\!   0.072\\
-0.092, & \!\!\!\!   0.141\\
-0.157, & \!\!\!\!   0.236\\
\end{bmatrix}
\end{equation*}
\vspace{0mm}
\caption{Illustration of Example~\ref{ex:algo-example1} for the \baseline algorithm.}\label{fig:algo-example1}
\vspace{-3mm}
\end{figure*}

\begin{example} \label{ex:algo-example1}
Given input graph $G$ in Fig.~\ref{fig:over-ppr} and input parameters $\kp=2$, $\alpha=0.15, \ell_1=20$, we run Algorithm \ref{alg:ApproxPPR} on $G$. It first applies \rsvd on the adjacency matrix $\AM \in \mathbb{R}^{9\times 9}$, which produces $\XM_1 \in \mathbb{R}^{9\times 2}$ and $\YM\in \mathbb{R}^{9\times 2}$ as shown in Fig.~\ref{fig:algo-example1}.

\baseline  first sets $\XM=\XM_1$. Then, in each of the following iterations, the algorithm updates $\XM$ to $0.85\cdot\PM\XM+\XM_1$.
After repeating this process for $\ell_1-1=19$ iterations, \approxppr scales $\XM$ by the weight $\alpha(1-\alpha)=0.1275$ and returns us $\XM$ and $\YM$ as in Fig.~\ref{fig:algo-example1}. The inner product between $\XM_u$ and $\YM_v$ approximates $\pi(u,v)$. For example, consider node pairs $\langle v_2, v_4\rangle$ and $\langle v_9, v_7\rangle$:
\begin{align*}
& \XM_{v_2}\YM_{v_4}^{\top}=[-0.18, 0.004]\cdot [-0.668, -0.359]^{\top}=0.119,\\
& \XM_{v_9}\YM_{v_7}^{\top}=[-0.157, 0.236]\cdot [-0.105, 0.633]^{\top}=0.166,
\end{align*}
which are close to $\pi(v_2,v_4)$ and $\pi(v_9,v_7)$ in Table \ref{tbl:over-ppr} respectively. \done
\end{example}

\header
{\bf Time Complexity.} By the analysis in Ref. \cite{musco2015randomized}, applying \rsvd on $\AM$ requires $O\left((m\kp+n{\kp}^2)\frac{\log{n}}{\epsilon}\right)$ time, where $\epsilon$ is a constant that controls the tradeoff between the efficiency and accuracy of SVD. In addition, Lines 2, 4, and 5 in Algorithm~\ref{alg:ApproxPPR} respectively run in $O(m \kp)$ time. Therefore, the overall time complexity of Algorithm~\ref{alg:ApproxPPR} is 
$$O\left(\left(\frac{\log{n}}{\epsilon} + \ell_1\right) m \kp + \frac{\log{n}}{\epsilon} n{\kp}^2 \right),$$ 
which equals $O\left(k (m+kn) \log{n}\right)$ when $\epsilon$ and $\ell_1$ are regarded as constants.


\section{Proposed \nrpb Algorithm}\label{sec:algo}
The \baseline algorithm presented in the previous section directly uses PPR as the proximity measure. However, as explained in Section \ref{sec:intro}, PPR by itself is not suitable for our purpose since it is a local measure, in the sense that PPR values are \textit{relative} with respect to the source node. Consequently, PPR values for different source nodes are essentially incomparable, which is the root cause of the counter-intuitive observation in the example of Fig.~\ref{fig:over-ppr} and Table \ref{tbl:over-ppr}.

In the proposed algorithm \nrp, we address the deficiency of PPR through a technique that we call \textit{node reweighting}. Specifically, for any two nodes $u$ and $v$, we aim to find forward and backward embeddings such that:
\begin{equation}\label{eq:nrp}
\XM_u \YM_v^\top \approx \wra_u \cdot \pi(u, v) \cdot \wla_v
\end{equation}
where $\pi(u, v)$ is the PPR value of $v$ with respect to node $u$ as source, $\wra_u$ and $\wla_v$ are weights assigned to $u$ and $v$, respectively. In other words, we let $\XM_u^\top \YM_v$ preserve a scaled version of $\pi(u, v)$. The goal of \nrp is then to find approximate node weights so that Eq. \eqref{eq:nrp} properly expresses the proximity between nodes $u$ and $v$. In \nrp, the node weights are learned  through an efficient optimization algorithm, described later in this section.
The proposed node reweighting overcomes the deficiency of PPR, which is confirmed in our experiments.

In the following, Section \ref{sec:over-weight} explains the choice of node weights in \nrp. Sections \ref{sec:algo-weight} and \ref{sec:algo-speedup} elaborate on the computation of node weights. Section~\ref{subsec:algo-all} summarizes the complete \nrp algorithm.

\subsection{Choice of Node Weights} \label{sec:over-weight}
As discussed before, the problem of PPR as a proximity measure is that it is a relative measure with respect to the source node. In particular, the PPR value does not take into account the number of out-going and in-coming edges that each node has. To address this issue, \nrp assigns to each node $u$ a {\it forward weight} $\wra_u$ and a {\it backward weight} $\wla_u$, and uses $\wra_u \cdot \pi(u, v) \cdot \wla_v$ instead of $\pi(u, v)$ to gauge the strength of connection from $u$ to $v$, as in Eq. \eqref{eq:nrp}. To compensate for the lack of node degree data in PPR values, we choose the forward and backward weights such that
\begin{equation} \label{eq:over-weight}
\begin{split}
&\textrm{$\forall{u}\in V$, } \sum_{\forall{v\in V\setminus u}} \left(\wra_u \cdot \pi(u, v) \cdot \wla_v\right) \approx d_{out}(u), \textrm{ and} \\
&\textrm{$\forall{v}\in V$, } \sum_{\forall{u\in V\setminus v}} \left(\wra_u \cdot \pi(u, v) \cdot \wla_v\right) \approx d_{in}(v).
\end{split}
\end{equation}
In other words, for any nodes $u, v \in G$, we aim to ensure that (i) the ``total strength'' of connections from $u$ to other nodes is roughly equal to the out-degree $d_{out}(u)$ of $u$, and (ii) the total strength of connection from other nodes to $v$ is roughly to equal the in-degree $d_{in}(v)$ of $v$. The rationale is that if $u$ has a large out-degree, then it is more likely to be connected to other nodes, and hence, the proximity from $u$ to other nodes should be scaled up accordingly. The case for a node $v$ with a large in-degree is similar. 
In Section~\ref{sec:exp}, we empirically show that this scaling approach significantly improves the effectiveness of our embeddings for not just link prediction but also other important graph analysis task such as graph reconstruction.

\subsection{Learning Node Weights}\label{sec:algo-weight}
Given the output $\XM$ and $\YM$ of \baseline (Algorithm~\ref{alg:ApproxPPR}), we use $\XM_v \YM_v^\top$ as an approximation of $\pi(u, v)$ for any two different nodes $u$ and $v$. Then we formulate an objective function $O$ for tuning node weights according to Eq.~\eqref{eq:over-weight}:
\begin{align}\label{eq:algo-obj}
O=\underset{\wra,\wla}{\min} & \sum_{v}\left\|\sum_{u\ne v} \left(\wra_u  \XM_u\YM_v^\top  \wla_{v}\right) - d_{in}(v)\right\|_{2} \nonumber \\
&+\sum_{u}\left\|\sum_{v \ne u} \left(\wra_u \XM_u\YM_v^\top \wla_{v}\right) - d_{out}(u)\right\|_{2} \\
&+\lambda \sum_{u} \left(\left\|\wra_u\right\|_{2} + \left\|\wla_u\right\|_{2}\right), \nonumber \\
& \!\!\!\!\!\!\!\!\!\! \text{subject to } \forall u\in V, \wra_u, \wla_u \ge \frac{1}{n}. \nonumber
\end{align}
To explain, recall that we use $\wra_u  \XM_u\YM_v^\top  \wla_{v}$ to quantify the strength of connection from $u$ to $v$, and hence, for any fixed $u$ (resp.\ $v$), $\sum_{u\ne v} \left(\wra_u  \XM_u\YM_v^\top  \wla_{v}\right)$ measures the total strength of connections from $u$ to other nodes (resp.\ from other nodes to $v$). Therefore, by minimizing $O$, we aim to ensure that the total strength of connections starting from (resp.\ ending at) each node $u$ is close to $u$'s out-degree (resp.\ in-degree), subject to a regularization term $\lambda \sum_{u} \left(\|\wra_u\|_{2} + \|\wla_u\|_{2}\right)$. In addition, we require that $\wra_u, \wla_u \ge \frac{1}{n}$ for all nodes $u$ to avoid negative node weights.

We derive an approximate solution for Eq.~\eqref{eq:algo-obj} using coordinate descent \cite{wright2015coordinate}: We start with an initial solution $\wra_v = d_{out}(v)$ and $\wla_v = 1$ for each node $v$, and then iteratively update each weight based on the other $2n-1$ weights. In particular, for any node $v^*$, the formula for updating $\wla_{v^*}$ is derived by taking the partial derivative of the objective function in Eq.~\eqref{eq:algo-obj} with respect to $\wla_{v^*}$:
\begin{align*}
\textstyle \frac{\partial O}{\partial \wla_{v^*}} & \textstyle= 2\Big[\left(\left(\sum_{u \ne v^*} \wra_u  \XM_{u}\right)\YM_{v^*}^\top\right)^2\wla_{v^*}\\
&\textstyle \ \ \ \ -d_{in}(v^*)\left(\sum_{u \ne v^*} \wra_u  \XM_{u}\right)\YM_{v^*}^\top\\
&\textstyle \ \ \ \ +\sum_{u}\left(\sum_{v\ne u,v\ne v^*} \wra_{u}  \XM_{u}\YM_{v}^\top \wla_{v}\right)\wra_{u} \XM_{u}\YM_{v^*}^\top\\
&\textstyle \ \ \ \ +\sum_{u \ne v^*}\left(\wra_{u}  \XM_{u}\YM_{v^*}^\top\right)^2\wla_{v^*}\\
&\textstyle \ \ \ \ -\left(\sum_{u}d_{out}(u)\wra_{u}  \XM_{u}\right)\YM_{v^*}^\top + \lambda\wla_{v^*}\Big]\\
&=2(a_3-a_2-a_1)+2(b_1+b_2+\lambda)\wla_{v^*},
\end{align*}
where
\begin{align}\label{eq:algo-back-update}
a_1=&\textstyle \left(\sum_{u}d_{out}(u)\wra_{u}  \XM_{u}\right)\YM_{v^*}^\top, \nonumber \\
a_2=&\textstyle d_{in}(v^*)\left(\sum_{u \ne v^*} \wra_u  \XM_{u}\right)\YM_{v^*}^\top, \nonumber \\
a_3=&\textstyle \sum_{u}\left(\sum_{v\ne u,v\ne v^*} \wra_{u}  \XM_{u}\YM_{v}^\top \wla_{v}\right)\wra_{u}  \XM_{u}\YM_{v^*}^\top,  \\
b_1=&\textstyle \sum_{u \ne v^*}\left(\wra_{u}  \XM_{u}\YM_{v^*}^\top\right)^2, \nonumber \\
b_2=&\textstyle \left(\left(\sum_{u \ne v^*} \wra_u  \XM_{u}\right)\YM_{v^*}^\top\right)^2. \nonumber
\end{align}

We identify the value of $\wla_{v^*}$ that renders the above partial derivative zero, {\it i.e.}, $\frac{\partial O}{\partial \wla_{v^*}}=0$. If the identified $\wla_{v^*}$ is smaller than $\frac{1}{n}$, then we set it to $\frac{1}{n}$ instead to avoid negativity. This leads to the following formula for updating backward weight $\wla_{v^*}$:
\begin{equation}\label{eq:back-weight-update}
 \wla_{v^*} = \max\left\{\frac{1}{n}, \frac{a_1+a_2-a_3}{b_1+b_2 + \lambda}\right\}
\end{equation}
The formula for updating $\wra_{u^*}$ is similar and included in \fwdref for brevity.

By Eq.~\eqref{eq:back-weight-update}, each update of $\wla_{v^*}$ requires computing $a_1$, $a_2$, $a_3$, $b_1$ and $b_2$. Towards this end, a straightforward approach is to compute these variables directly based on their definitions in Eq.~\eqref{eq:algo-back-update}. This, however, leads to tremendous overheads. In particular, computing $a_1$, $a_2$, and $b_2$ requires a linear scan of $\XM_u$ for each node $u$, which requires $O(n\kp)$ time. Deriving $b_1$ requires computing $\wra_{u}\XM_u \YM^\top_{v^*}$ for each node $u$, which incurs $O(n{\kp}^2)$ overhead. Furthermore, computing $b_3$ requires calculating $\wra_{u}\XM_u \YM^\top_{v} \wla_{v}$ for all $u \ne v \ne v^*$, which takes $O(n^2 {\kp}^2)$ time. Therefore, each update of $\wla_{v^*}$ takes $O(n^2{\kp}^2)$, which leads to a total overhead of $O(n^3 {\kp}^2)$ for updating all $\wla_{v^*}$ once. Apparently, this overhead is prohibitive for large graphs. To address this deficiency, in Section~\ref{sec:algo-speedup}, we present a solution that reduces the overhead to $O(n {\kp}^2)$ instead of $O(n^3{\kp}^2)$.


\subsection{Accelerating Weight Updates}\label{sec:algo-speedup}

We observe that the updates of different node weights share a large amount of common computation. For example, for any node $v^*$, deriving $a_1$ always requires computing $\sum_{u}d_{out}(u)\wra_{u}  \XM_{u}$. Intuitively, if we are able to reuse the result of such common computation for different nodes, then the overheads of our coordinate descent algorithm could be significantly reduced. In what follows, we elaborate how we exploit this idea to accelerate the derivation of $a_1, a_2, a_3, b_1$, and $b_2$.


\header
{\bf Computation of $\mathbf{a_1,a_2,b_2}$.}
By the definitions of $a_1, a_2, b_2$ in Eq.~\eqref{eq:algo-back-update},
\begin{align}\label{eq:algo-a1a2b2}
 & a_1=\boldsymbol{\xi} \YM_{v^*}^\top, \; a_2=d_{in}(v^*) (\boldsymbol{\chi}-\wra_{v^*}  \XM_{v^*}) \YM_{v^*}^\top, \nonumber \\
 & \text{and }  b_2=\left(\left(\boldsymbol{\chi}-\wra_{v^*}  \XM_{v^*}\right) \YM_{v^*}^\top\right)^2,  \\
 & \text{where } \boldsymbol{\xi}=\sum_{u}d_{out}(u)\wra_{u}  \XM_{u}, \; \text{ and }\boldsymbol{\chi}=\sum_{u} \wra_u  \XM_{u}. \nonumber
\end{align}
Eq.~\eqref{eq:algo-a1a2b2} indicates that the $a_1$ values of all nodes $v^*\in V$ share a common $\boldsymbol{\xi}$, while $a_2$ and $b_2$ of each node $v^*$ have $\boldsymbol{\chi}$ in common. Observe that both $\boldsymbol{\xi}$ and $\boldsymbol{\chi}$ are independent of any backward weight. Motivated by this, we propose to first compute $\boldsymbol{\xi}\in \mathbb{R}^{1 \times \kp}$ and $\boldsymbol{\chi} \in \mathbb{R}^{1 \times \kp}$, which takes $O(n \kp)$ time. After that, we can easily derive $a_1, a_2$, and $b_2$ for any node with precomputed $\boldsymbol{\xi}$ and $\boldsymbol{\chi}$.  
In that case, each update of $a_1$, $a_2$, and $b_2$ takes only $O(\kp)$ time, due to Eq.~\eqref{eq:algo-a1a2b2}. This leads to $O(n\kp)$ (instead of $O(n^2 \kp)$) total computation time of $a_1$, $a_2$, and $b_2$ for all nodes.

\header
{\bf Computation of $\mathbf{a_3}$.} Note that
\begin{align*}
a_3&= \sum_{u}\left(\sum_{v} \wra_{u}  \XM_{u}\YM_{v}^\top \wla_{v}\right)\wra_{u}  \XM_{u}\YM_{v^*}^\top\\
&\ \ \ \ -\sum_{u}\left( \wra_{u}  \XM_{u}\YM_{v^*}^\top \wla_{v^*}\right)\wra_{u}  \XM_{u}\YM_{v^*}^\top\\
&\ \ \ \ -\sum_{v}\left( \wra_{v}  \XM_{v}\YM_{v}^\top \wla_{v}\right)\wra_{v}  \XM_{v}\YM_{v^*}^\top\\
& \ \ \ \ +\left( \wra_{v^*}  \XM_{v^*}\YM_{v^*}^\top \wla_{v^*}\right)\wra_{v^*}  \XM_{v^*}\YM_{v^*}^\top,
\end{align*}
which can be rewritten as:
\begin{align}\label{eq:algo-a3}
a_3=&\boldsymbol{\rho}_1 \mathbf{\Lambda}\YM_{v^*}^\top - \wla_{v^*} \YM_{v^*}\mathbf{\Lambda}\YM_{v^*}^\top - \boldsymbol{\rho}_2\YM_{v^*}^\top \nonumber \\
& + \wla_{v^*}  \left(\XM_{v^*}\YM_{v^*}^\top\right)^{2} \wra_{v^*}^2, \nonumber \\
&\text{where } \mathbf{\Lambda}=\sum_{u} \wra_{u}^2 (\XM_{u}^\top  \XM_{u}), \; \boldsymbol{\rho}_1=\sum_{v}\wla_{v} \YM_{v},  \\
&\text{and } \boldsymbol{\rho}_2=\sum_{v} \left(\wra_{v}^2 \cdot \wla_{v}  \left(\XM_v\YM_v^\top\right)  \XM_{v}\right). \nonumber
\end{align}

Observe that $\mathbf{\Lambda}$ is independent of any backward weight. Thus, it can be computed once and reused in the computation of $a_3$ for all nodes. Meanwhile, both $\boldsymbol{\rho}_1$ and $\boldsymbol{\rho}_2$ are dependent on all of the backward weights, and hence, cannot be directly reused if we are to update each backward weight in turn. However, we note that $\boldsymbol{\rho}_1$ and $\boldsymbol{\rho}_2$ can be {\it incrementally} updated after the change of any single backward weight. Specifically, suppose that we have computed $\boldsymbol{\rho}_1$ and $\boldsymbol{\rho}_2$ based on Eq.~\eqref{eq:algo-a3}, and then we change the backward weight of $v^*$ from $\wla_{v^*}^\prime$ as $\wla_{v^*}$. In that case, we can update $\boldsymbol{\rho}_1$ and $\boldsymbol{\rho}_2$ as:
%
\begin{equation}\label{eq:update-rho1-rho2}
\begin{split}
& \boldsymbol{\rho}_1 = \boldsymbol{\rho}_1+ \left(\wla_{v^*} - \wla_{v^*}^\prime\right)\YM_{v^*},\\
& \boldsymbol{\rho}_2 = \boldsymbol{\rho}_2+ \left(\wla_{v^*} - \wla_{v^*}^\prime\right)\wra_{v^*}^2  \left(\XM_{v^*}\YM_{v^*}^\top\right)  \XM_{v^*}.
\end{split}
\end{equation}
Since $\wla_{v^*}, \wla_{v^*}^\prime \in \mathbb{R}$ and $\XM_{v^*}, \YM_{v^*} \in \mathbb{R}^{1 \times \kp}$, each of such updates takes only $O(\kp)$ time.

The initial values of $\boldsymbol{\rho}_1$ and $\boldsymbol{\rho}_2$ can be computed in $O(n\kp)$ time based on Eq.~\eqref{eq:algo-a3}, while $\mathbf{\Lambda}$ can be calculated in $O(nk^{\prime 2})$ time. Given $\mathbf{\Lambda}$, $\boldsymbol{\rho}_1$, and $\boldsymbol{\rho}_2$, we can compute $a_3$ for any node $v^*$ in $O(k^{\prime 2})$ time based on Eq.~\eqref{eq:algo-a3}. Therefore, the total time required for computing $a_3$ for all nodes is $O(nk^{\prime 2})$, which is an significant reduction from the $O(n^3k^{\prime 2})$ time required by the naive solution described in Section~\ref{sec:algo-weight}.

\header
{\bf Approximation of $\mathbf{b_1}$.} We observe that the value of $b_1$ is insignificant compared to $b_2$. Thus, we propose to approximate its value instead of deriving it exactly, so as to reduce computation cost. By the inequality of arithmetic and geometric means, we have:
\begin{equation}\label{eq:algo-b1}
\textstyle \frac{1}{\kp} b_1 \le \sum_{u\ne v^*}\wra_{u}^2 (\sum_{r=1}^{k^\prime}\XM_u[r]^2\YM_{v^*}[r]^2) \le  b_1.
\end{equation}
Let $\boldsymbol{\phi}$ be a length-$\kp$ vector where the $r$-th ($r \in [1, \kp]$) element is
\begin{equation}\label{eq:algo-phi}
\textstyle \boldsymbol{\phi}[r] = \sum_{u}\wra_{u}^2 \XM_u[r]^2.
\end{equation}
We compute $\boldsymbol{\phi}$ in $O(n \kp)$ time, and then, based on Eq.~\eqref{eq:algo-b1}, we approximate $b_1$ for each node in $O(\kp)$ time with
\begin{equation}\label{eq:algo-b1_2}
\textstyle b_1\approx  \frac{\kp}{2}\sum_{r=1}^{k^\prime}\YM_{v^*}[r]^2 (\boldsymbol{\phi}[r]-\wra_{v^*}^2 \XM_{v^*}[r]^2).
\end{equation}
Therefore, the total cost for approximating $b_1$ for all nodes is $O(n \kp)$.

\IncMargin{1ex}
\begin{algorithm}[t]
\begin{small}
	\caption{$\mathsf{update Bwd Weights}$}
	\label{alg:algo-updateBwdWeights}
	\BlankLine
	\KwIn{$G$, $\kp$, $\wrab$, $\wlab$, $\XM$, $\YM$.}
	\KwOut{$\wlab$}
	Compute $\boldsymbol{\xi}, \boldsymbol{\chi}, \boldsymbol{\rho_1}, \boldsymbol{\rho_2}, \mathbf{\Lambda}$, and $\mathbf{\Phi}$ based on Eq.~\eqref{eq:algo-a1a2b2}, \eqref{eq:algo-a3}, and \eqref{eq:algo-phi}\;
	\For{$r \gets 1$ to $\kp$}{
		$\boldsymbol{\phi}[r]=\sum_{u}\wra_{u}^2 \XM_{u}[r]^2$\; 
	}
	\For{$v^* \in V$ in random order}{
		Compute $a_1, a_2, a_3, b_1, b_2$ by Eq.~\eqref{eq:algo-a1a2b2}, \eqref{eq:algo-a3}, and \eqref{eq:algo-b1_2}\;
		$\wla_{v^*}^\prime= \wla_{v^*}$\;
		$\wla_{v^*}=\max\left\{\frac{1}{n}, \frac{a_1+a_2-a_3}{b_1+b_2 + \lambda}\right\}$\;
		$\boldsymbol{\rho}_1 = \boldsymbol{\rho}_1+ \left(\wla_{v^*} - \wla_{v^*}^\prime\right)\YM_{v^*}$\;
		$\boldsymbol{\rho}_2 = \boldsymbol{\rho}_2+ \left(\wla_{v^*} - \wla_{v^*}^\prime\right)\wra_{v^*}^2  \left(\XM_{v^*}\YM_{v^*}^\top\right)  \XM_{v^*}$
	}
	\Return $\wlab$\;
\end{small}
\end{algorithm}

\header
{\bf Summary.} As a summary, Algorithm~\ref{alg:algo-updateBwdWeights} presents the pseudo-code of our method for updating the backward weight of each node. The algorithm first computes $\boldsymbol{\xi}, \boldsymbol{\chi}, \boldsymbol{\rho_1}, \boldsymbol{\rho_2}, \mathbf{\Lambda}, \boldsymbol{\phi}$ in $O(n k^{\prime 2})$ time (Lines 1-3). After that, it examines each node's backward weight in random order, and computes $a_1, a_2, a_3, b_1, b_2$ by Eq.~\eqref{eq:algo-a1a2b2}, \eqref{eq:algo-a3}, and \eqref{eq:algo-b1_2},
which takes $O\left(k^{\prime 2}\right)$ time per node (Line 5). Given $a_1, a_2, a_3, b_1, b_2$, the algorithm updates the backward weight examined, and then updates $\boldsymbol{\rho_1}$ and $\boldsymbol{\rho_2}$ in $O(\kp)$ time (Lines 7-9). The total time complexity of Algorithm~\ref{alg:algo-updateBwdWeights} is $O(n k^{\prime 2})$, which is significantly better than the $O(n^3 k^{\prime 2})$-time method in Section~\ref{sec:algo-weight}. We illustrate Algorithm~\ref{alg:algo-updateBwdWeights} with an example.

\begin{figure}[!t]
\begin{align*}
&\boldsymbol{\xi} = [-8.1453,\   -7.6509],\ \boldsymbol{\chi}=[-3.5227,\   -3.2933],\\
&\boldsymbol{\rho}_1=[-4.2126,\   -3.7234],\ \boldsymbol{\rho}_2=[-1.2659,\   -1.1678],\\
&\boldsymbol{\Lambda}=\begin{bmatrix}
1.4478, &  1.3308\\
1.3308, &  1.2575
\end{bmatrix},\  \boldsymbol{\phi}=[1.4478,\    1.2575].
\end{align*}
\vspace{-3mm}
\caption{Illustration for Example~\ref{ex:algo-example2}}\label{fig:algo-example2}
\vspace{-2mm}
\end{figure}

\begin{example} \label{ex:algo-example2}
Suppose that we invoke Algorithm~\ref{alg:algo-updateBwdWeights} given graph $G$ in Fig.~\ref{fig:over-ppr}, $\kp=2$, $\XM$ and $\YM$ from Example~\ref{ex:algo-example1}, and the following $\wla$ and $\wra$:
\begin{align*}
\wla = \left[ 1, 1, 1, 1, 1, 1, 1, 1, 1 \right], \;\;\;
\wra = \left[ 3, 3, 4, 3, 4, 2, 2, 2, 1 \right].
\end{align*}


The algorithm first computes $\boldsymbol{\xi},\boldsymbol{\chi},\boldsymbol{\rho}_1,\boldsymbol{\rho}_2,\boldsymbol{\Lambda}$ and $\boldsymbol{\phi}$ according to Eq.~\eqref{eq:algo-a1a2b2}, \eqref{eq:algo-a3}, and \eqref{eq:algo-phi}. Fig.~\ref{fig:algo-example2} shows the results. 

Then, we update each backward weight in a random order with the above precomputed values. Let's pick $\wla_{v_1}$ for the first update. According to Eq.~\eqref{eq:algo-a1a2b2}, \eqref{eq:algo-a3} and \eqref{eq:algo-b1_2}, we do not need to perform summations over all $9$ nodes as in Eq.~\eqref{eq:algo-back-update} but some multiplications between a $2\times 2$ matrix and a length-$2$ vector, as well as inner products between length-$2$ vectors, yielding the following results fast:
\begin{equation*}
\begin{split}
& a_1=\boldsymbol{\xi}\YM^{\top}_{v_1}=7.7968, \\
& a_2= 2(\boldsymbol{\chi}-2\XM_{v_1}) \YM_{v_1}^\top=5.903,\\
& a_3=\boldsymbol{\rho}_1 \mathbf{\Lambda}\YM_{v_1}^\top -\YM_{v_1}\mathbf{\Lambda}\YM_{v_1}^\top - \boldsymbol{\rho}_2\YM_{v_1}^\top=8.1324,\\
& b_1=\sum_{r=1}^{2}\YM_{v_1}[r]^2 (\boldsymbol{\phi}[r]-\wra_{v_1}^2 \XM_{v_1}[r]^2)=0.9683,\\
& b_2=\left(\left(\boldsymbol{\chi}-2\XM_{v_1}\right) \YM_{v_1}^\top\right)^2=8.7113.
\end{split}
\end{equation*}
\vspace{-1mm}
Let $\lambda=0$. The backward weight for $v_1$ is updated as
$$ \wla_{v_1}=\max\left\{\frac{1}{9}, \frac{a_1+a_2-a_3}{b_1+b_2}\right\}=0.5752,$$
and then $\boldsymbol{\rho}_1$ and $\boldsymbol{\rho}_2$ are updated accordingly with the updated $\wla_{v_1}$ based on Eq.~\eqref{eq:update-rho1-rho2} before proceeding to the next backward weight.\done
%
\end{example}
\header\textbf{Remark.} The forward weights $\wra_{v^*}$ can be learned using an algorithm very similar to Algorithm~\ref{alg:algo-updateBwdWeights}, with the same space and time complexities.
For brevity, we include the details in \fwdref. 
\subsection{Complete \nrpb Algorithm and Analysis}\label{subsec:algo-all}
Algorithm~\ref{alg:nrp} presents the pseudo-code for constructing embeddings with \nrp. Given a graph $G$, embedding dimensionality $k$, random walk decay factor $\alpha$, thresholds $\ell_1,\ell_2$ and relative error threshold $\epsilon$, it first generates the initial embedding matrices $\XM$ and $\YM$ using Algorithm~\ref{alg:ApproxPPR} (Lines 1-2, see Section \ref{sec:algo-ppr} for details). After that, it initializes the forward and backward weights for each node (Lines 3-4) and then applies coordinate descent to refine the weights (Lines 5-7).
In particular, in each epoch of the coordinate descent, it first invokes Algorithm~\ref{alg:algo-updateBwdWeights} to update each backward weight once (Line 6), and then applies a similar algorithm to update each forward weight once (Line 7, see Algorithm $\mathsf{updateFwdWeights}$ in \fwdref). The total number of epochs is controlled by $\ell_2$, which we set to $O(\log n)$ for efficiency. After the coordinate descent terminates, \nrp multiplies the forward (resp.\ backward) embedding of each node by its forward (resp.\ backward) weight to obtain the final embeddings (Lines 8-9).

\begin{algorithm}[t]
\begin{small}
	\caption{\nrp}
	\label{alg:nrp}
	\BlankLine
	\KwIn{Graph $G$, embedding dimensionality $k$, thresholds $\ell_1,\ell_2$, random walk decay factor $\alpha$ and error  threshold $\epsilon$}
	\KwOut{Embedding matrices $\XM$ and $\YM$.}
	$k^\prime \gets k/2$\;
	$[\XM, \YM]\gets \mathsf{ApproxPPR}(\AM, \DM^{-1}, \PM, \alpha, k^\prime, \ell_1,\epsilon)$\;
    \For{$v \in V$}{
        $\wra_v = d_{out}(v), \; \wla_v = 1$\;
    }		
	\For{$l \gets 1$ to $\ell_2$}{
		$\wlab=\mathsf{updateBwdWeights}(G, k^\prime, \wrab, \wlab, \XM, \YM)$\;
		$\wrab=\mathsf{updateFwdWeights}(G, k^\prime, \wrab, \wlab, \XM, \YM)$\;
	}
	\For{$v \in V$}{
		$\XM_{v} = \wra_{v}\cdot \XM_{v}, \;\; \YM_{v} = \wla_{v}\cdot \YM_{v}$\;
	}
	\Return $\XM$, $\YM$\;
\end{small}
\end{algorithm}

\header
{\bf Complexity Analysis.}
\nrp has three main steps: Algorithm \ref{alg:ApproxPPR}, Algorithm \ref{alg:algo-updateBwdWeights}, and Algorithm $\mathsf{updateFwdWeights}$. 
By the analysis of time complexity in Section~\ref{sec:algo-ppr}, Algorithm \ref{alg:ApproxPPR} runs in $O\left(k (m+kn) \log{n}\right)$ time, and its space overhead is determined by the number of non-zero entries in the matrices, which is $O(m+nk)$. For Algorithm \ref{alg:algo-updateBwdWeights} and Algorithm $\mathsf{updateFwdWeights}$, each epoch takes $O(n k^{\prime 2})$ time, as analysed in Section~\ref{sec:algo-speedup}. Hence, the time complexities of Algorithm~\ref{alg:algo-updateBwdWeights} and Algorithm $\mathsf{updateFwdWeights}$ are both $O(n k^{\prime 2})$ when the number of epochs $\ell_2$ is a constant. In addition, the space costs of Algorithm \ref{alg:algo-updateBwdWeights} and Algorithm $\mathsf{updateFwdWeights}$ depend on the size of $\boldsymbol{\xi}, \boldsymbol{\chi}, \boldsymbol{\rho_1}, \boldsymbol{\rho_2}, \mathbf{\Lambda}, \boldsymbol{\phi}$ and the number of weights, which is bounded by $O(n\kp)$. As a result, the time complexity of Algorithm~\ref{alg:nrp} is $O\left( k(m+kn)\log n\right)$ and its space complexity is $O(m+nk)$.

\vspace{2mm}
\section{Experiments}\label{sec:exp}
We experimentally evaluate our proposed method, {\it i.e.}, \nrp, against 18 existing methods, including 4 classic ones and 14 recent ones, on three graph analysis tasks: link prediction, graph reconstruction, and node classification. We also study the efficiency of all methods and analyze the parameter choices of \nrp. All experiments are conducted using a single thread on a Linux machine powered by an Intel Xeon(R) E5-2650 v2@2.60GHz CPU and 96GB RAM.

\begin{table}[t]
\centering
\renewcommand{\arraystretch}{1.2}
\begin{small}
\caption{Dataset statistics ($K=10^3$, $M=10^6$, $B=10^9$).} \label{tbl:exp-data}
\vspace{1mm}
\begin{tabular}{|l|r|r|c|c|}
	\hline
	{\bf Name} & \multicolumn{1}{c|}{$|V|$} & \multicolumn{1}{c|}{$|E|$} & \multicolumn{1}{c|}{\bf Type}  & \multicolumn{1}{c|}{\bf \#labels}\\
	\hline
	{\em Wiki}      & 4.78K    &184.81K     & directed & 40 \\
	\hline
	{\em BlogCatalog}      & 10.31K    &333.98K     & undirected & 39 \\
	\hline
	{\em Youtube}   & 1.13M & 2.99M & undirected & 47 \\
	\hline
	{\em TWeibo}   & 2.32M & 50.65M & directed & 100 \\
	\hline
	{\em Orkut}   & 3.1M & 234M & undirected & 100 \\
	\hline
	{\em Twitter} & 41.6M & 1.2B & directed & - \\
	\hline
	{\em Friendster} & 65.6M & 1.8B & undirected & - \\
	\hline
\end{tabular}
\end{small}
\vspace{-1mm}
\end{table}

\subsection{Experimental Settings}\label{sec:exp-set}
\header
{\bf Datasets.} We experiment with seven real networks that are used in previous work~\cite{node2vec16,HOPE16,verse18}, including two billion-edge networks: directed network {\em Twitter} \cite{kwak2010twitter} and undirected network {\em Friendster} \cite{yang2015defining}. Table~\ref{tbl:exp-data} shows the dataset statistics. For \emph{Wiki, BlogCatalog, Youtube}, and \emph{Orkut}, we use the node labels suggested in previous work \cite{node2vec16,deepwalk14,verse18}. For {\em TWeibo}, we collect its node tags from \cite{kddcup2012tweibo} and only keep the top 100 tags in the network, following the practice in \cite{verse18}. 

\header
{\bf Competitors.} We evaluate \nrp against eighteen existing methods, including four classic methods ({\it i.e.},  \deepwalk, \nodetovec, \linealg and \dngr) and fourteen recent methods, many of which have not been compared against each other in previous work. To our knowledge, we are the first to systematically evaluate such a large number of existing network embedding techniques. 
We categorize the eighteen existing methods into four groups as follows:
\begin{enumerate}[leftmargin=3ex,topsep=3pt,itemsep=0ex,partopsep=1ex,parsep=1ex]
\item factorization-based methods: \arope \cite{AROPE18}, \randne \cite{RandNE18}, \netsmf \cite{qiu2019netsmf}, \prone \cite{zhang2019prone}, and \strap \cite{yin2019scalable};
\item random-walk-based methods: \deepwalk \cite{deepwalk14}, \linealg \cite{LINE15}, \nodetovec \cite{node2vec16}, \pbg \cite{lerer2019pytorch}, \app \cite{APP17}, and \versealg \cite{verse18};
\item neural-network-based methods: \dngr \cite{DNGR16}, \drne \cite{tu2018deep}, \graphgan \cite{wang2018graph}, and \ga \cite{abu2018watch};
\item other methods: \rare \cite{gu2018social}, \nethiex \cite{ma2018hierarchical} and \graphwave \cite{donnat2018}.
\end{enumerate}


\header
{\bf Parameter Settings.} 
For \nrp, we set $\ell_1=20, \ell_2=10,\alpha=0.15,\epsilon=0.2$, and $\lambda=10$. Note that $\ell_1=20$ means up to 20-order proximities can be preserved in the embeddings, 
and most forward and backward weights converge with $\ell_2=10$ epochs. For fair comparison, the random walk decay factor, $\alpha$, is set to $0.15$, in all PPR-based methods, including \versealg, \app, \strap, and \nrp. 
We use the default parameter settings of all competitors as suggested in their papers. 
For instance, the error threshold $\delta$ in \strap is set to $10^{-5}$ as suggested in \cite{yin2019scalable}. 
We obtain the source codes of all competitors from their respective authors. Unless otherwise specified, we set the embedding dimensionality $k$ of each method to $128$.

Note that \arope, \randne, \nethiex, \graphwave, \netsmf and \prone are designed for undirected graphs only.
For a thorough evaluation, we still report their performance on the directed graphs, {\it i.e.}, {\em Wiki}, {\em TWeibo}, and {\em Twitter}, by omitting the direction of each edge when feeding the graphs as input to these methods.

In addition, the following methods are designed for some specific tasks: {\it e.g.}, \graphwave and \drne for structural role discovery;
\nodetovec, \deepwalk, \linealg, \dngr, \netsmf and \prone for node classification or network visualization.
For completeness,
we evaluate all methods over three commonly used tasks, namely, link prediction, node classification, and graph reconstruction. 
We exclude a method if it cannot report results within 7 days.



\vspace{1mm}
\subsection{Link Prediction}

\begin{figure}[!t]
\centering
\captionsetup[subfloat]{captionskip=-1mm}
\begin{small}
\begin{tabular}{cc}
\multicolumn{2}{c}{\hspace{-4mm}\includegraphics[height=8mm]{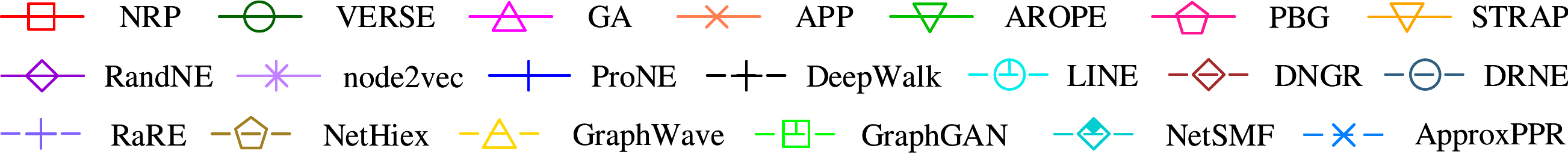}\vspace{2mm}}  \\
\hspace{-6mm}\subfloat[{\em Wiki}]{\includegraphics[width=0.55\linewidth]{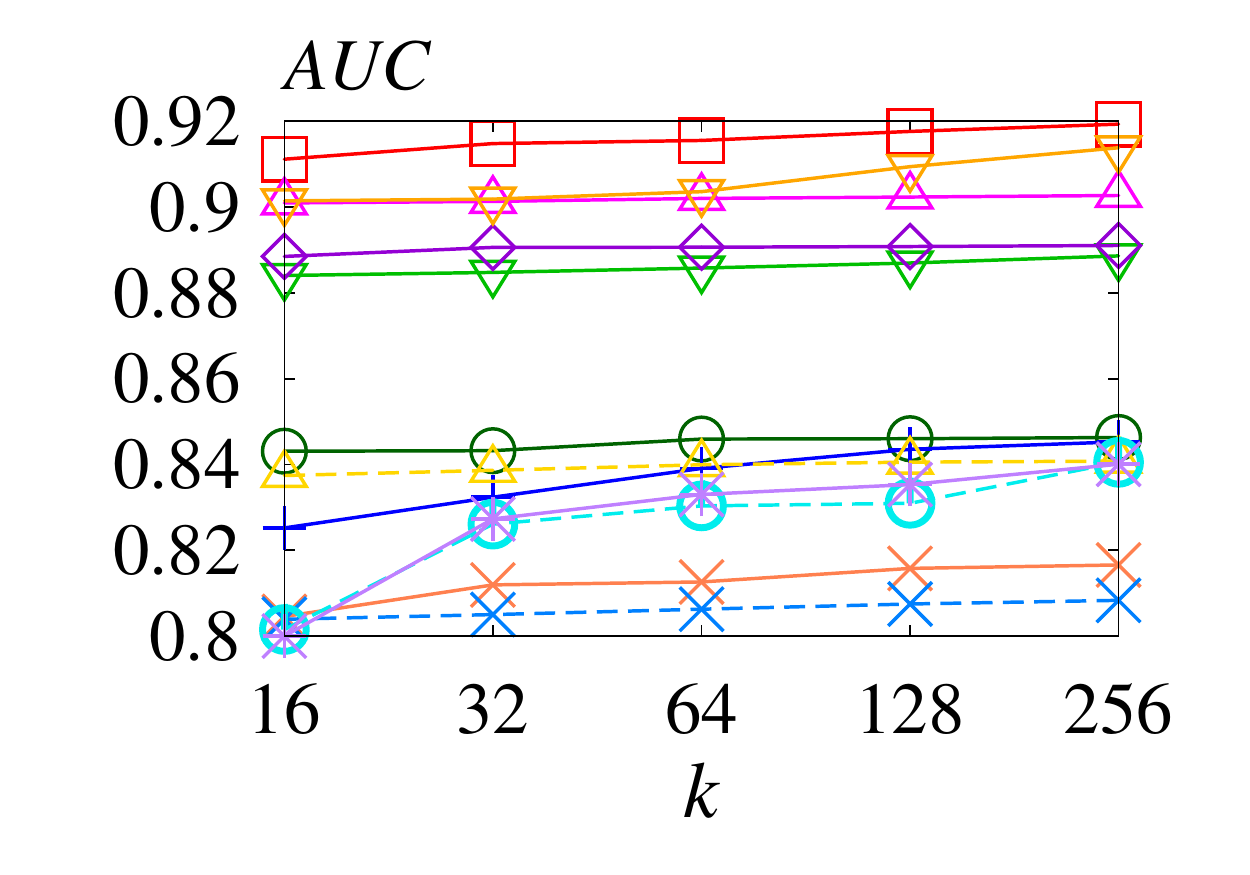}\label{fig:auc-link-wiki}} &
\hspace{-6mm}\subfloat[{\em BlogCatalog}]{\includegraphics[width=0.55\linewidth]{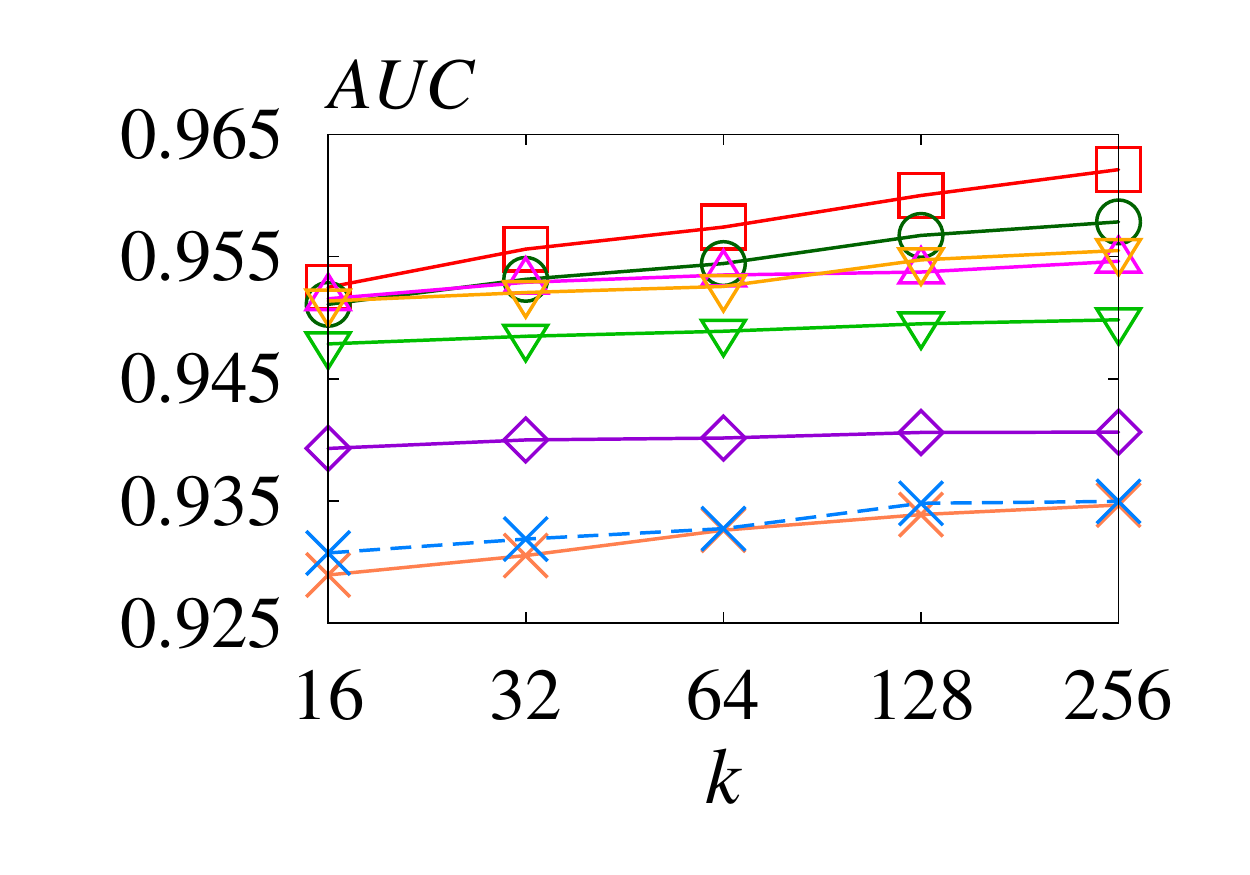}\label{fig:auc-link-ppi}}
\\[4mm]
\hspace{-6mm}\subfloat[{\em TWeibo}]{\includegraphics[width=0.55\linewidth]{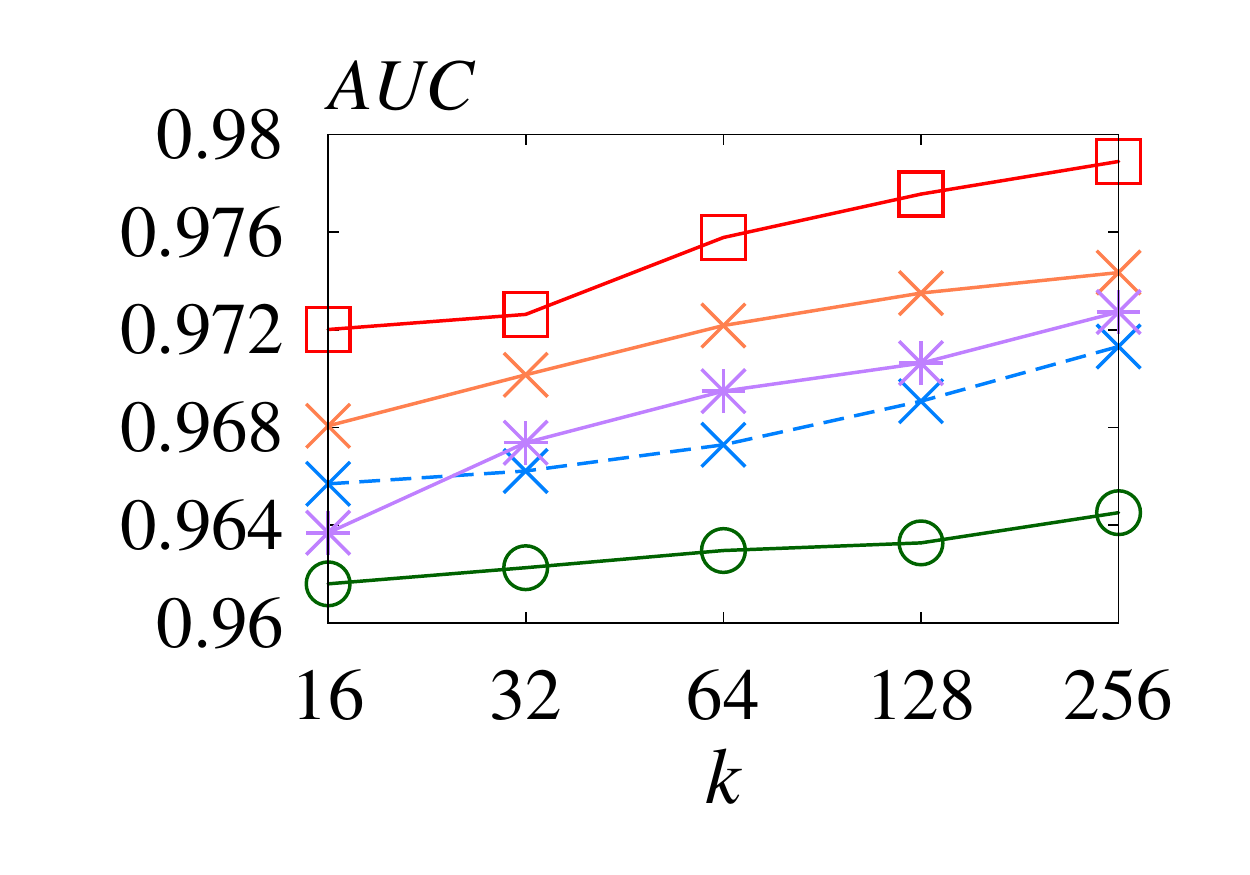}\label{fig:auc-link-twb}} &
\hspace{-6mm}\subfloat[{\em Orkut}]{\includegraphics[width=0.55\linewidth]{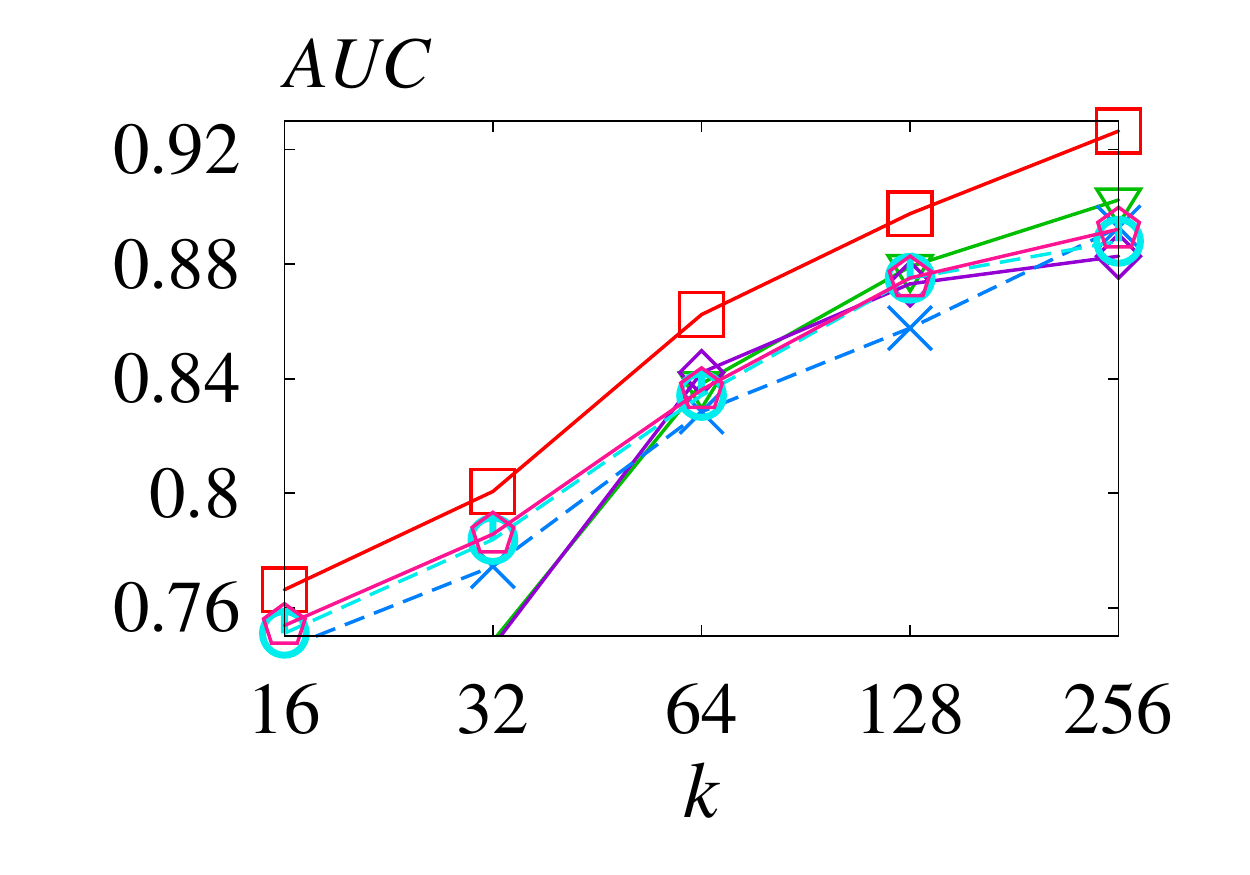}\label{fig:auc-link-okt}}
\\[4mm]
\hspace{-6mm}\subfloat[{\em Twitter}]{\includegraphics[width=0.55\linewidth]{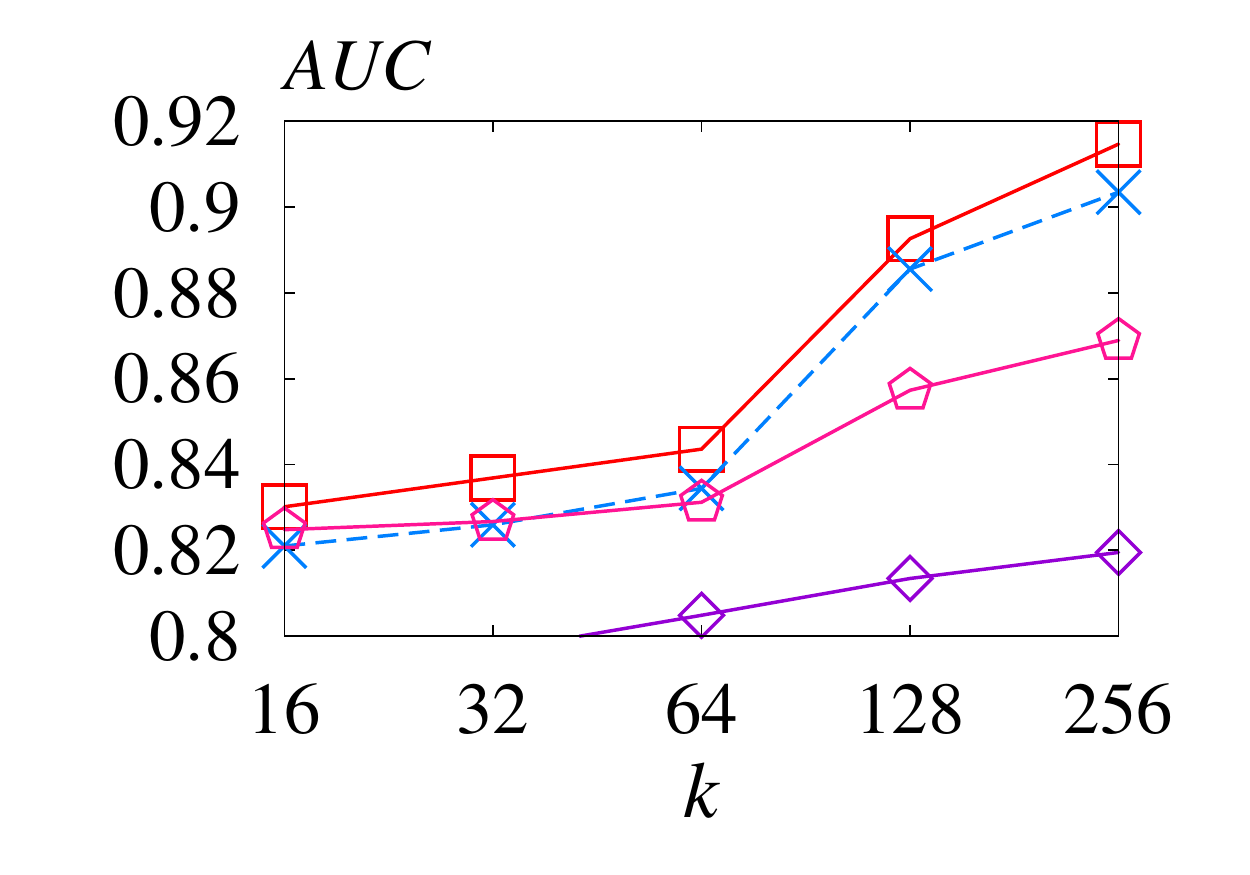}\label{fig:auc-link-twt}} & 
\hspace{-6mm}\subfloat[{\em Friendster}]{\includegraphics[width=0.55\linewidth]{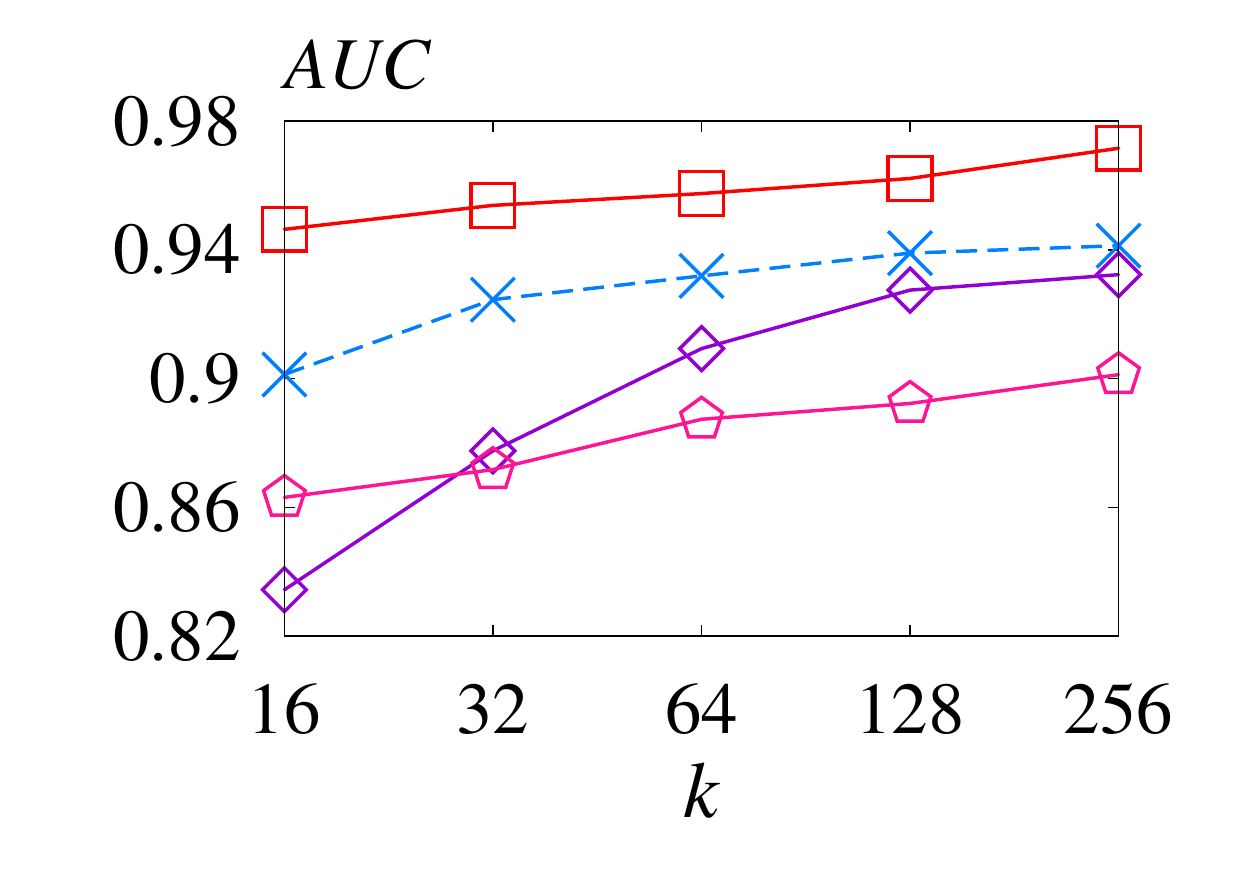}\label{fig:auc-link-fr}}
\\[2mm]
\end{tabular}
\caption{Link prediction results vs. embedding dimensionality $k$ (best viewed in color).} \label{fig:auc-link}
\end{small}
\vspace{-1mm}
\end{figure}

Link prediction aims to predict which node pairs are likely to form edges. Following previous work \cite{AROPE18}, we first remove $30\%$ randomly selected edges from the input graph $G$, and then construct embeddings on the modified graph $G'$. After that, we form a testing set $E_{test}$ consisting of (i) the node pairs corresponding to the $30\%$ removed edges, and (ii) an equal number of node pairs that are not connected by any edge in $G$. Note that on directed graphs, each node pair $(u, v)$ is ordered, {\it i.e.}, we aim to predict whether there is a directed edge from $u$ to $v$.

Given a method's embeddings, we compute a score for each node pair $(u, v)$ in the testing set based on embedding vectors of $u$ and $v$, and then evaluate the method's performance by the \textit{Area Under Curve (AUC)} of the computed scores. Following their own settings, for \arope, \randne, \nethiex, \netsmf and \prone, the score for $(u, v)$ is computed as the inner product $u$ and $v$'s embedding vectors; for \nrp, \approxppr, \app, \ga, and \strap, the score equals the inner product of $u$'s forward vector and $v$'s backward vector. For \rare, we apply the probability function described in \cite{RaRE18} for computing the score for $(u,v)$. For \deepwalk, \linealg, \nodetovec, \dngr, \drne, \graphgan, and \graphwave, we use the {\it edge features} approach \cite{ma2018hierarchical}: (i) for each node pair $(u,v)$ in $G$, concatenate $u$'s and $v$'s embeddings into a length-$2k$ vector; (ii) sample a training set of node pairs $E'_{train}$ (with same size as  $E_{test}$), such that half of the node pairs are from $G'$ and the other half are node pairs not connected in $G$; (iii) feed the length-$2k$ vectors of node pairs in $E'_{train}$ into a logistic regression classifier; (iv) then use the classifier to obtain the scores of node pairs in $E_{test}$ for link prediction. For \versealg and \pbg, the inner product approach only works for undirected graphs, since \versealg and \pbg generate only one embedding vector per node, due to which the inner product approach cannot differentiate $(u, v)$ from $(v, u)$. Therefore, on directed graphs, we also employ the aforementioned edge features approach for \versealg and \pbg.

Fig.~\ref{fig:auc-link} shows AUC of each method when $k$ varies from $16$ to $256$. \nrp consistently outperforms all competitors, by a significant margin of up to $3\%$ on {\em Orkut} and {\em Friendster}, and a large margin of $0.5\%$ to $2\%$ on other graphs. Compared with the best competitor, {\it i.e.}, \arope, \nrp achieves a considerable gain of $1.9\%$ on {\em Orkut} when $k=128$.
Note that \nrp outperforms all the PPR-based competitors, including \approxppr, \app, \versealg and \strap, over all datasets, which confirms the efficacy of our reweighting scheme in \nrp, and validates  our analysis of the conventional PPR deficiency in Section~\ref{sec:intro}. 
Moreover, we observe that \versealg is worse  on directed graphs, {\it i.e.,} {\em Wiki} and {\em TWeibo}, although it is the best competitor on  undirected graph {\em BlogCatalog}. This is because that \versealg generates only one embedding vector per node, making it fail to capture the asymmetric transitivity ({\it i.e.,} direction of edges) in directed graphs \cite{HOPE16,APP17}, which is critical for link prediction. 
Our method, \nrp, instead generates two embedding vectors per node and successfully distinguishes the edge directions and thus is more promising.
\strap and \ga cannot efficiently handle large graphs ({\it i.e.}, {\em Youtube}, {\em TWeibo}, {\em Orkut}, {\em Twitter} and {\em Friendster}), since they require the materialization of a large $n\times n$ matrix, which is extremely costly in terms of both space and time; in contrast, \nrp does not require to do so. \nrp also consistently outperforms \arope by about $2\%$ absolute improvement on all graphs. For the other competitors, their performance is also less than satisfactory, as shown in the figures. In summary, for link prediction, \nrp yields considerable performance improvements compared with the state-of-the-art methods, over graphs with various sizes.

\subsection{Graph Reconstruction}

\begin{figure*}[!t]
\centering
\captionsetup[subfloat]{captionskip=-0.5mm}
\begin{small}
\begin{tabular}{cccc}
\multicolumn{4}{c}{\hspace{-3mm}\includegraphics[height=6.4mm]{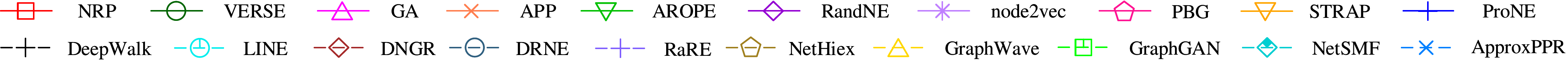}}\vspace{1mm}  \\
\hspace{-4mm}\subfloat{\includegraphics[width=0.25\linewidth]{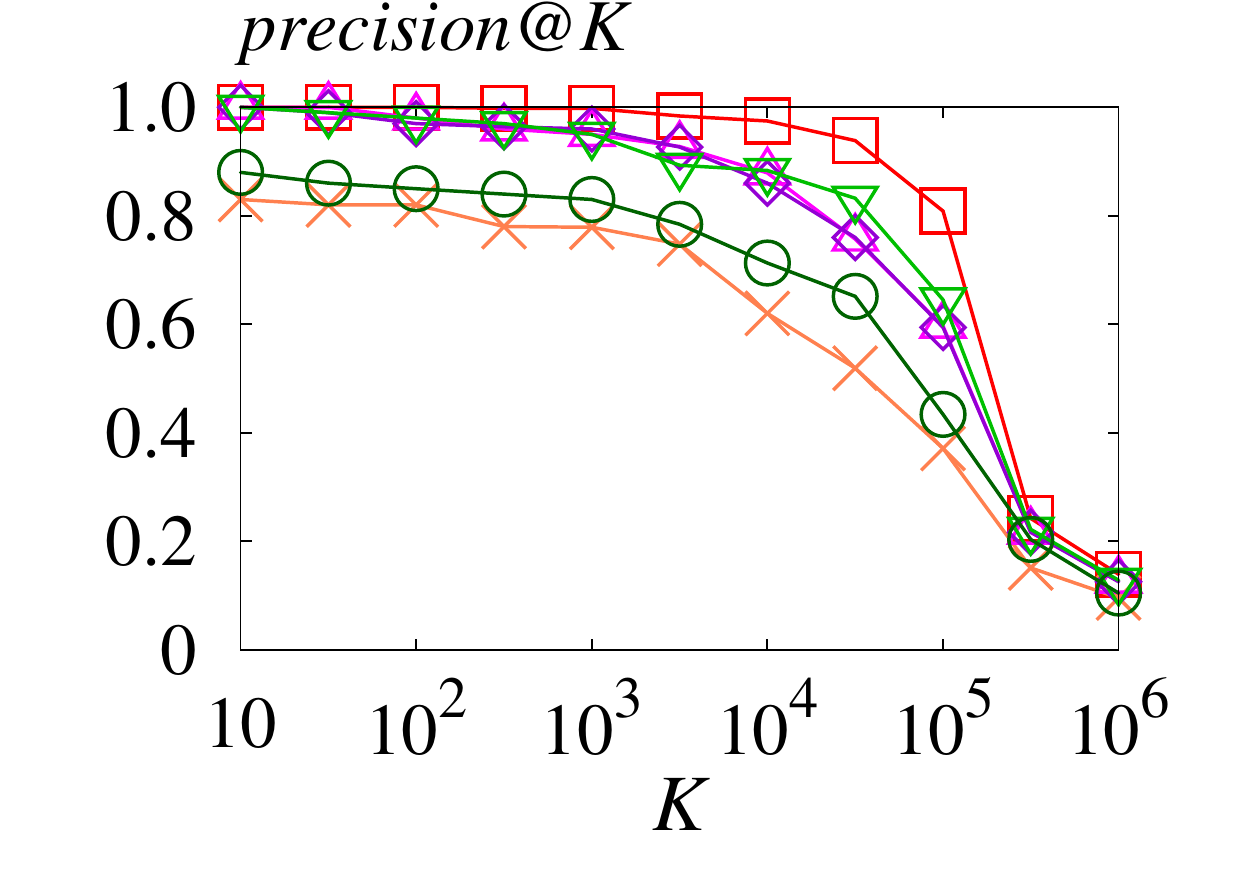} \label{fig:pre-net-wiki}} &
\addtocounter{subfigure}{-1}\hspace{-3mm}\subfloat{\includegraphics[width=0.25\linewidth]{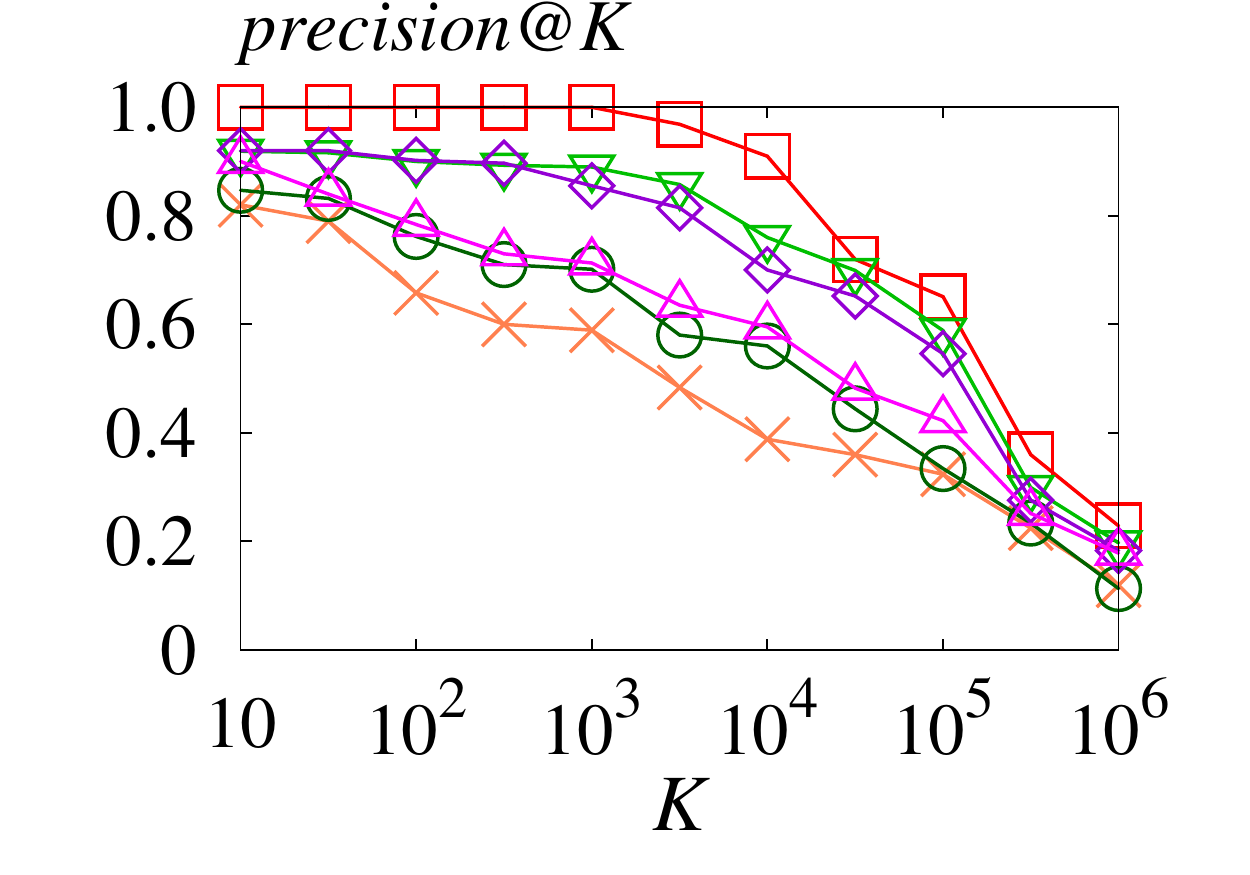}\label{fig:pre-net-ppi}} &
\addtocounter{subfigure}{-1}\hspace{-3mm}\subfloat{\includegraphics[width=0.25\linewidth]{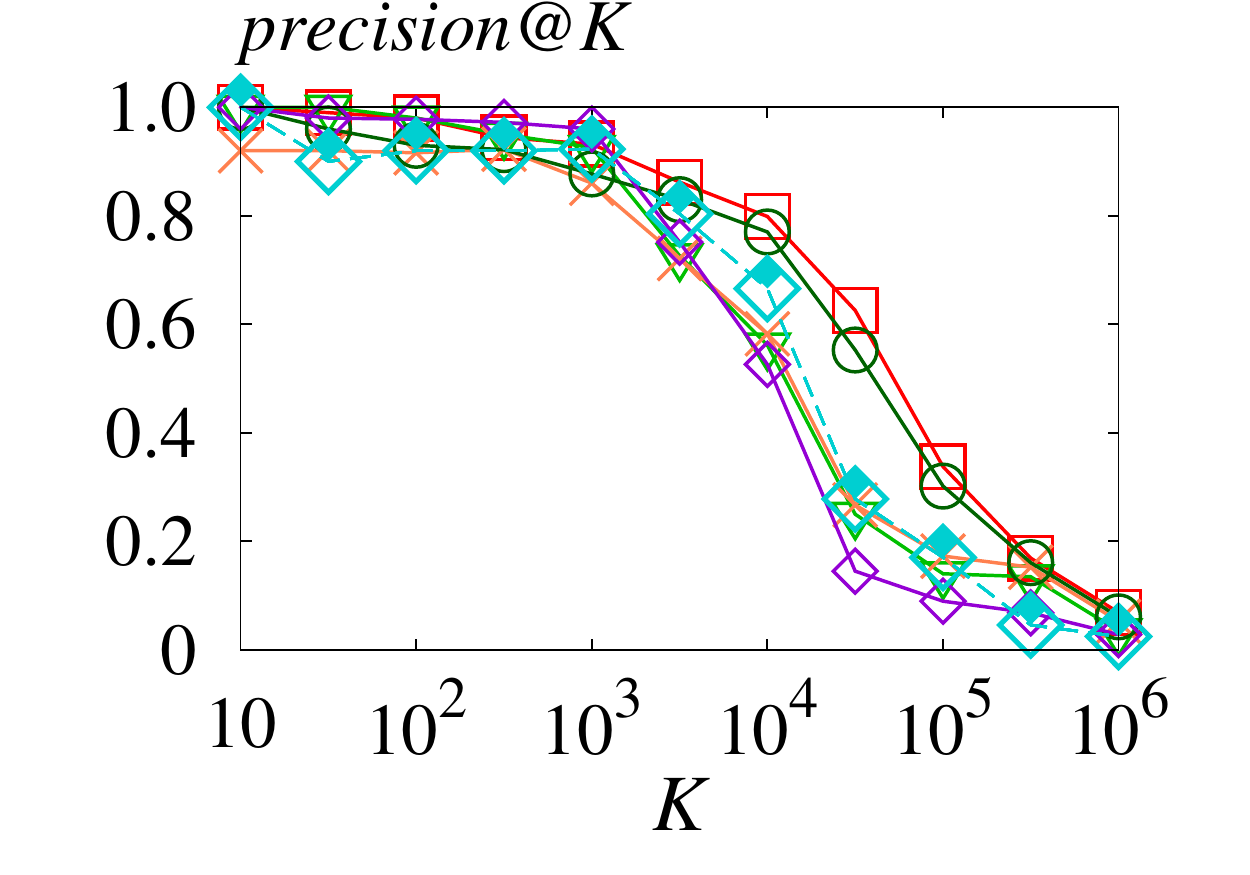}\label{fig:pre-net-ytb}} &
\addtocounter{subfigure}{-1}\hspace{-3mm}\subfloat{\includegraphics[width=0.25\linewidth]{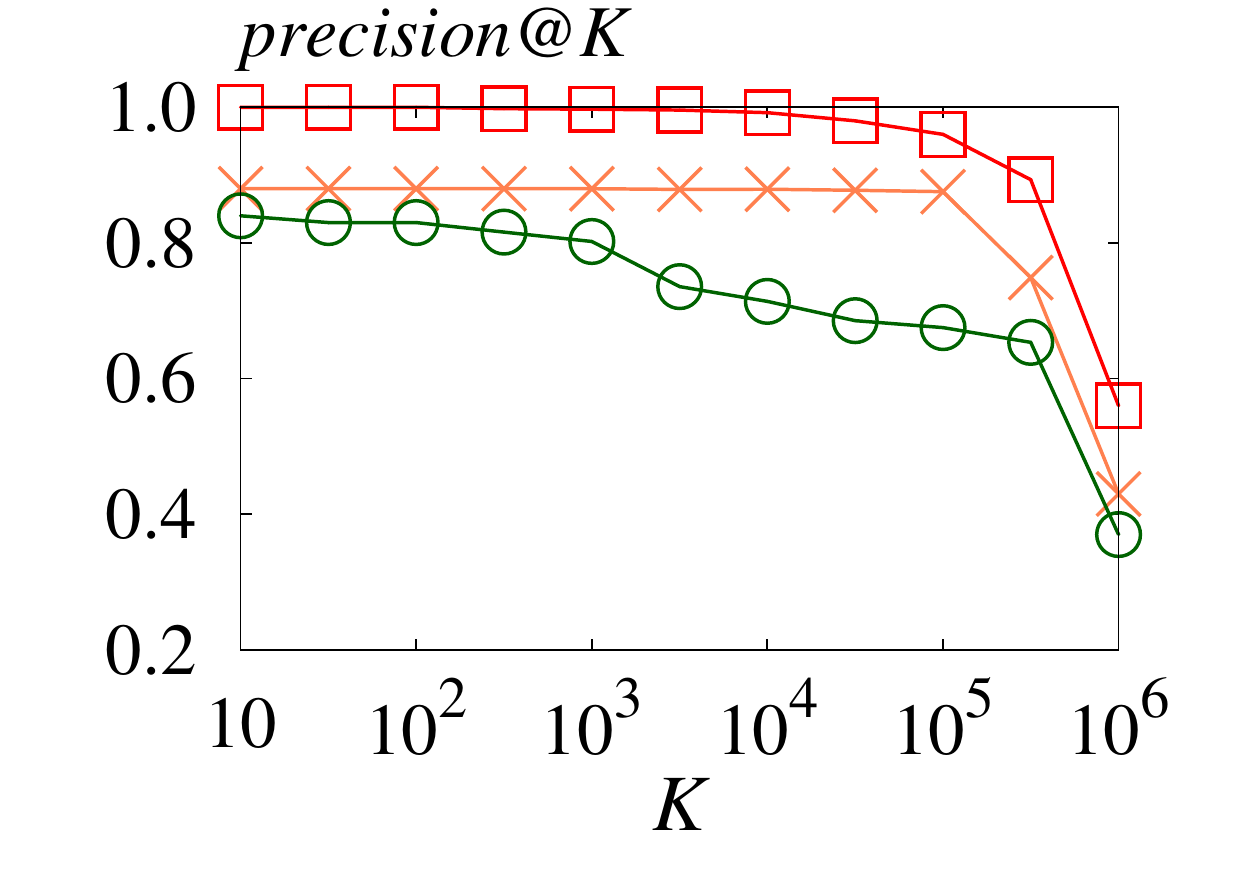}\label{fig:pre-net-twb}}
\\[ 1mm]
\addtocounter{subfigure}{-1}
\hspace{-4mm}\subfloat[{\em Wiki}]{\includegraphics[width=0.25\linewidth]{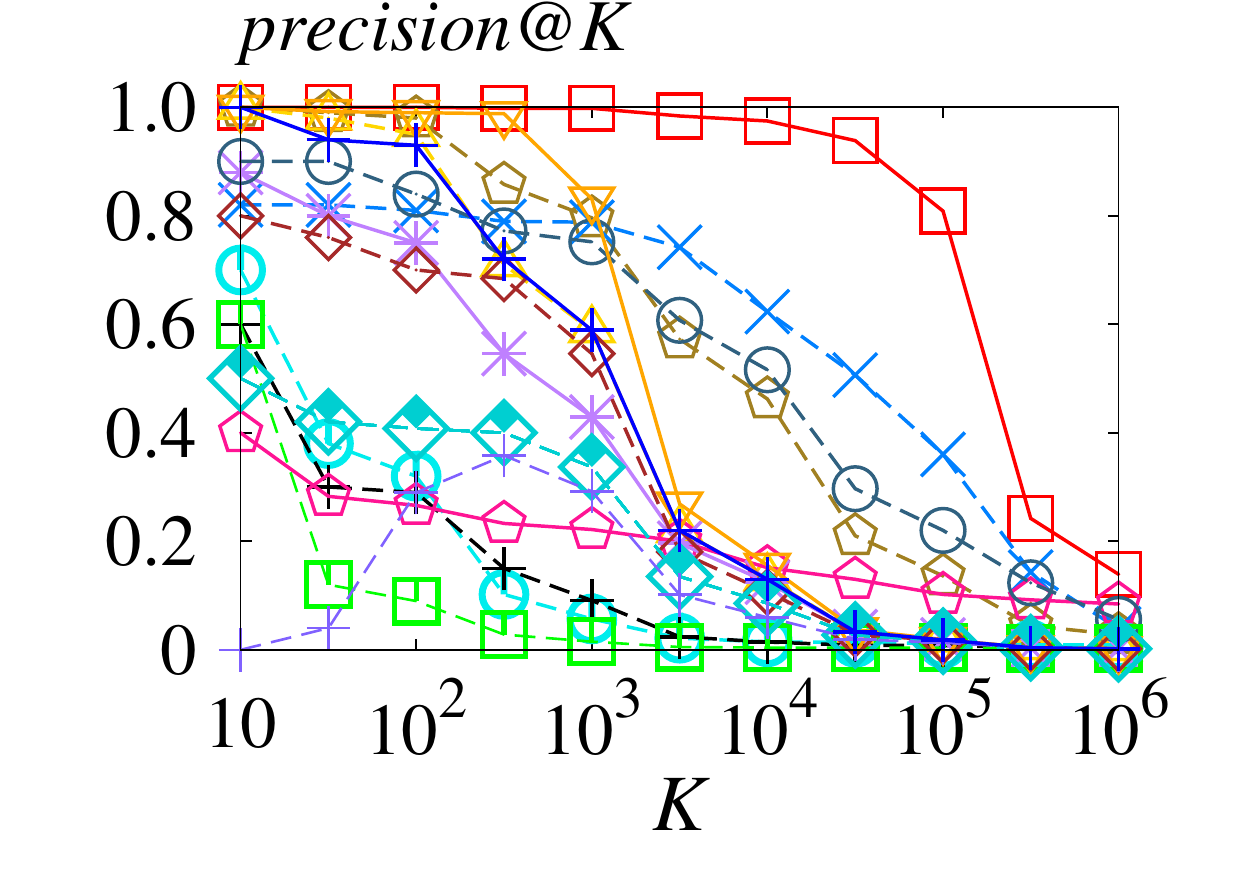} \label{fig:pre-net-wiki2}} &
\hspace{-3mm}\subfloat[{\em BlogCatalog}]{\includegraphics[width=0.25\linewidth]{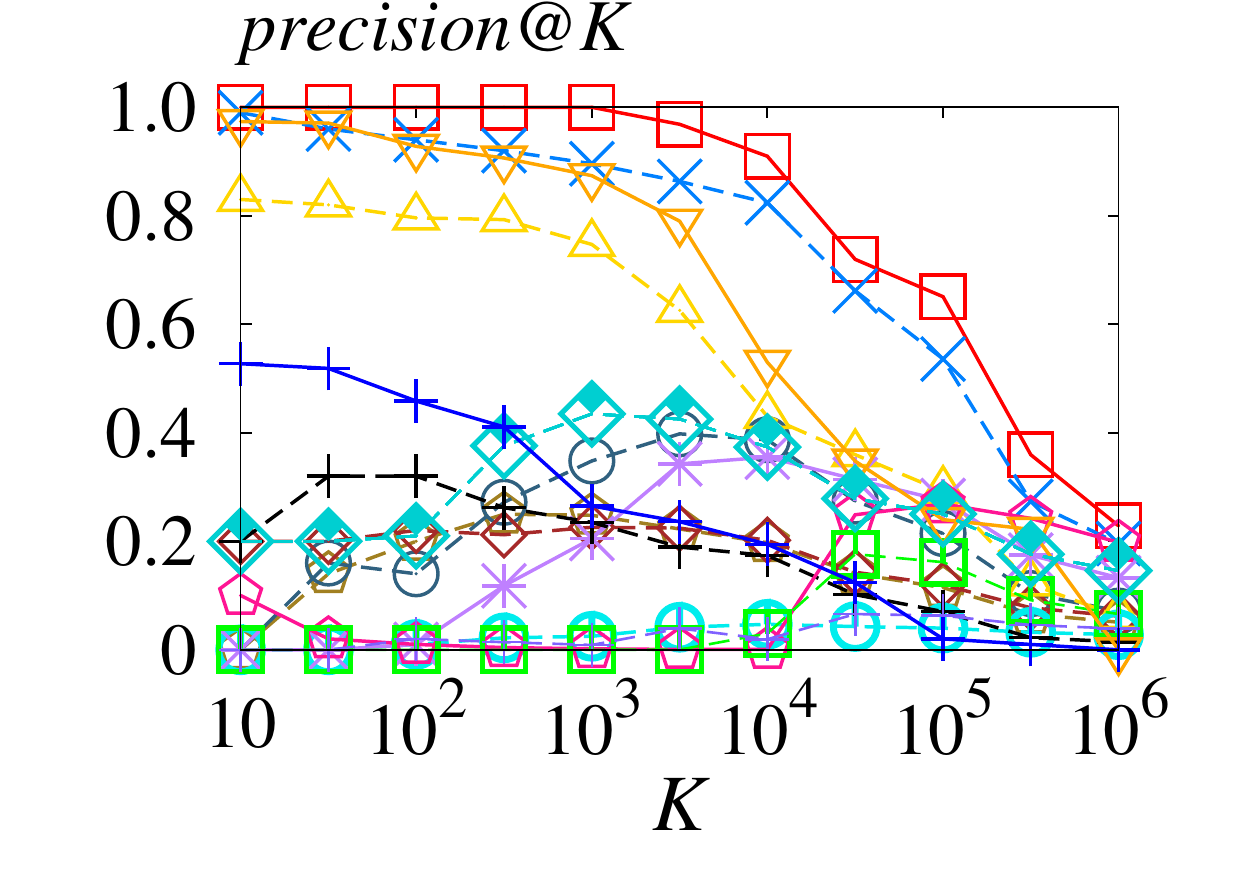}\label{fig:pre-net-ppi2}} &
\hspace{-3mm}\subfloat[{\em Youtube}]{\includegraphics[width=0.25\linewidth]{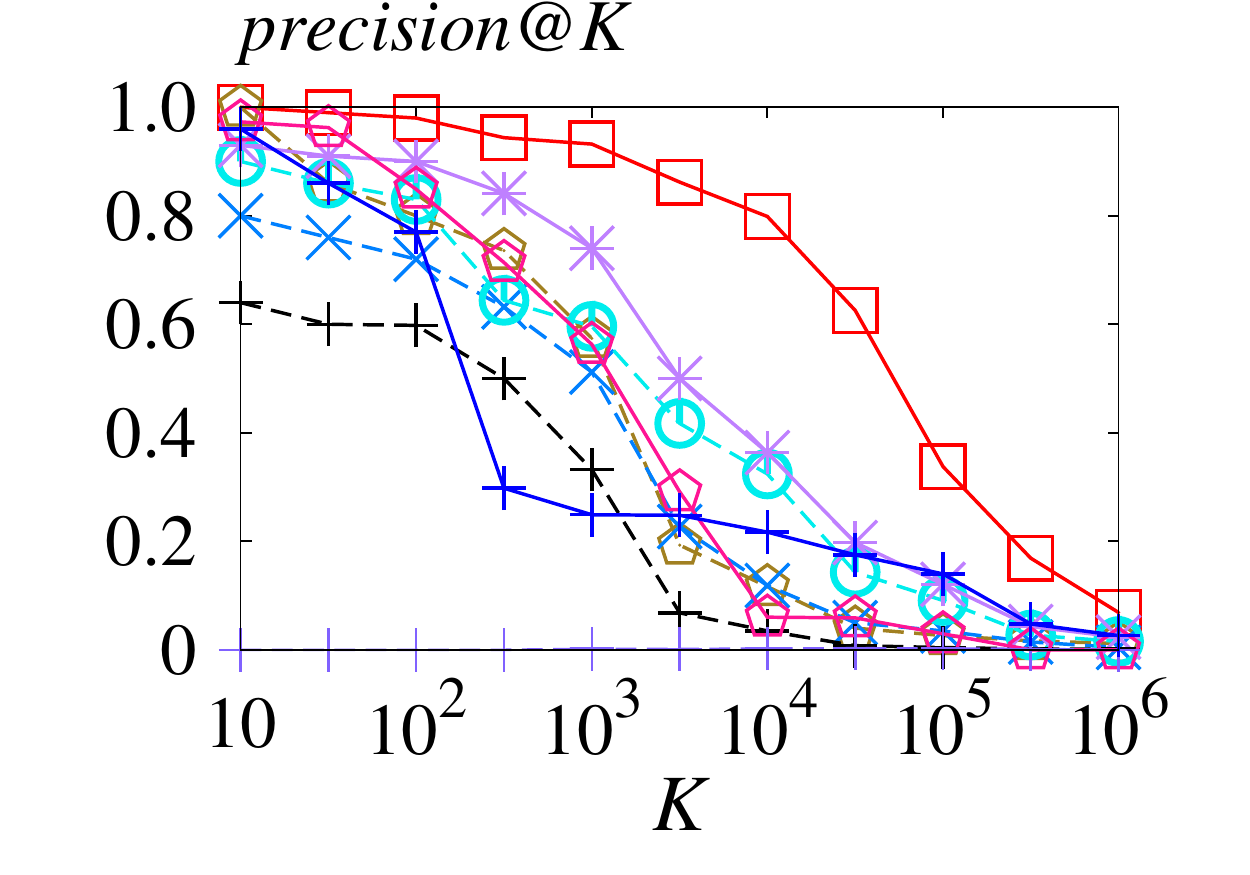}\label{fig:pre-net-ytb2}} &
\hspace{-3mm}\subfloat[{\em Tweibo}]{\includegraphics[width=0.25\linewidth]{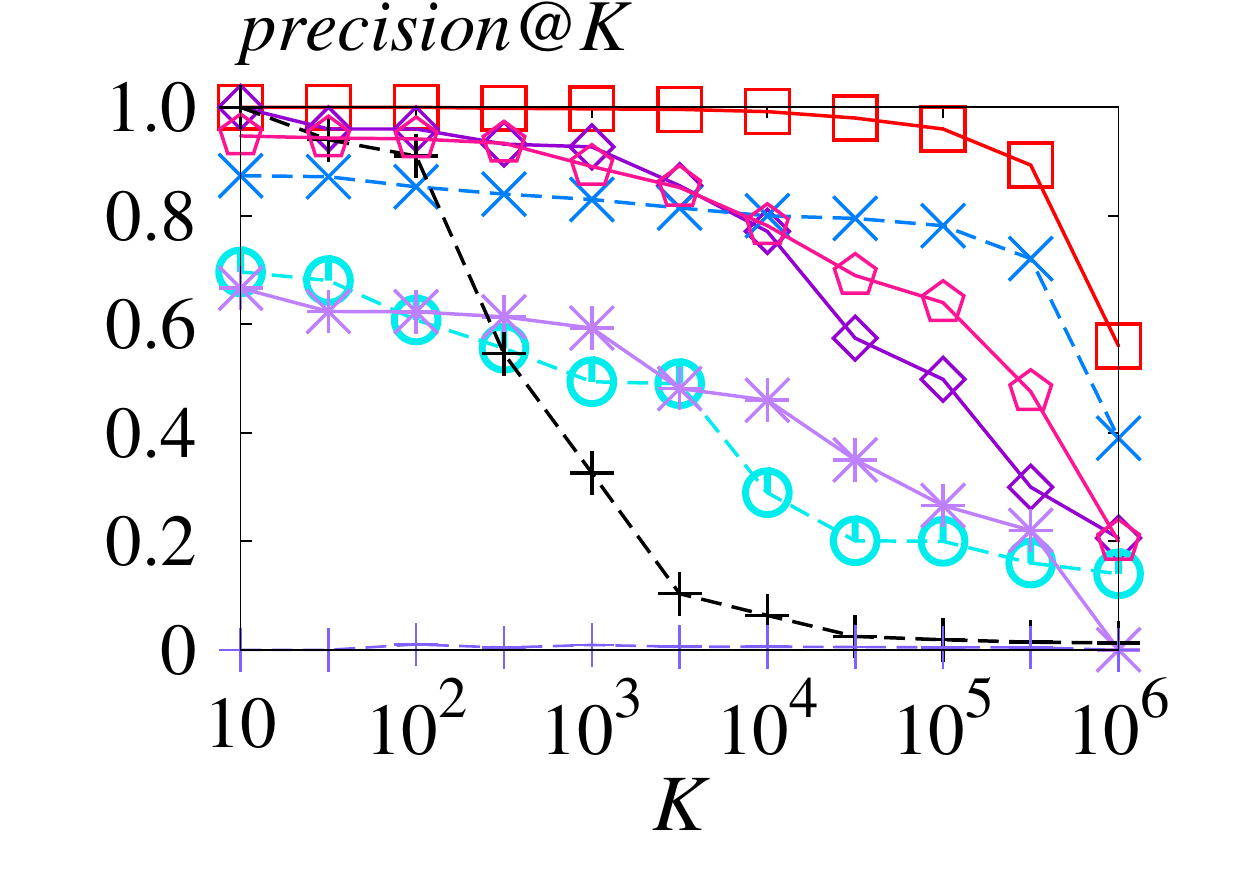}\label{fig:pre-net-twb2}}
\\[2mm]
\end{tabular}
\caption{Graph reconstruction results vs.\ $K$ (best viewed in color).} \label{fig:pre-net}
\end{small}
\vspace{0mm}
\end{figure*}

\begin{figure*}[!t]
\centering
\captionsetup[subfloat]{captionskip=-0.5mm}
\begin{small}
\begin{tabular}{cccc}
\hspace{-5mm}\subfloat[{\em Wiki}]{\includegraphics[width=0.26\linewidth]{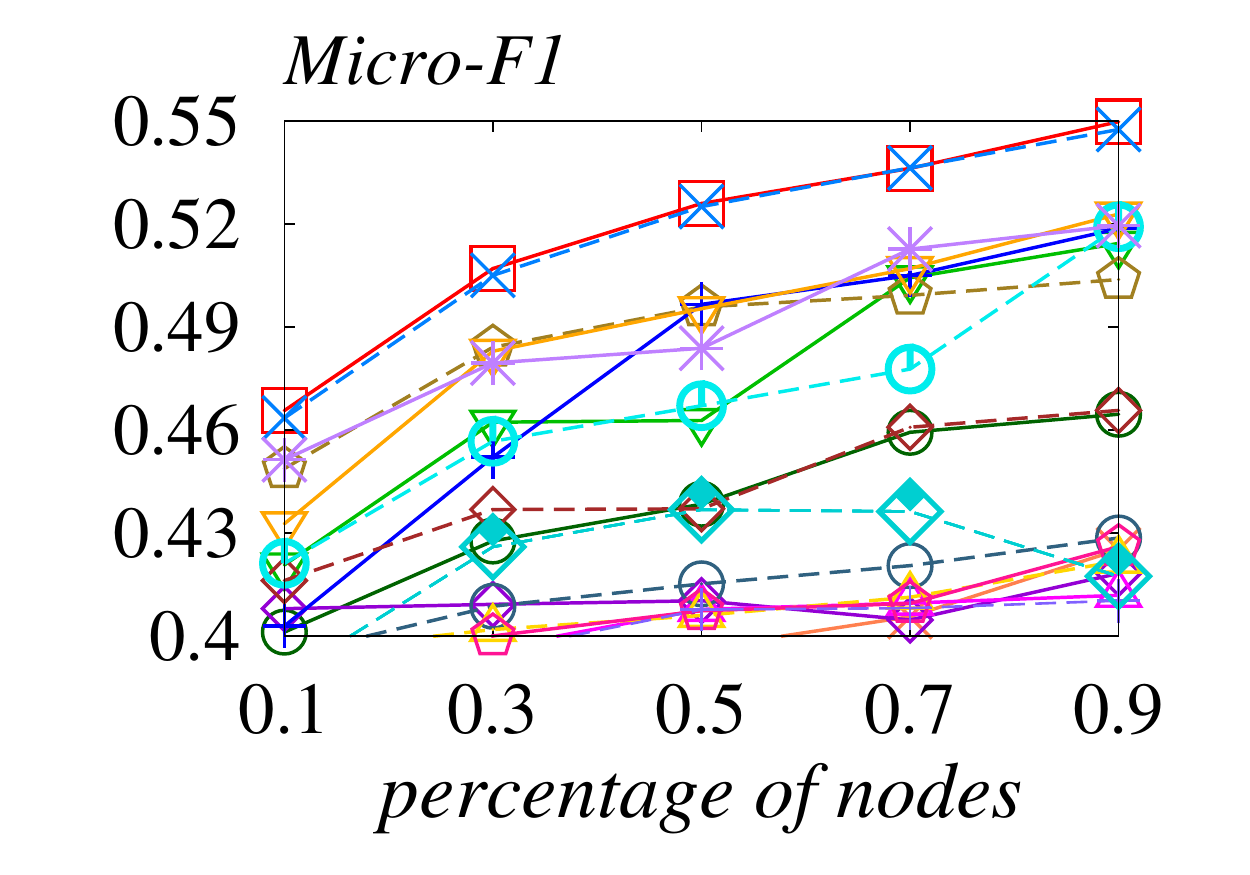}\label{fig:acc-class-wiki}} &
\hspace{-4mm}\subfloat[{\em BlogCatalog}]{\includegraphics[width=0.26\linewidth]{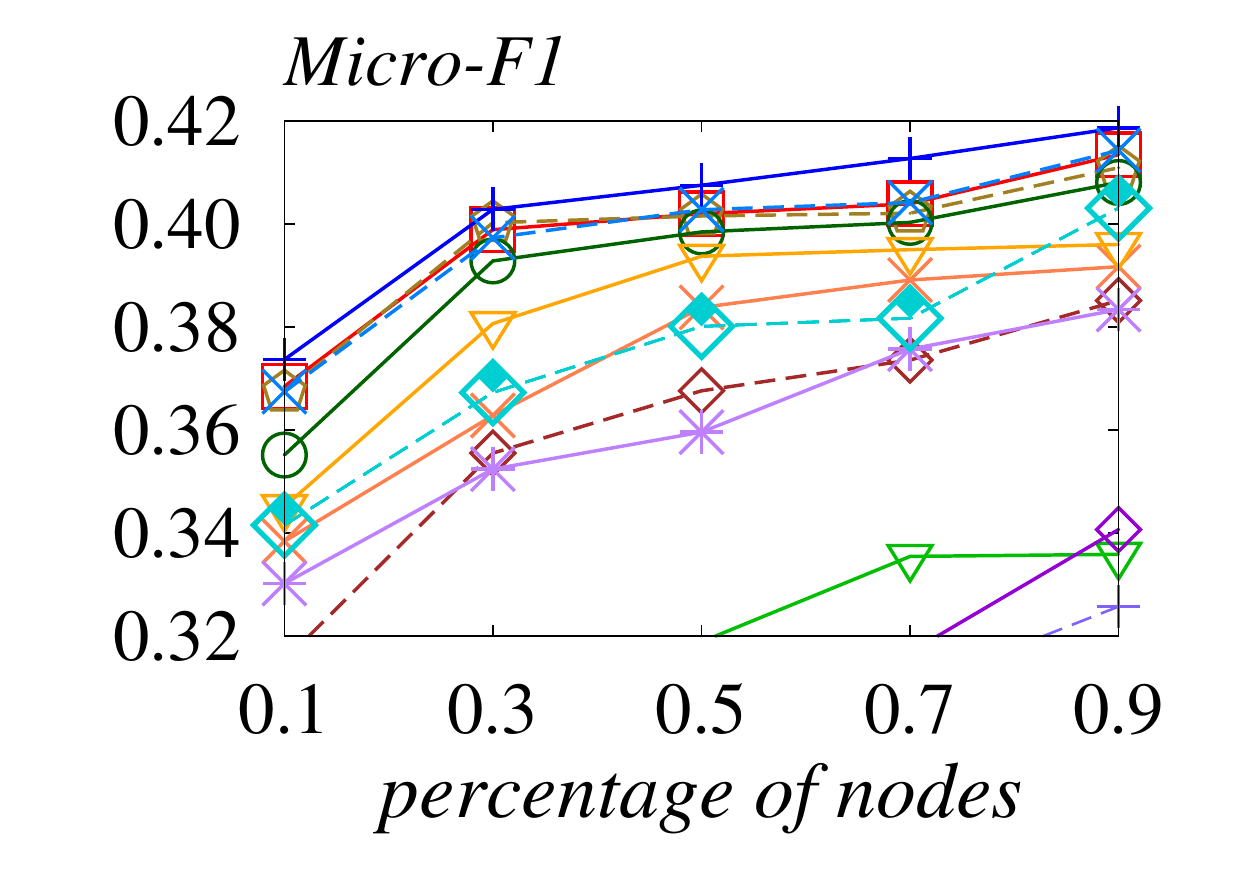}\label{fig:acc-class-blog}} &
\hspace{-4mm}\subfloat[{\em Youtube}]{\includegraphics[width=0.26\linewidth]{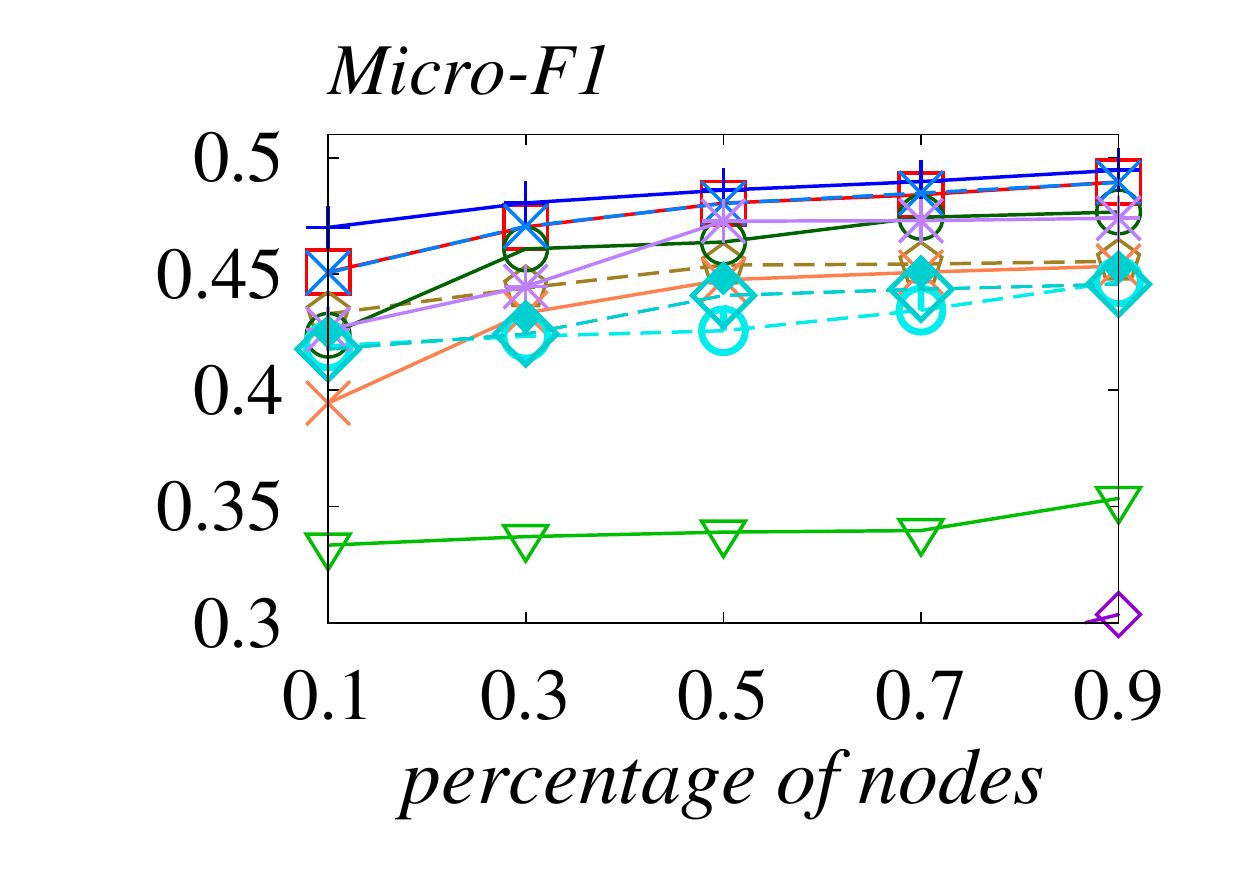}\label{fig:acc-class-ytb}} &
\hspace{-4mm}\subfloat[{\em TWeibo}]{\includegraphics[width=0.26\linewidth]{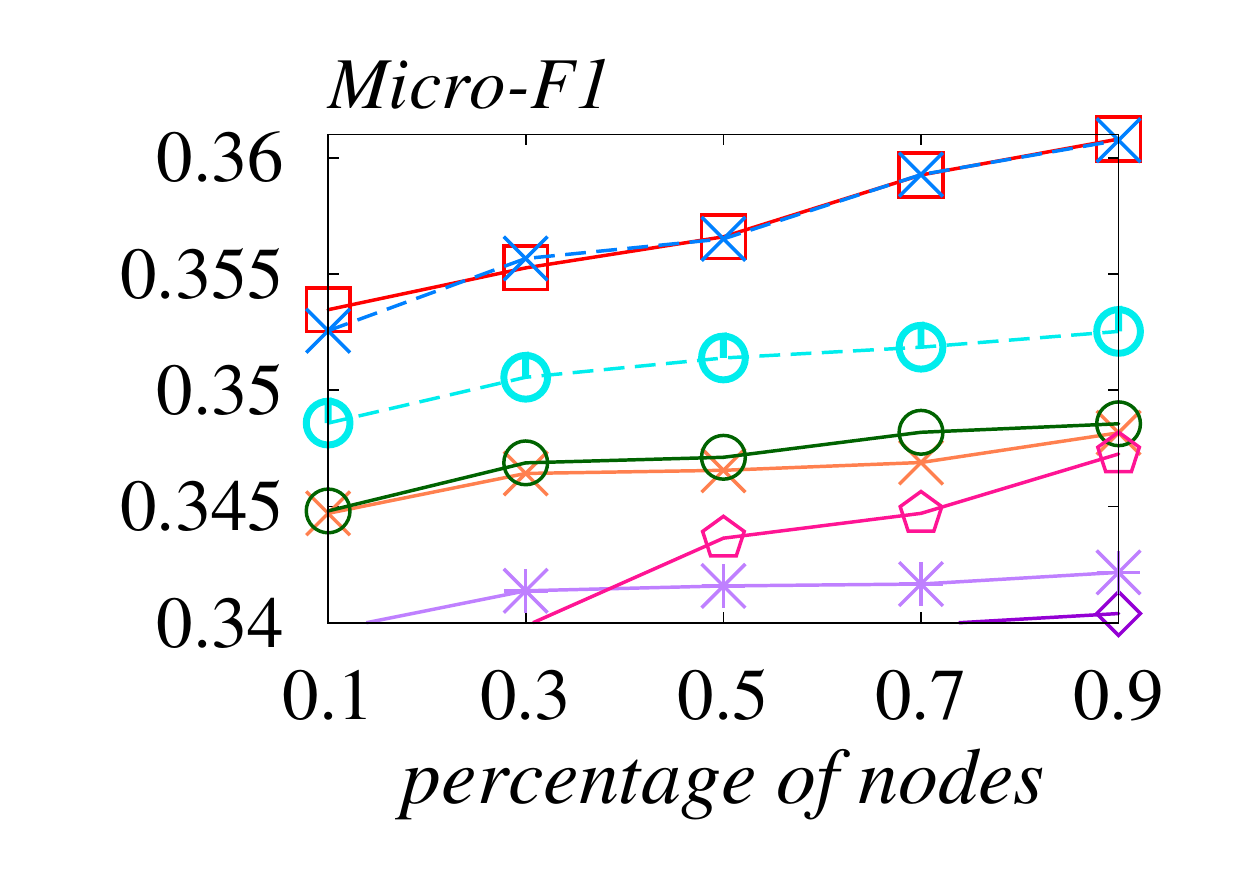}\label{fig:acc-class-twb}}
\\[2mm]
\end{tabular}
\caption{Node classification results (best viewed in color).} \label{fig:acc-class}
\end{small}
\vspace{-2mm}
\end{figure*}

Following previous work, for this task, we (i) take a set $S$ of node pairs from the input graph $G$, (ii) compute the score of each pair using the same approach as in link prediction, and then (iii) examine the top-$K$ node pairs to identify the fraction of them that correspond to the edges in $G$. This fraction is referred to as the {\em precision@K} of the method considered. On {\it Wiki} and {\em BlogCatalog}, we let $S$ be the set of all possible node pairs. Meanwhile, on {\it Youtube} and {\em TWeibo}, following previous work \cite{AROPE18,RandNE18}, we construct $S$ by taking a $1\%$ sample of the ${n \choose 2}$ possible pairs of nodes. We exclude the results on {\em Orkut} and {\em Twitter} since $1\%$ of all the possible node pairs from these two graphs are excessively large.

Fig.~\ref{fig:pre-net} shows the performance of all methods for graph reconstruction, varying $K$ from $10$ to $10^6$. For readability, we split the results of each dataset into two sub-figures in vertical, and each sub-figure compares \nrp against a subset of the competitors. \nrp outperforms all competitors consistently on all datasets. \nrp remains highly accurate when $K$ increases to $10^4$ or even $10^{5}$, while the precisions of other methods, especially \ga, \arope, \randne, \app, \versealg and \strap, drop significantly. Specifically, \nrp achieves at least $90\%$ precision when $K$ reaches $10^4$ on {\em Wiki}, {\em Blogcatalog} and {\em TWeibo}, which means at least $10\%$ absolute improvement over state-of-the-art methods. In addition, on {\em Youtube}, \nrp  achieves $2$-$8\%$  absolute improvement over the best competitors, including \versealg. The superiority of \nrp over the other PPR-based methods, {\it i.e.}, \approxppr, \app, \versealg and \strap, in graph reconstruction demonstrates the power of our reweighting scheme. Meanwhile, the improvements over all other methods like \ga, \arope and \randne implies that \nrp accurately captures the structural information of the input graph via PPR.

\subsection{Node Classification}
Node classification aims to predict each node's label(s) based on its embeddings. Following previous work \cite{verse18}, we first construct network embeddings from the input graph $G$, and use the embeddings and labels of a random subset of the nodes to train a one-vs-all logistic regression classifier, after which we test the classifier with the embeddings and labels of the remaining nodes. In particular, for \nrp, \approxppr, \app, \ga, and \strap, we first normalize the forward and backward vectors, respectively, of each node $v$, and then concatenate them as the feature representation of $v$ before feeding it to the classifier. 
Note that the embeddings produced by \nrp are weighted versions of that produced by \approxppr, and thus, they have the same feature representation for each node $v$ after the normalization, for the task of node classification.

Fig.~\ref{fig:acc-class} shows the Micro-F1 score achieved by each method when the percentage of nodes used for training varies from $10\%$ to $90\%$ ({\it i.e.}, $0.1$ to $0.9$ in the figures). The Macro-F1 results are qualitatively similar and thus omitted for the interest of space.
\nrp consistently outperforms all competitors on {\em Wiki} and {\em TWeibo}, and has comparable performance to \prone on {\em BlogCatalog} and {\em Youtube}.
Specifically, on {\it Wiki}, \nrp achieves an impressive improvement of at least $3\%$ in Micro-F1 over existing methods and about $1\%$ lead on {\em TWeibo}, which is considerable in contrast to that of our competitors. This demonstrates that \nrp can accurately capture the graph structure via PPR. 
On {\em BlogCatalog} and {\em Youtube}, \nrp, \nethiex, \versealg and \prone all achieve comparable performance. \prone is slightly better than \nrp, but note that \prone can only handle undirected graphs and is specifically designed for node classification task by employing graph spectrum and graph partition techniques. \nethiex also requires the input graphs to be undirected. \versealg cannot achieve the same high-quality performance on directed graphs (Fig.~\ref{fig:acc-class-wiki} and \ref{fig:acc-class-twb}) as it does on undirected graphs (Fig.~\ref{fig:acc-class-blog} and \ref{fig:acc-class-ytb}).
The reason is that \versealg only generates one embedding vector per node, and neglects the directions of edges in the directed graphs, while our method \nrp can preserve the directions.
Typically, \nrp achieves consistent and outstanding performance for node classification task over all the real-world graphs.


\subsection{Efficiency}

\begin{figure}[!t]
\centering
\captionsetup[subfloat]{captionskip=-0.4mm}
\begin{small}
\begin{tabular}{cc}
\multicolumn{2}{c}{
\hspace{-4mm} \includegraphics[height=8.2mm]{./figures/algo-legend-cikm-tr-2-eps-converted-to.pdf}}\vspace{2mm}  \\
\hspace{-5mm}\subfloat[{\em Wiki}]{\includegraphics[width=0.53\linewidth]{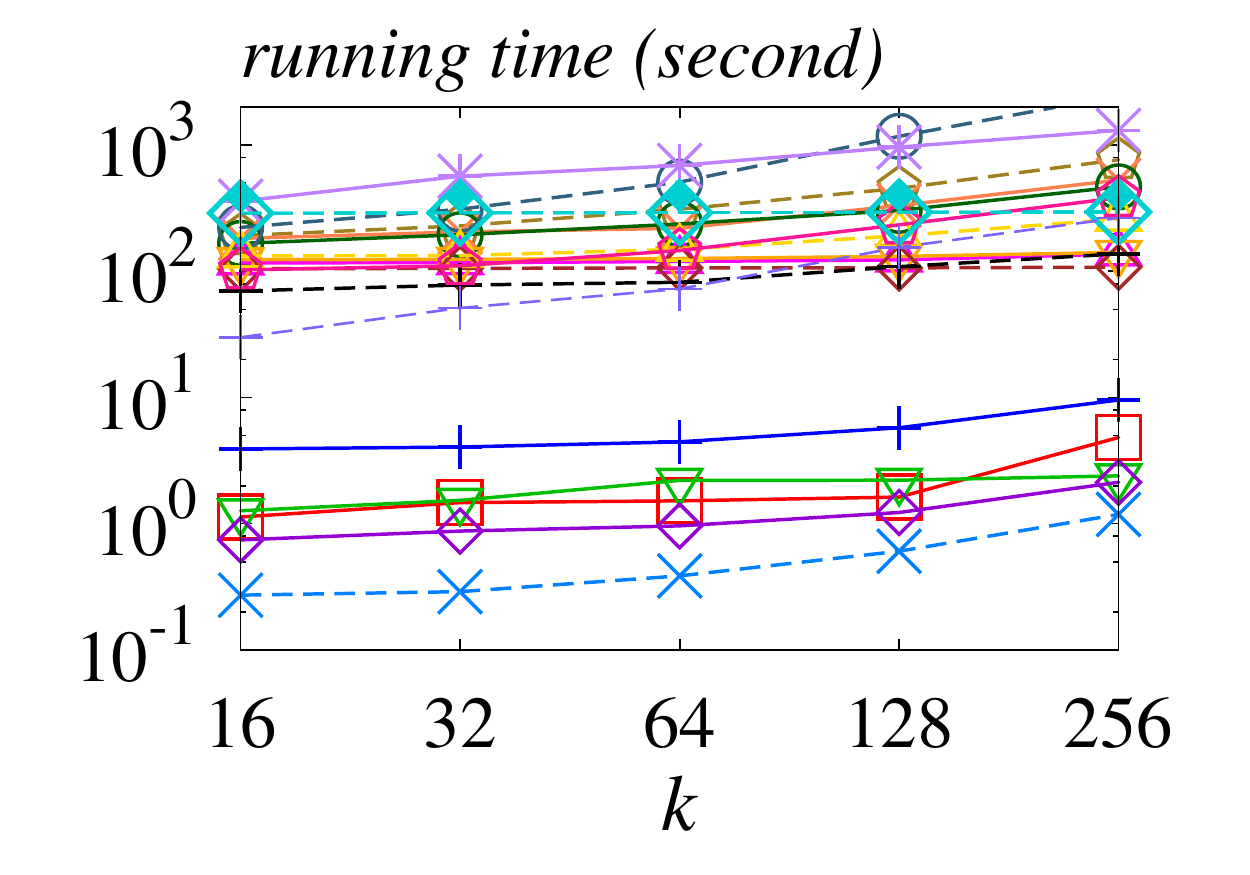}\label{fig:time-wiki}} &
\hspace{-4mm}\subfloat[{\em BlogCatalog}]{\includegraphics[width=0.53\linewidth]{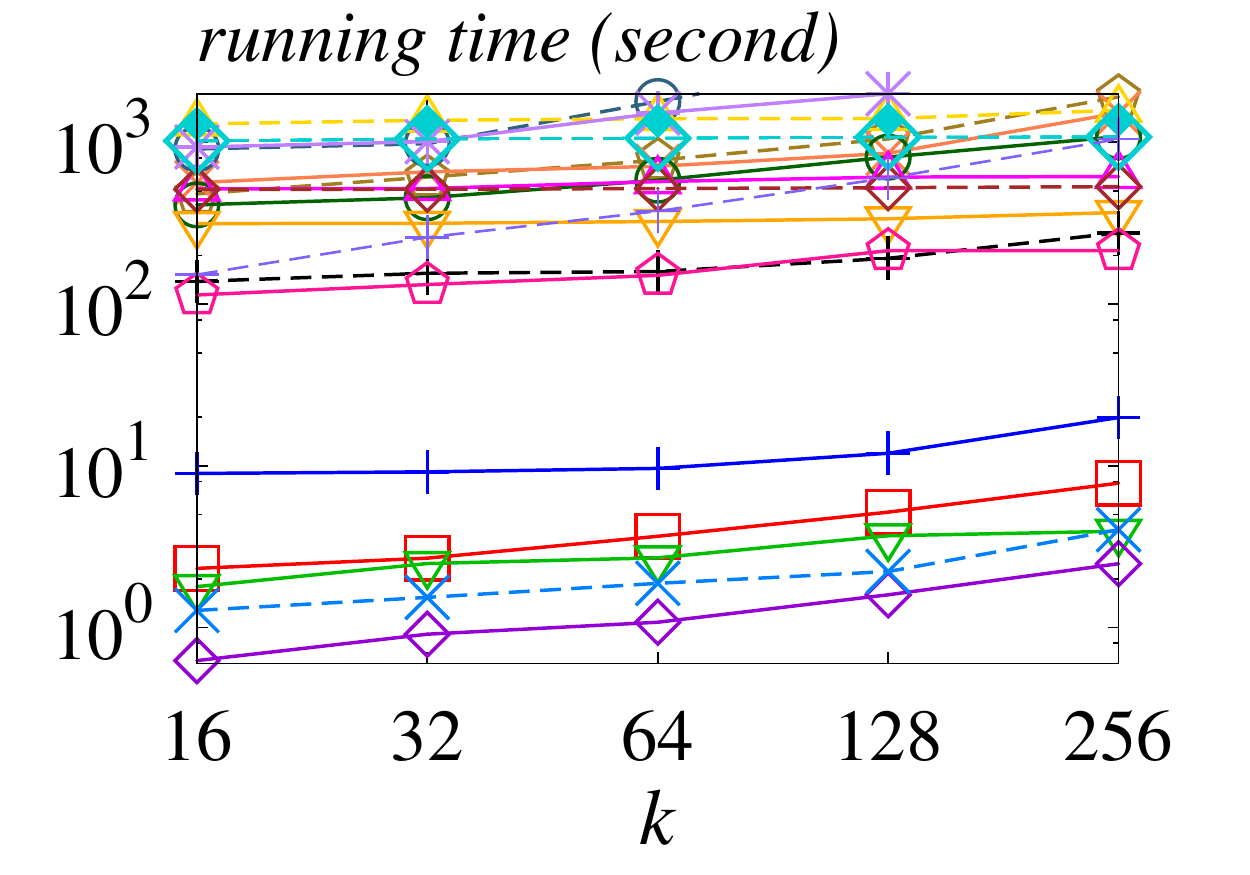}\label{fig:time-ppi}} 
\\[5mm]
\hspace{-4mm}\subfloat[{\em TWeibo}]{\includegraphics[width=0.53\linewidth]{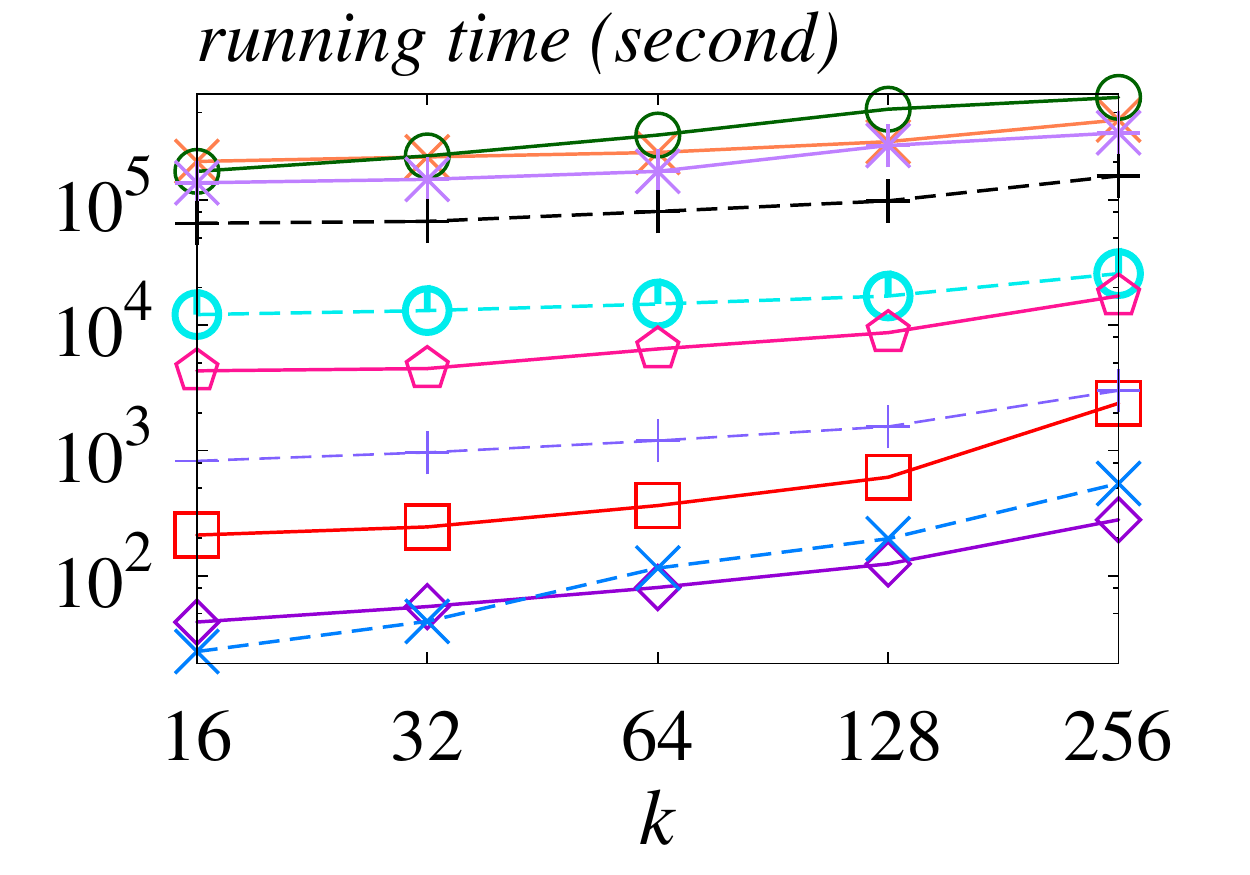}\label{fig:time-twb}} &
\hspace{-5mm}\subfloat[{\em Orkut}]{\includegraphics[width=0.53\linewidth]{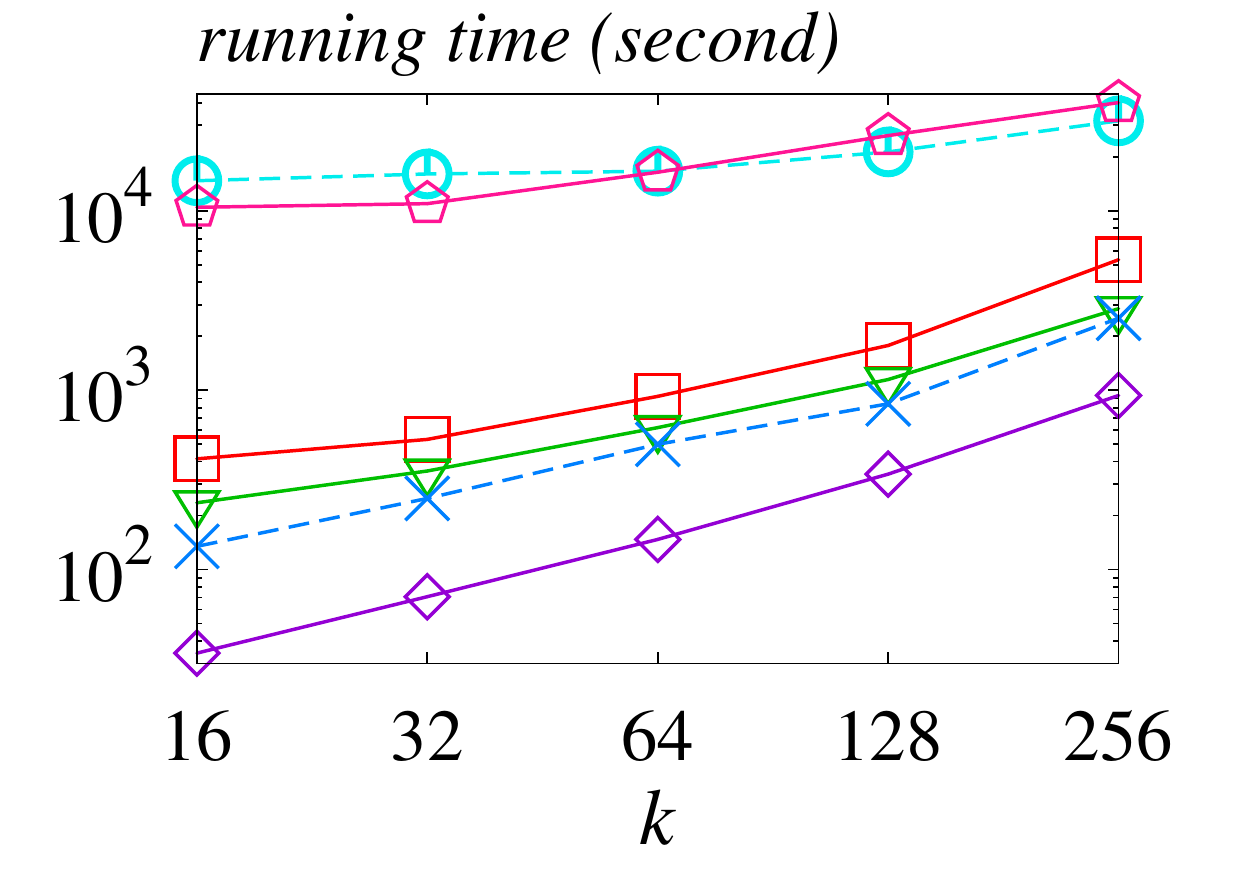}\label{fig:time-okt}}
\\[5mm]
\hspace{-4mm}\subfloat[{\em Twitter}]{\includegraphics[width=0.53\linewidth]{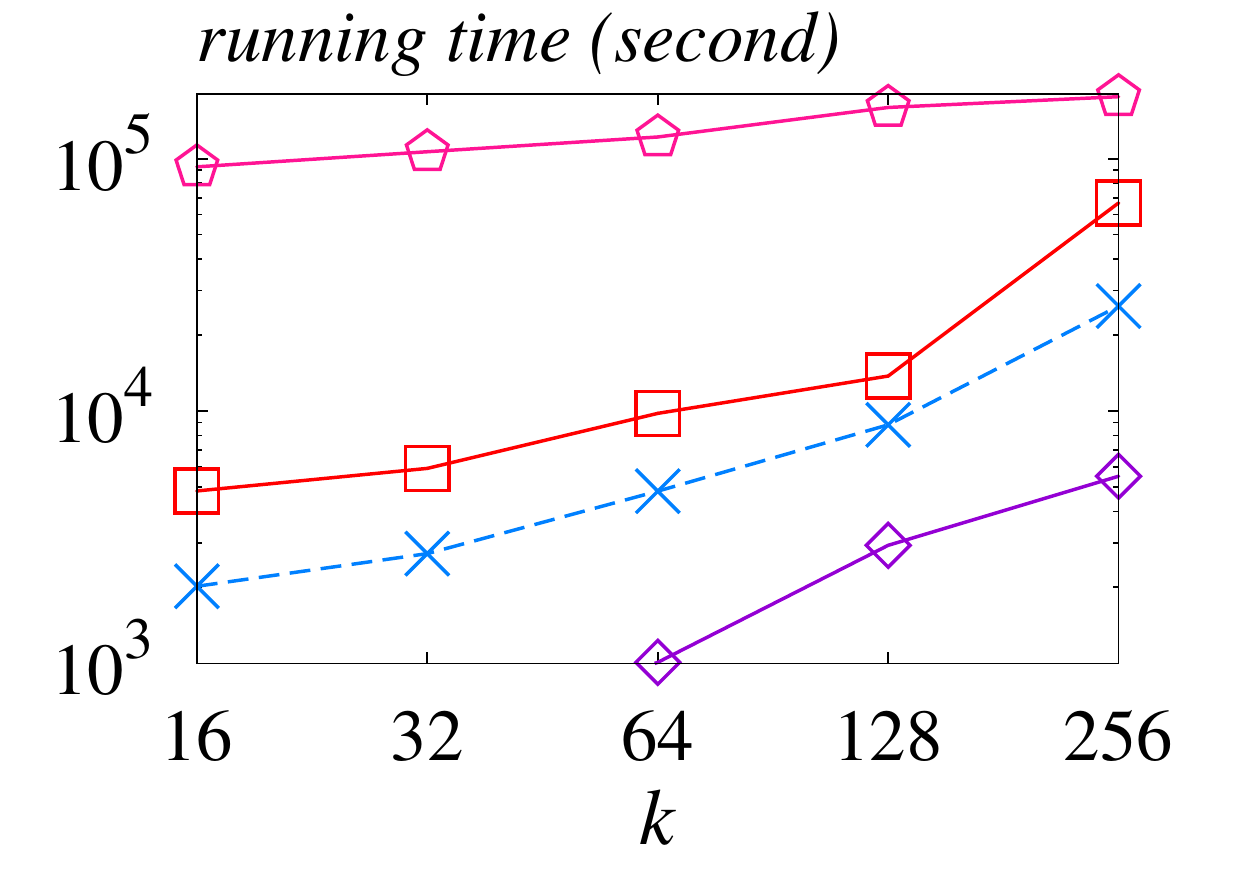}\label{fig:time-twt}} &
\hspace{-4mm}\subfloat[{\em Friendster}]{\includegraphics[width=0.53\linewidth]{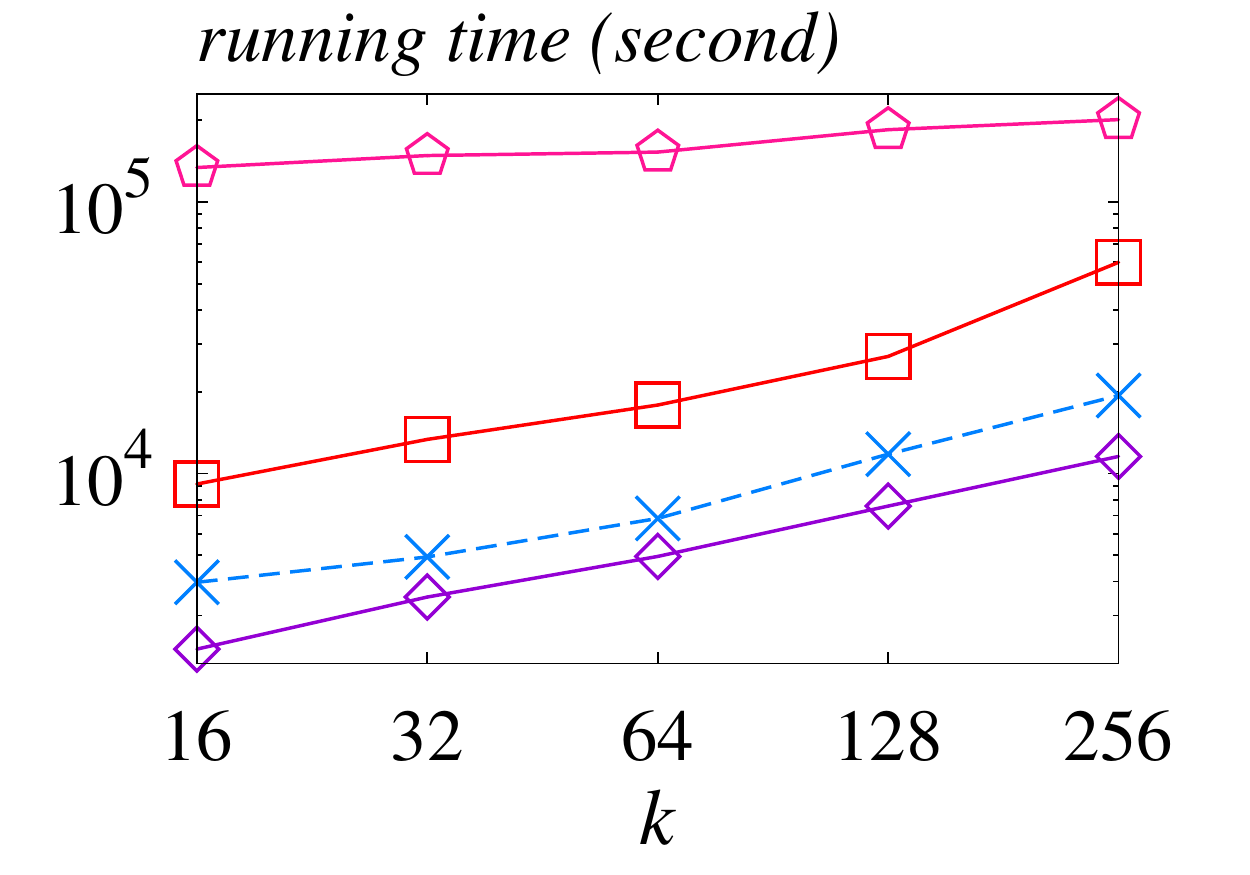}\label{fig:time-fr}}
\\[2mm]
\end{tabular}
\caption{Running time vs. embedding dimensionality $k$ (best viewed in color).} \label{fig:time}
\end{small}
\vspace{-2mm}
\end{figure}

Fig.~\ref{fig:time} plots the time required by each method to construct embeddings, when $k$ is varied from 16 to 256. Note that the Y-axis is in log-scale, and that the reported time excludes the overheads for loading datasets and outputting embeddings. We also omit any method with processing time exceeding $7$ days. For a fair comparison, all methods are ran with a single thread. 


Among all methods tested, \nrp strikes the best balance between effectiveness and efficiency, and is up to 2 orders of magnitude faster than all methods except \approxppr, \prone, \randne and \arope. 
In addition, as illustrated in Fig.~\ref{fig:auc-link}, \ref{fig:pre-net}, and \ref{fig:acc-class}, \randne and \arope are both less effective compared to \nrp for the three tasks. Furthermore, the results of \randne, \arope, and \prone on directed graphs ({\it i.e.}, {\em Wiki}, {\em TWeibo} and {\em Twitter}) are all inferior to those of \nrp, as shown in Fig.~\ref{fig:auc-link}, \ref{fig:pre-net}, and \ref{fig:acc-class}. This is because \randne, \arope, and \prone are specifically designed for undirected graphs instead of directed graphs. 
\prone is also inferior to \nrp in terms of link prediction and graph reconstruction on undirected graphs. Although \approxppr runs faster than \nrp, it is less effective than the latter in terms of link prediction and graph reconstruction, due to the deficiency of conventional PPR discussed in Section \ref{sec:intro}. 
Neither \ga nor \strap is able to scale to large graphs, which again manifests the power of our scalable PPR computation. 
The remaining methods either rely on expensive training phases ({\it e.g.,} \deepwalk and \versealg), or require constructing a huge matrix ({\it e.g.,} \netsmf), thereby failing to handle large graphs efficiently as well.
\subsection{Parameter Analysis} \label{sec:parameteranalysis}
\begin{figure}[!t]
\centering
\captionsetup[subfloat]{captionskip=-0.4mm}
\begin{small}
\begin{tabular}{cc}
\multicolumn{2}{c}{
\hspace{-4mm} \includegraphics[height=2.7mm]{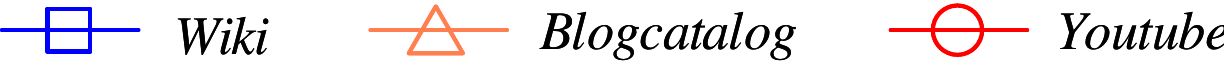}}\vspace{1mm}  \\
\hspace{-6mm}\subfloat[Varying $\alpha$]{\includegraphics[width=0.55\linewidth]{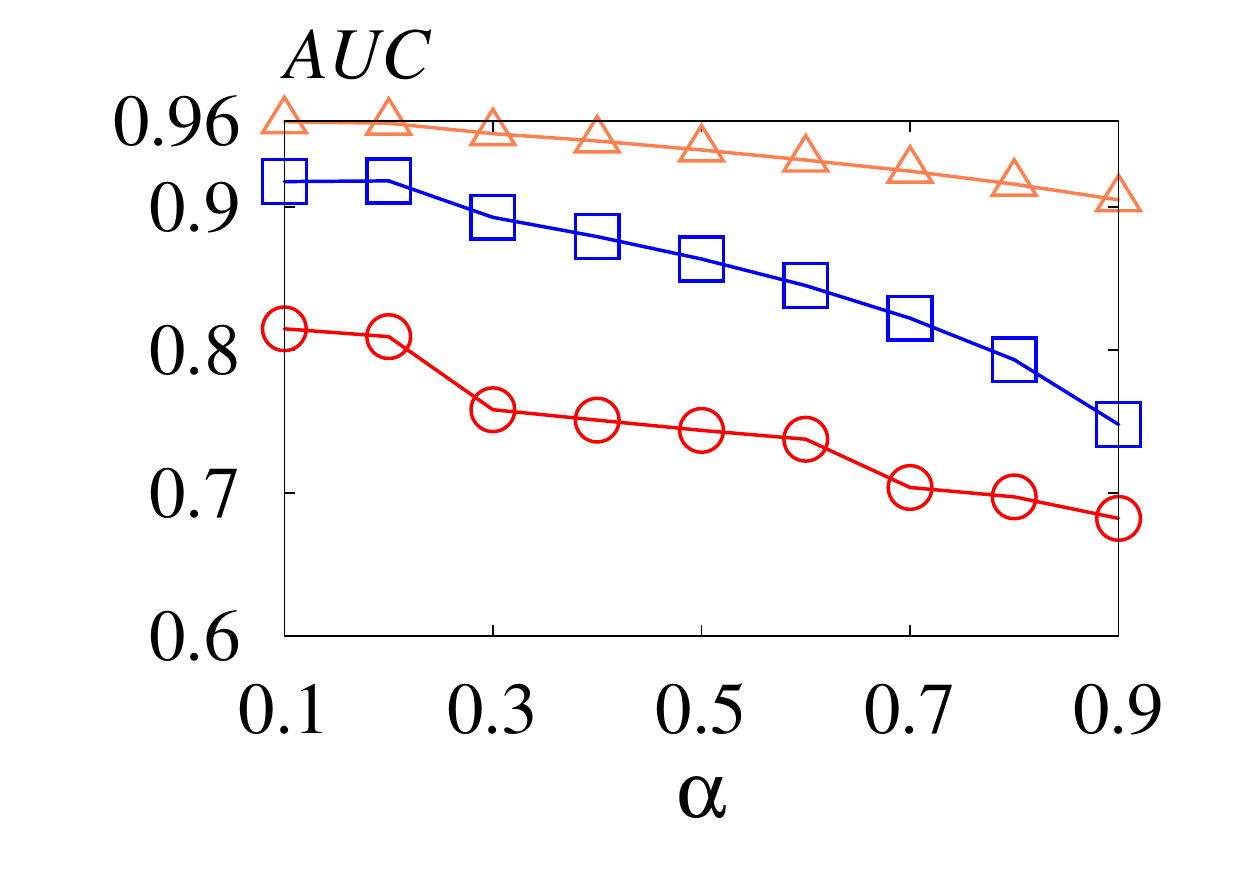}\label{fig:para-alpha}} &
\hspace{-6mm}\subfloat[Varying $\epsilon$]{\includegraphics[width=0.55\linewidth]{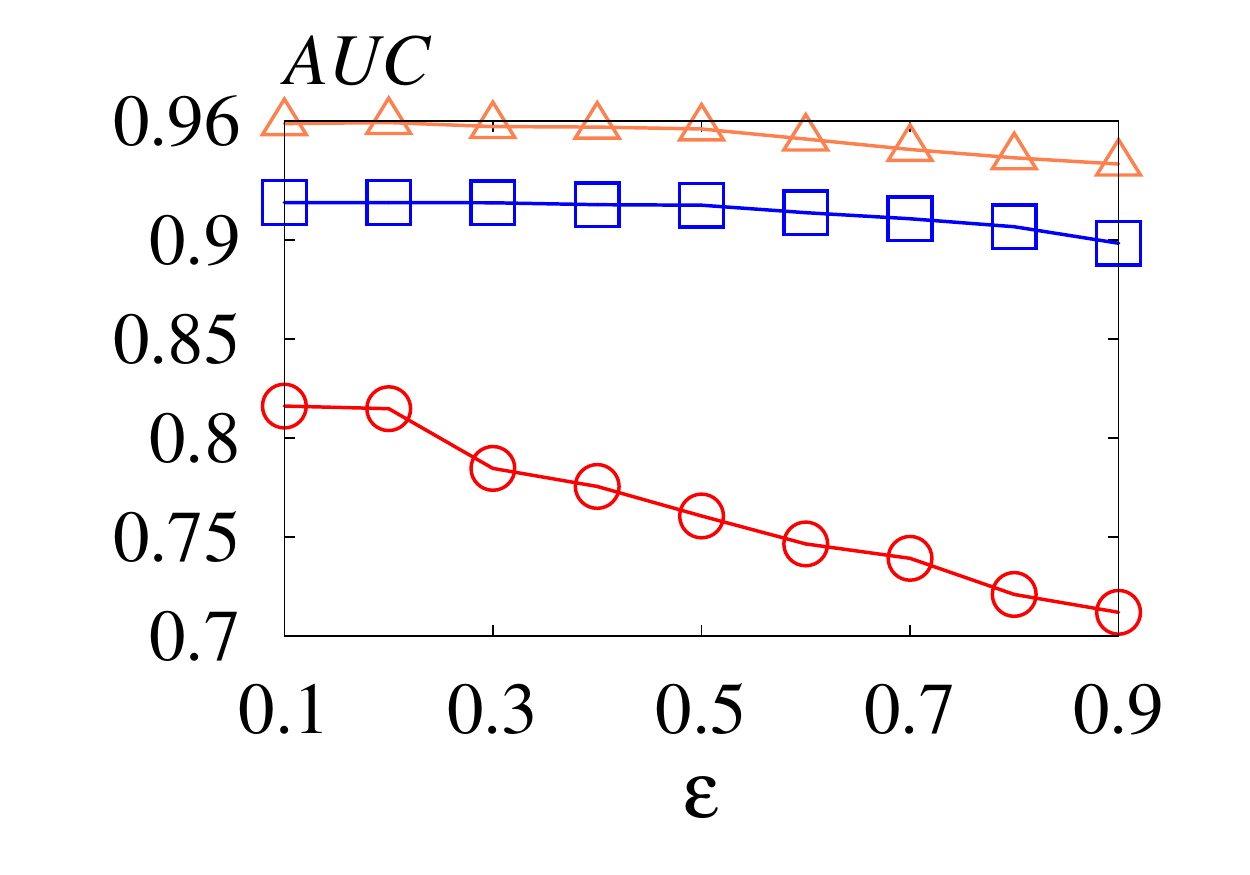}\label{fig:para-eps}}
\\[6mm]
\hspace{-6mm}\subfloat[Varying $\ell_1$]{\includegraphics[width=0.55\linewidth]{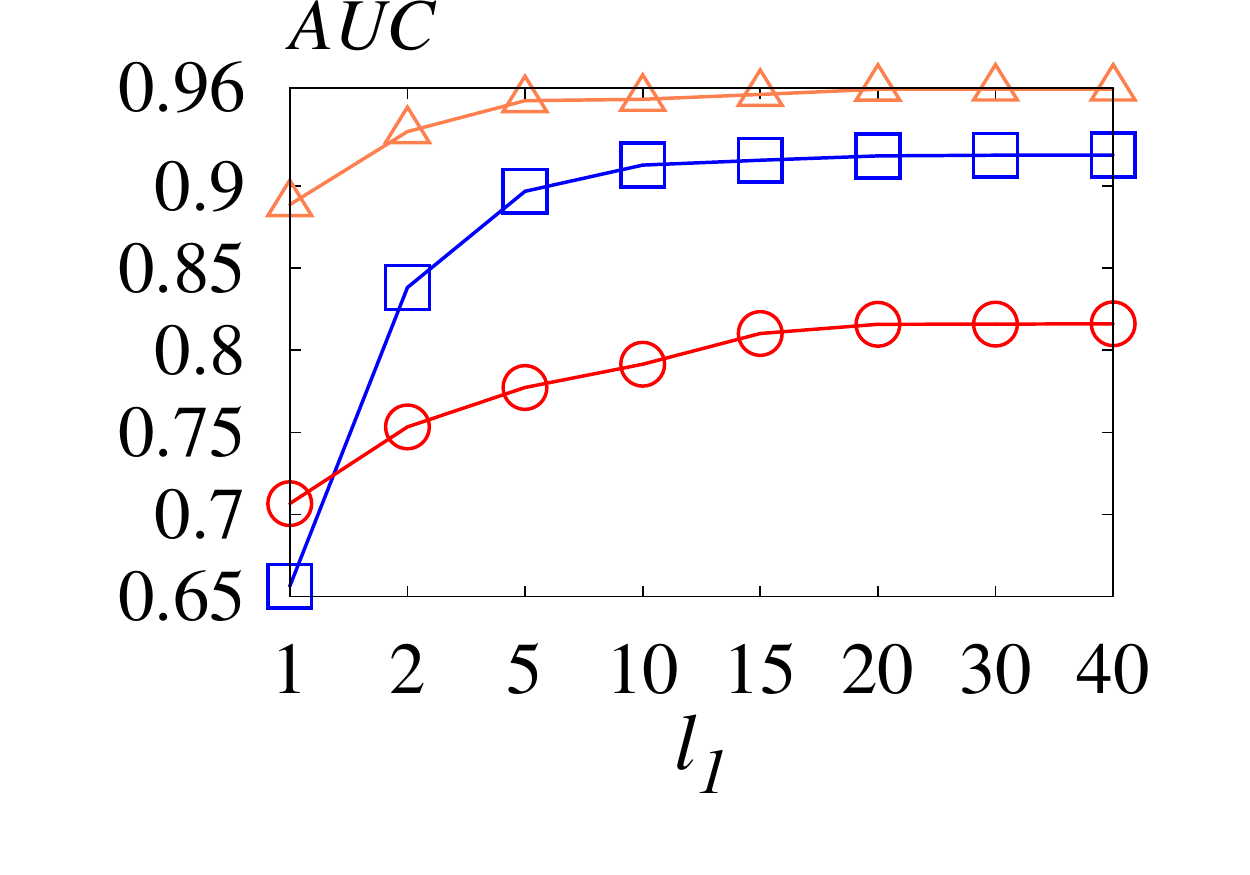}\label{fig:para-ell1}} &
\hspace{-6mm}\subfloat[Varying $\ell_2$]{\includegraphics[width=0.55\linewidth]{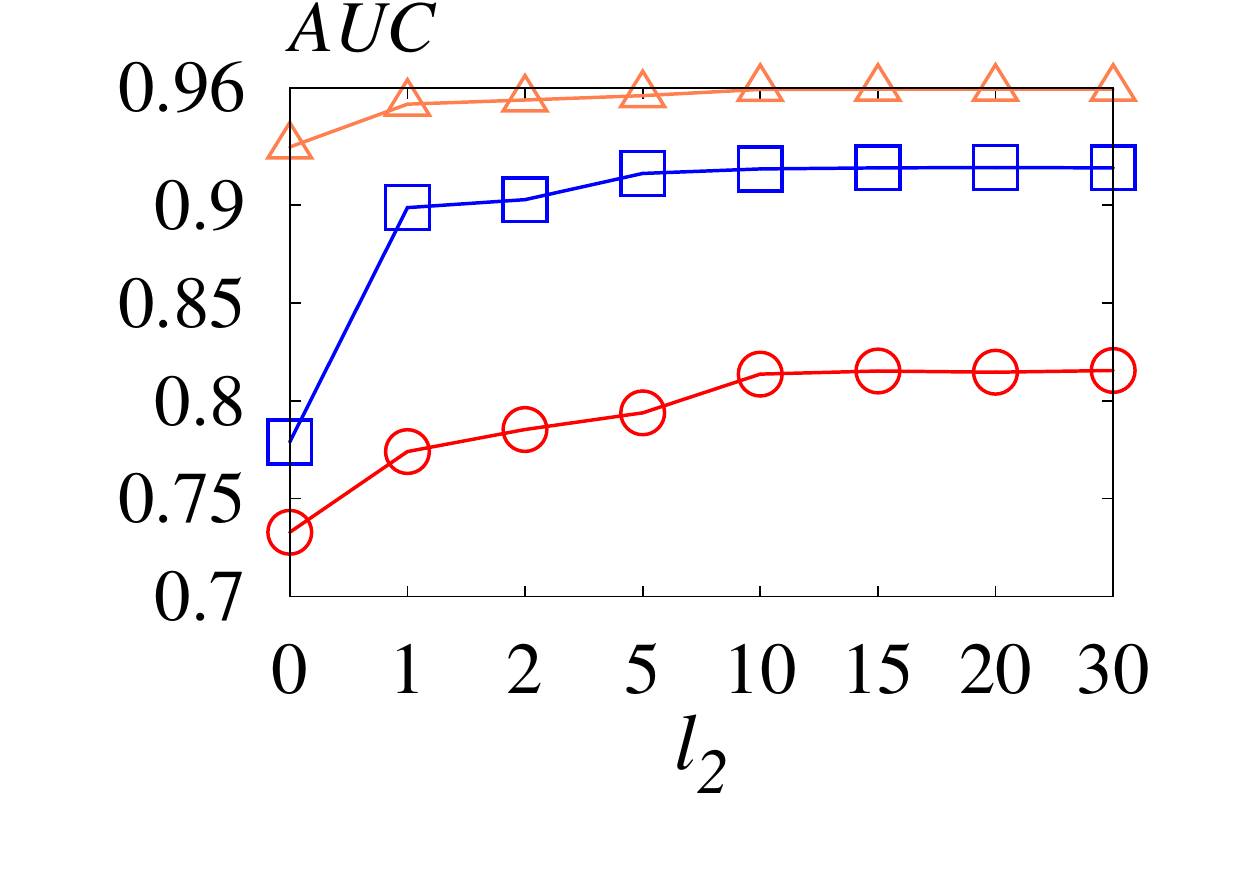}\label{fig:para-ell2}} 
\\[2mm]
\end{tabular}
\caption{Link prediction results with varying parameters (best viewed in color).} \label{fig:para}
\end{small}
\vspace{-3mm}
\end{figure}

We study the effect of varying the parameters in \nrp, including $\alpha$, $\epsilon$, $\ell_1$ and $\ell_2$, for link prediction on {\em Wiki}, {\em Blogcatalog} and {\em Youtube} datasets. 
Note that $\alpha$ is the decay factor in PPR (Eq. \eqref{eq:over-ppr} in Section \ref{sec:over-basic}); $\epsilon$ is the error threshold of \rsvd used in our PPR approximation (Algorithm \ref{alg:ApproxPPR}); $\ell_1$ is the number of iterations for computing PPR (Algorithm \ref{alg:ApproxPPR}); $\ell_2$ is the number of epochs for reweighting node embeddings (Algorithm \ref{alg:nrp}).
The AUC results are shown in Fig.~\ref{fig:para}, when one of the parameters is varied, the others are kept as default values in Section \ref{sec:exp-set}.

Fig. \ref{fig:para-alpha} displays the AUC of \nrp when we vary the decay factor $\alpha$ from $0.1$ to $0.9$. As $\alpha$ increases, the performance downgrades since only limited local neighborhoods of nodes are preserved and high-order proximities are failed to be captured in the embeddings, which is consistent with the observation in \cite{verse18,yin2019scalable}. When $\alpha=0.1$ or $0.2$, the AUC score is the highest, which holds on all the three datasets. And thus our choice of $\alpha=0.15$ makes sure that the best efficacy is achieved.

The AUC result of \nrp when varying $\epsilon$ from $0.1$ to $0.9$ is depicted in Fig.~\ref{fig:para-eps}. According to Theorem \ref{thrm:ppr}, $\epsilon$ influences the accuracy of our PPR approximation.
As shown in Fig.~\ref{fig:para-eps}, when $\epsilon$ is increased (\textit{i.e.}, the error caused by \rsvd is larger), the AUC performance of the embedding decreases, especially on {\em Youtube} dataset.
Therefore, we set $\epsilon$ to $0.2$, which has the same excellent performance as $0.1$ but is computationally more efficient.

In Fig.~\ref{fig:para-ell1}, observe that the AUC of \nrp grows significantly when we vary $\ell_1$ from $1$ to $15$, and keeps stable and excellent for larger $\ell_1$ from $15$ to $40$, which holds for all the three datasets. Recall that the accuracy of our PPR approximation is affected by $\ell_1$ as well and when $\ell_1$ increases, the approximate PPR scores are more accurate. According to Fig.~\ref{fig:para-ell1}, our choice of $\ell_1=20$ is proper and robust. 

Fig.~\ref{fig:para-ell2} shows the AUC of \nrp when we vary $\ell_2$ from $0$ to $30$.
The AUC increases significantly when $\ell_2$ is increased from $0$ to $10$, and then keeps stable for larger $\ell_2$ values, which is consistent across the three datasets.
When $\ell_2=0$, it is equivalent to disable our reweighting scheme and only use the conventional PPR for embedding, which is significantly inferior to the case that our \nrp reweighting scheme is enabled, {\it e.g.}, when $\ell_2=10$. 
Specifically, on {\em Wiki}, the AUC is increased from $0.78$ to $0.91$ when  $\ell_2$ is varied from $0$ to $10$.
This validates our insight about the drawback of vanilla PPR for embeddings and demonstrates the power of our proposed reweighting scheme.
Further, Fig.~\ref{fig:para-ell2} also shows that our reweighting scheme converges quickly when the epochs are increased.

\section{Conclusion}\label{sec:ccl}
This paper presents \nrp, a novel, efficient and effective approach for homogeneous network embedding. \nrp constructs embedding vectors based on personalized PageRank values and reweights each node's embedding vectors based on an objective function concerning the in-/out- degree of each node. We show that \nrp runs in time almost linear to the size of the input graph, and that it requires less than four hours to process a graph with 1.2 billion edges. Extensive experiments on real data also demonstrate that \nrp considerably outperforms the state of the arts in terms of the accuracy link prediction, graph reconstruction and on node classification tasks. As for future work, we plan to study how to extend \nrp to handle attributed graphs. 
%

\section*{APPENDIX}
\subsection*{A \ \ Proof of Theorem \ref{thrm:ppr}}\label{sec:proof}
\begin{proof} We need the following theorem,
\begin{theorem}[Eckart–Young Theorem \cite{golub1996matrix}]\label{lem:eym}
Suppose $\AM_{\kp}$ is the rank $\kp$ approximation to $\AM$ produced by exact SVD, then
\begin{equation}\label{eq:svd-dapprox}
\min_{rank(\widehat{\AM})\le \kp}{\|\AM-\widehat{\AM}\|_2}=\|\AM-\AM_{\kp}\|_2=\sigma_{\kp+1},
\end{equation}
where $\sigma_{\kp+1}$ is the $(\kp+1)$-th largest singular value of $\AM$.
\end{theorem}
Recall that $\XM_1\YM^{\top}=\DM^{-1}\UM\Sigma\VM^{\top}$, where $\UM,\boldsymbol{\Sigma},\VM$ are produced by \rsvd. Then, by Theorem 1 of \rsvd \cite{musco2015randomized} and Eckart–Young theorem \cite{golub1996matrix}, we have
\begin{align}
\|\AM-\UM\boldsymbol{\Sigma}\VM^{\top}\|_{2} = \|\AM-\DM\XM_1\YM^{\top}\|_{2}\le (1+\epsilon)\sigma_{\kp+1},
\end{align}
where $\sigma_{\kp+1}$ is the $\kp+1$ largest singular value of $\AM$.
According to \cite{golub1996matrix}, the following inequalities hold
\begin{equation*}
\|\AM-\DM\XM_1\YM^{\top}\|_{max} \le \|\AM-\DM\XM_1\YM^{\top}\|_{2}\le (1+\epsilon)\sigma_{\kp+1},
\end{equation*}
\begin{equation*}
\|\AM-\DM\XM_1\YM^{\top}\|_{1} \le \sqrt{n}\|\AM-\DM\XM_1\YM^{\top}\|_{2}\le \sqrt{n}(1+\epsilon)\sigma_{\kp+1},
\end{equation*}
which indicates that, for any node pair $(u,v)\in V\times V$,
\begin{align}\label{eq:svd-error}
\textstyle |\PM[u,v]-(\XM_1\YM^{\top})[u,v]|&=\textstyle \left|\frac{\AM[u,v]}{d(u)}-(\XM_1\YM^{\top})[u,v]\right|\nonumber \\
&\textstyle \le \frac{1}{d(u)}(1+\epsilon)\sigma_{\kp+1},
\end{align}
and for any node $u\in V$,
\begin{align}\label{eq:svd-error-2}
\sum_{u\in V}{|\PM[u,v]-(\XM_1\YM^{\top})[u,v]|}&= \sum_{u\in V}{\left|\frac{\AM[u,v]}{d(u)}-(\XM_1\YM^{\top})[u,v]\right|}\nonumber \\
&\textstyle \le \sqrt{n}(1+\epsilon)\sigma_{\kp+1}.
\end{align}
By Lines 2-5 in Algorithm \ref{alg:ApproxPPR},
\begin{equation}
\textstyle \XM\YM^{\top}=\alpha(1-\alpha)\XM_{\ell_1}\YM^{\top}=\sum_{i=1}^{\ell_1}{\PM^{i-1}\XM_1\YM^{\top}}.
\end{equation}
By the definition of $\PiMp$ in Eq. \eqref{eq:algo-ppr},
\begin{align}\label{eq:ppr-err-2}
&\textstyle |\PiMp[u,v]-(\XM\YM^{\top})[u,v]| \nonumber\\
&=\left|\sum_{i=1}^{\ell_1}{\alpha(1-\alpha)^i\sum_{w\in V}{\PM^{i-1}[u,w]\cdot\left(\PM[w,v]-(\XM\YM^{\top})[w,v]\right)}} \right|
\end{align}
With Eq. \eqref{eq:svd-error}, \eqref{eq:svd-error-2} and \eqref{eq:ppr-err-2}, for every node pair $(u,v)\in V\times V$ with $v\neq v$ and every node $u\in V$, the follow inequalities hold,
\begin{equation}\label{eq:ppr-err-3}
\begin{split}
&\textstyle |\PiMp[u,v]-(\XM\YM^{\top})[u,v]|\le \sigma_{\kp+1}(1+\epsilon)\sum_{i=1}^{\ell_1}{\alpha(1-\alpha)^i}.\\
&\sum_{v\in V}{|\PiMp[u,v]-(\XM\YM^{\top})[u,v]|}\le \sqrt{n}\sigma_{\kp+1}(1+\epsilon)\sum_{i=1}^{\ell_1}{\alpha(1-\alpha)^i}.
\end{split}
\end{equation}
In addition, according to Eq. \eqref{eq:over-ppr} and \eqref{eq:algo-ppr}, for every node pair $(u,v)\in V\times V$ with $v\neq v$, we have
\begin{align}\label{eq:ppr-err-1}
\textstyle |\PiM[u,v]-\PiMp[u,v]|&\textstyle \le \sum_{v\in V}{|\PiM[u,v]-\PiMp[u,v]|}\nonumber\\
&\textstyle \le 1-\sum_{i=0}^{\ell_1}{\alpha(1-\alpha)^i}.
\end{align}
Combining Eq. \eqref{eq:ppr-err-3} and \eqref{eq:ppr-err-1} obtains the following results, for every node pair $(u,v)\in V\times V$ with $v\neq v$,
\begin{align*}
&\left|\PiM[u,v]-\XM\YM^{\top}(u,v)\right| \\
&\le \left|\PiM[u,v]-\PiMp[u,v]\right| + \left|\PiMp[u,v]-(\XM\YM^{\top})[u,v]\right|\\
& \le (1+\epsilon)\sigma_{\kp+1}(1-\alpha)\left(1-(1-\alpha)^{\ell_1}\right)+(1-\alpha)^{\ell_1+1},
\end{align*}
and for every node $u\in V$,
\begin{align*}
&\sum_{v\in V}{\left|\PiM[u,v]-\XM\YM^{\top}(u,v)\right|} \\
&\le \sum_{v\in V}{\left|\PiM[u,v]-\PiMp[u,v]\right|} + \sum_{v\in V}{\left|\PiMp[u,v]-(\XM\YM^{\top})[u,v]\right|}\\
& \le \sqrt{n}(1+\epsilon)\sigma_{\kp+1}(1-\alpha)\left(1-(1-\alpha)^{\ell_1}\right)+(1-\alpha)^{\ell_1+1},
\end{align*}
which completes our proof.
\end{proof}

\subsection*{B \ \ Updating Forward Weights}\label{sec:fwd_weight}
For any node $u^*$, the formula for updating $\wra_{u^*}$ is derived by (i) taking the partial derivative of the objective function in Eq.~\eqref{eq:algo-obj} with respect to $\wra_{u^*}$, 
\begin{align*}
\textstyle \frac{\partial O}{\partial \wra_{u^*}} & \textstyle= 2\Big[\left(\XM_{u^*}\sum_{v\ne u^*} \wla_{v}\YM_v^\top\right)^2\wra_{u^*}\\
&\textstyle \ \ \ \ -d_{out}(u^*)\XM_{u^*}\sum_{v\ne u^*}\wla_{v}\YM_v^\top\\
&\textstyle \ \ \ \ +\sum_{v}(\sum_{u\ne v, u\ne u^*} \wra_{u} \XM_{u}\YM_v^\top  \wla_{v})\XM_{u^*}\YM_v^\top \wla_{v}\\
&\textstyle \ \ \ \ +\sum_{v\ne u^*}(\XM_{u^*}\YM_v^\top  \wla_{v})^2\wra_{u^*}\\
&\textstyle \ \ \ \ -\XM_{u^*} \sum_{v}d_{in}(v)\wla_{v}\YM_v^\top + \lambda\wra_{u^*}\Big]\\
&=2(a_3^\prime-a_2^\prime-a_1^\prime)+2(b_1^\prime+b_2^\prime+\lambda)\wra_{u^*},
\end{align*}
and then (ii) identifying the value of $\wra_{u^*}$ that renders the partial derivative equal to zero. In addition, if the identified $\wra_{u^*}$ is smaller than $\frac{1}{n}$, then we set it to $\frac{1}{n}$ instead. Then the forward weight learning rule is as in Eq.~(\ref{equ:forward_update_rule}):
\begin{equation}\label{equ:forward_update_rule}
\begin{split}
 \wra_{u^*} =& \max\left\{\frac{1}{n}, \frac{a_1^\prime+a_2^\prime-a_3^\prime}{b_1^\prime+b_2^\prime + \lambda}\right\} \text{, where }\\
 a_1^\prime=& \XM_{u^*} \sum_{v}d_{in}(v)\wla_{v}\YM_v^\top,\\
 a_2^\prime=& d_{out}(u^*)\XM_{u^*}\sum_{v\ne u^*}\wla_{v}\YM_v^\top,\\
 a_3^\prime=& \sum_{v}(\sum_{u\ne v, u\ne u^*} \wra_{u} \XM_{u}\YM_v^\top  \wla_{v})\XM_{u^*}\YM_v^\top \wla_{v},\\
 b_1^\prime=& \sum_{v\ne u^*}(\XM_{u^*}\YM_v^\top  \wla_{v})^2,\\
 b_2^\prime=& (\XM_{u^*}\sum_{v\ne u^*} \wla_{v}\YM_v^\top)^2.
\end{split}
\end{equation}
By Eq.~\eqref{equ:forward_update_rule}, each update of $\wra_{u^*}$ requires computing $a_1^\prime, a_2^\prime, a_3^\prime, b_1^\prime$ and $b_2^\prime$, which are similar to the computation of $a_1, a_2, a_3, b_1$ and $b_2$ in Section \ref{sec:algo-weight}. Hence, it takes $O(n^2{\kp}^2)$ time to update $\wra_{u^*}$ once, which leads to a total overhead of $O(n^3 {\kp}^2)$ for updating all $\wra_{u^*}$ once.

In the following, we present the solution to accelerate the computation of $a_1^\prime,a_2^\prime,a_3^\prime,b_1^\prime,b_2^\prime$ for forward weight $\wra_{u^*}$. Since the techniques for updating forward weight $\wra_{u^*}$ are similar to that for backward weights, for brevity, we use the same symbols to represent the intermediate computations of forward weights as those of backward weights.

\header
{\bf Computation of $\mathbf{a_1^\prime,a_2^\prime,b_2^\prime}$.} By the definitions of $a_1, a_2, b_2$ in Eq.~\eqref{equ:forward_update_rule},
\begin{equation}\label{eq:algo-a1a2b2_fwd}
\begin{split}
&a_1^\prime=\XM_{u^*}\boldsymbol{\xi}^\top, \; a_2^\prime=d_{out}(u^*)\XM_{u^*} (\boldsymbol{\chi}-\wla_{u^*}\YM_{u^*})^\top,\\
&\text{and } b_2^\prime=\left(\XM_{u^*} (\boldsymbol{\chi}-\wla_{u^*}\YM_{u^*})^\top\right)^2, \\
& \text{where } \boldsymbol{\xi}=\sum_{v}d_{in}(v)\wla_{v}\YM_v, \; \text{ and }\boldsymbol{\chi}=\sum_{v}\wla_{v}\YM_v.
\end{split}
\end{equation}
Eq.~\eqref{eq:algo-a1a2b2_fwd} indicates that the $a_1^\prime$ values of all nodes $u^*\in V$ share a common $\boldsymbol{\xi}$, while $a_2^\prime$ and $b_2^\prime$ of each node $u^*$ have $\boldsymbol{\chi}$ in common. Observe that both $\boldsymbol{\xi}$ and $\boldsymbol{\chi}$ are independent of any forward weight. Motivated by this, we propose to first compute $\boldsymbol{\xi}\in \mathbb{R}^{1 \times \kp}$ and $\boldsymbol{\chi} \in \mathbb{R}^{1 \times \kp}$, which takes $O(n \kp)$ time. After that, we can easily derive $a_1^\prime, a_2^\prime$, and $b_2^\prime$ for any node with precomputed $\boldsymbol{\xi}$ and $\boldsymbol{\chi}$.  
In that case, each update of $a_1^\prime$, $a_2^\prime$, and $b_2^\prime$ takes only $O(\kp)$ time, due to Eq.~\eqref{eq:algo-a1a2b2_fwd}. This leads to $O(n\kp)$ (instead of $O(n^2 \kp)$) total computation time of $a_1^\prime,a_2^\prime$, and $b_2^\prime$ for all nodes.

\header
{\bf Computation of $\mathbf{a_3^\prime}$.} 
Note that
\begin{align*}
a_3^\prime&=\textstyle \sum_{v}\left(\sum_{u} \wra_{u}  \XM_{u}\YM_{v}^\top \wla_{v}\right)\wla_{v}  \XM_{u^*}\YM_{v}^\top\\
&\textstyle\ \ \ \ -\sum_{v}\left( \wra_{u^*}  \XM_{u^*}\YM_{v}^\top \wla_{v}\right)\wla_{v}  \XM_{u^*}\YM_{v}^\top\\
&\textstyle\ \ \ \ -\sum_{v}\left( \wra_{v}  \XM_{v}\YM_{v}^\top \wla_{v}\right)\wla_{v}  \XM_{u^*}\YM_{v}^\top\\
&\textstyle \ \ \ \ +\left( \wra_{u^*}  \XM_{u^*}\YM_{u^*}^\top \wla_{u^*}\right)\wla_{u^*}  \XM_{u^*}\YM_{u^*}^\top,
\end{align*}
which can be rewritten as:
\begin{equation}\label{eq:algo-a3_fwd}
\begin{split}
a_3^\prime  = & \boldsymbol{\rho}_1 \mathbf{\Lambda}\XM_{u^*}^\top - \wra_{u^*} \XM_{u^*}\mathbf{\Lambda}\XM_{u^*}^\top - \boldsymbol{\rho}_2\XM_{u^*}^\top  \\
& {} + \wla_{u^*}^2  \left(\XM_{u^*}\YM_{u^*}^\top\right)^{2} \wra_{u^*},\\
& \!\!\!\!\!\!\!\!\!\!\!\! \text{where } \mathbf{\Lambda}=\sum_{v} \wla_{v}^2 (\YM_{v}^\top  \YM_{v}), \; \boldsymbol{\rho}_1=\sum_{u}\wra_{u} \XM_{u},\\
& \:\:\:\text{and } \boldsymbol{\rho}_2=\sum_{v} \left(\wra_{v} \cdot \wla_{v}^2  \left(\XM_v\YM_v^\top\right)  \YM_{v}\right).
\end{split}
\end{equation}

Observe that $\mathbf{\Lambda}$ is independent of any forward weight. Thus, it can be computed once and reused in the computation of $a_3^\prime$ for all nodes. Meanwhile, both $\boldsymbol{\rho}_1$ and $\boldsymbol{\rho}_2$ dependent on all of the foward weights, and hence, cannot be directly reused if we are to update each foward weight in turn. However, we note that $\boldsymbol{\rho}_1$ and $\boldsymbol{\rho}_2$ can be {\it incrementally} updated after the change of any single forward weight. Specifically, suppose that we have computed $\boldsymbol{\rho}_1$ and $\boldsymbol{\rho}_2$ based on Eq.~\eqref{eq:algo-a3_fwd}, and then we change the forward weight of $u^*$ from $\wra_{u^*}^\prime$ as $\wra_{u^*}$. In that case, we can update $\boldsymbol{\rho}_1$ and $\boldsymbol{\rho}_2$ as:
\begin{equation}\label{eq:update-rho1-rho2_fwd}
\begin{split}
&\textstyle \boldsymbol{\rho}_1 = \boldsymbol{\rho}_1+ \left(\wra_{u^*} - \wra_{u^*}^\prime\right)\XM_{u^*},\\
&\textstyle \boldsymbol{\rho}_2 = \boldsymbol{\rho}_2+ \left(\wra_{u^*} - \wra_{u^*}^\prime\right)\wla_{u^*}^2  \left(\XM_{u^*}\YM_{u^*}^\top\right)  \YM_{u^*}.
\end{split}
\end{equation}
Each of such updates takes only $O(\kp)$ time, since $\wra_{u^*}, \wra_{u^*}^\prime \in \mathbb{R}$ and $\XM_{u^*}, \YM_{u^*} \in \mathbb{R}^{1 \times \kp}$.

The initial values of $\boldsymbol{\rho}_1$ and $\boldsymbol{\rho}_2$ can be computed in $O(n\kp)$ time based on Eq.~\eqref{eq:algo-a3_fwd}, while $\mathbf{\Lambda}$ can be calculated in $O(nk^{\prime 2})$ time. Given $\mathbf{\Lambda}$, $\boldsymbol{\rho}_1$, and $\boldsymbol{\rho}_2$, we can compute $a_3^\prime$ for any node $u^*$ in $O(k^{\prime 2})$ time based on Eq.~\eqref{eq:algo-a3_fwd}. Therefore, the total time required for computing $a_3^\prime$ for all nodes is $O(nk^{\prime 2})$, which is an significant reduction from the $O(n^3k^{\prime 2})$ time required by the naive solution in Equation \eqref{equ:forward_update_rule}.

\header
{\bf Approximation of $\mathbf{b_1}^\prime$.} We observe that the value of $b_1^\prime$ is insignificant compared to $b_2^\prime$. Thus, we propose to approximate its value instead of deriving it exactly, so as to reduce computation cost. By the inequality of arithmetic and geometric means, we have:
\begin{equation}\label{eq:algo-b1_fwd}
 \frac{1}{\kp} b_1^\prime \le \sum_{v\ne u^*}\wla_{v}^2 (\sum_{r=1}^{k^\prime}\XM_{u^*}[r]^2\YM_v[r]^2) \le  b_1^\prime.
\end{equation}
Let $\boldsymbol{\phi}$ be a length-$\kp$ vector where the $r$-th ($r \in [1, \kp]$) element is
\begin{equation}\label{eq:algo-phi-fwd}
\textstyle \boldsymbol{\phi}[r] = \sum_{v}\wla_{v}^2 \YM_v[r]^2.
\end{equation}
We compute $\boldsymbol{\phi}$ in $O(n \kp)$ time, and then, based on Eq.~\eqref{eq:algo-b1_fwd}, we approximate $b_1^\prime$ for each node in $O(\kp)$ time with
\begin{equation}\label{eq:algo-b1_2_fwd}
 b_1^\prime\approx  \frac{\kp}{2}\sum_{r=1}^{k^\prime}\XM_{u^*}[r]^2 \left(\boldsymbol{\phi}[r]-\wla_{u^*}^2 \YM_{u^*}[r]^2\right).
\end{equation}
Therefore, the total cost for approximating $b_1^\prime$ for all nodes is $O(n \kp)$.

\begin{algorithm}[t]
\begin{small}
\caption{$\mathsf{update Fwd Weights}$}
\label{alg:algo-updateFwdWeights}
\BlankLine
\KwIn{$G$, $\kp$, $\wrab$, $\wlab$, $\XM$, $\YM$.}
\KwOut{$\wrab$}
Compute $\boldsymbol{\xi}, \boldsymbol{\chi}, \boldsymbol{\rho_1}, \boldsymbol{\rho_2}, \mathbf{\Lambda}$ based on Eq.~\eqref{eq:algo-a1a2b2_fwd}, \eqref{eq:algo-a3_fwd}\;
\For{$r \gets 1$ to $\kp$}{
	$\boldsymbol{\phi}[r]=\sum_{v}\wla_{v}^2 \YM_v[r]^2$\; 
}
\For{$u^* \in V$ in random order}{
	Compute $a_1^\prime, a_2^\prime, a_3^\prime, b_1^\prime, b_2^\prime$ by Eq.~\eqref{eq:algo-a1a2b2_fwd}, \eqref{eq:algo-a3_fwd}, and \eqref{eq:algo-b1_2_fwd}\;
	$\wra_{u^*}^\prime= \wra_{u^*}$\;
	$\wra_{u^*}=\max\left\{\frac{1}{n}, \frac{a_1^\prime+a_2^\prime-a_3^\prime}{b_1^\prime+b_2^\prime + \lambda}\right\}$\;
	$\boldsymbol{\rho}_1 = \boldsymbol{\rho}_1+ \left(\wra_{u^*} - \wra_{u^*}^\prime\right)\XM_{u^*}$\;
	$\boldsymbol{\rho}_2 = \boldsymbol{\rho}_2+ \left(\wra_{u^*} - \wra_{u^*}^\prime\right)\wla_{u^*}^2  \left(\XM_{u^*}\YM_{u^*}^\top\right)  \YM_{u^*}$
}
\Return $\wrab$\;
\end{small}
\end{algorithm}
Algorithm \ref{alg:algo-updateFwdWeights} illustrates the pseudo-code for updating forward weights, which is analogous to Algorithm \ref{alg:algo-updateBwdWeights}. Based on the above analysis, it is easy to verify that it has the same time complexity and space overhead as Algorithm \ref{alg:algo-updateBwdWeights}. 

\subsection*{C \ \ Additional Experiments}\label{sec:fwd_weight}
\begin{figure}[!t]
\centering
\captionsetup[subfloat]{captionskip=-0.5mm}
\begin{small}
\begin{tabular}{cc}
\multicolumn{2}{c}{\hspace{-4mm}\includegraphics[height=8.2mm]{./figures/algo-legend-cikm-tr-2-eps-converted-to.pdf}\vspace{1mm}}  \\
\hspace{-2mm}\subfloat[{\em VK}]{\includegraphics[width=0.5\linewidth]{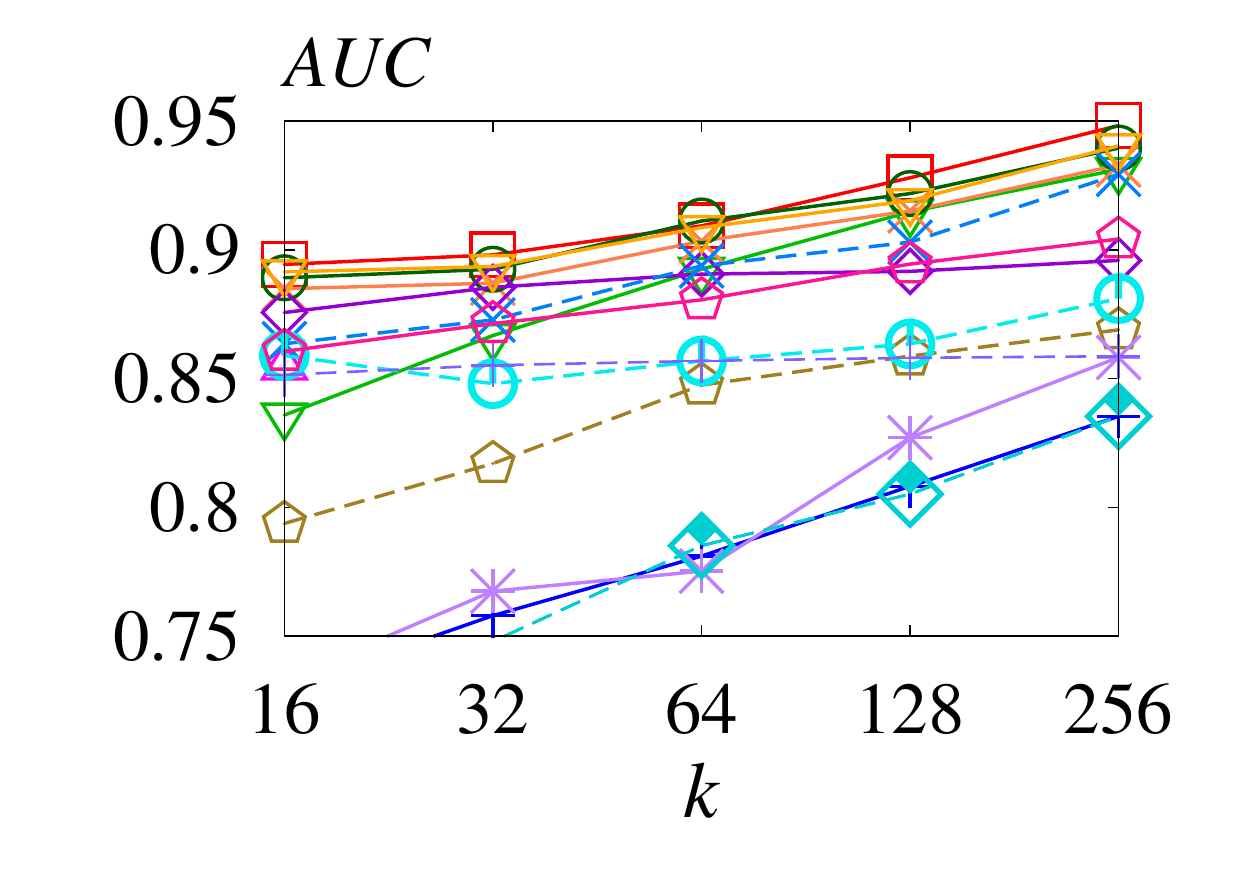}\label{fig:link-vk}} &
\hspace{-2mm}\subfloat[{\em Digg}]{\includegraphics[width=0.5\linewidth]{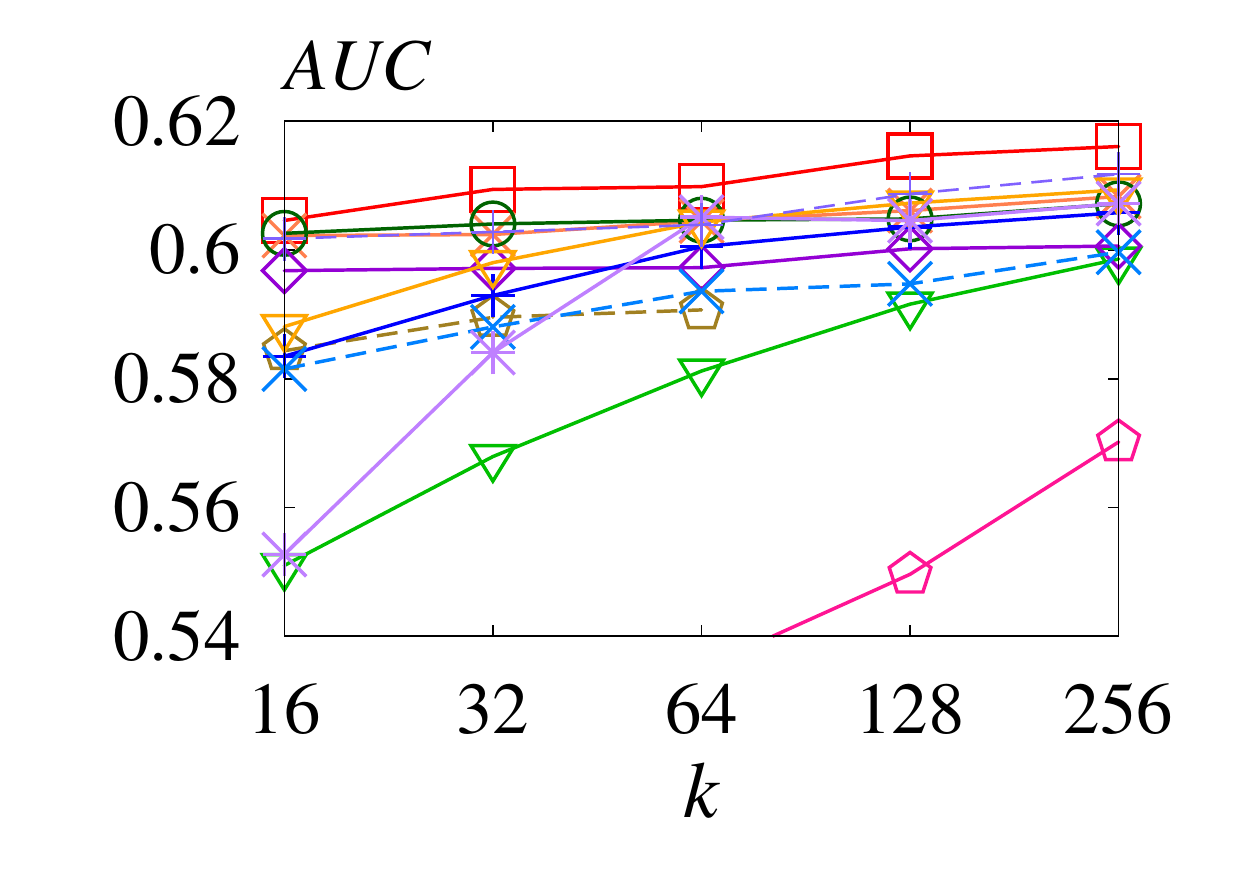}\label{fig:link-digg}}
\\[-1mm]
\end{tabular}
\caption{Link prediction performance on dynamic graphs (best viewed in color).} \label{fig:dyna-link}
\end{small}
\vspace{-2mm}
\end{figure}

\header
{\bf Link Prediction on Evolving Grpahs.}
In this set of experiments, we evaluate the link performance of all methods on real-world datasets with real new links, {\it i.e.}, evolving graphs. Table \ref{tbl:exp-data2} shows the statistics of the datasets.  Specifically, {\em VK} \cite{verse18} and {\em Digg} \cite{hogg2012social} are two real-world social networks, where each node represents a user and a link represents the friendship or following relationship. For {\em VK}, $|E_{old}|$ denotes the social network snapshot of {\em VK} in 2016 and $|E_{new}|$ is the set of new links ({\it i.e.}, friendships) in 2017. In terms of {\em Digg}, $|E_{old}|$ is the snapshot of the social network in 2008 and $|E_{new}|$ consists of new links ({\it i.e.}, following relationships) in 2009. We run all network embedding methods on $|E_{old}|$ and and then employ the learned embeddings to predict the new links $|E_{new}|$. Figure \ref{fig:dyna-link} plots the AUC results of all methods on {\em VK} and {\em Digg}. It can be observed that \nrp achieves similar performance as PPR-based methods \strap, \versealg and \app on undirected graph {\em VK}. On directed graph {\em Digg}, \nrp outperforms all competitors by at least a large margin of $0.7\%$. The experimental results indicate the effectiveness of \nrp in predicting "real new links" on real-world datasets. 
\begin{table}[t]
\centering
\renewcommand{\arraystretch}{1.2}
\begin{small}
\begin{tabular}{|l|r|r|c|c|c|}
\hline
{\bf Name} & \multicolumn{1}{c|}{$|V|$} & \multicolumn{1}{c|}{$|E|$} & \multicolumn{1}{c|}{\bf $|E_{old}|$} & \multicolumn{1}{c|}{\bf $|E_{new}|$} & \multicolumn{1}{c|}{\bf Type}\\
\hline
{\em VK}      & 78.59K    & 5.35M & 2.68M  & 2.67M & undirected \\
\hline
{\em Digg}      & 279.63K & 1.73M  & 1.03M & 701.59K  & directed  \\
\hline
\end{tabular}
\caption{Dataset statistics ($K=10^3$, $M=10^6$).} \label{tbl:exp-data2}
\end{small}
\vspace{0mm}
\end{table}

\begin{figure}[!t]
\centering
\captionsetup[subfloat]{captionskip=-0.5mm}
\begin{small}
\begin{tabular}{cc}
\hspace{-2mm}\subfloat[Varying the number of nodes]{\includegraphics[width=0.5\linewidth]{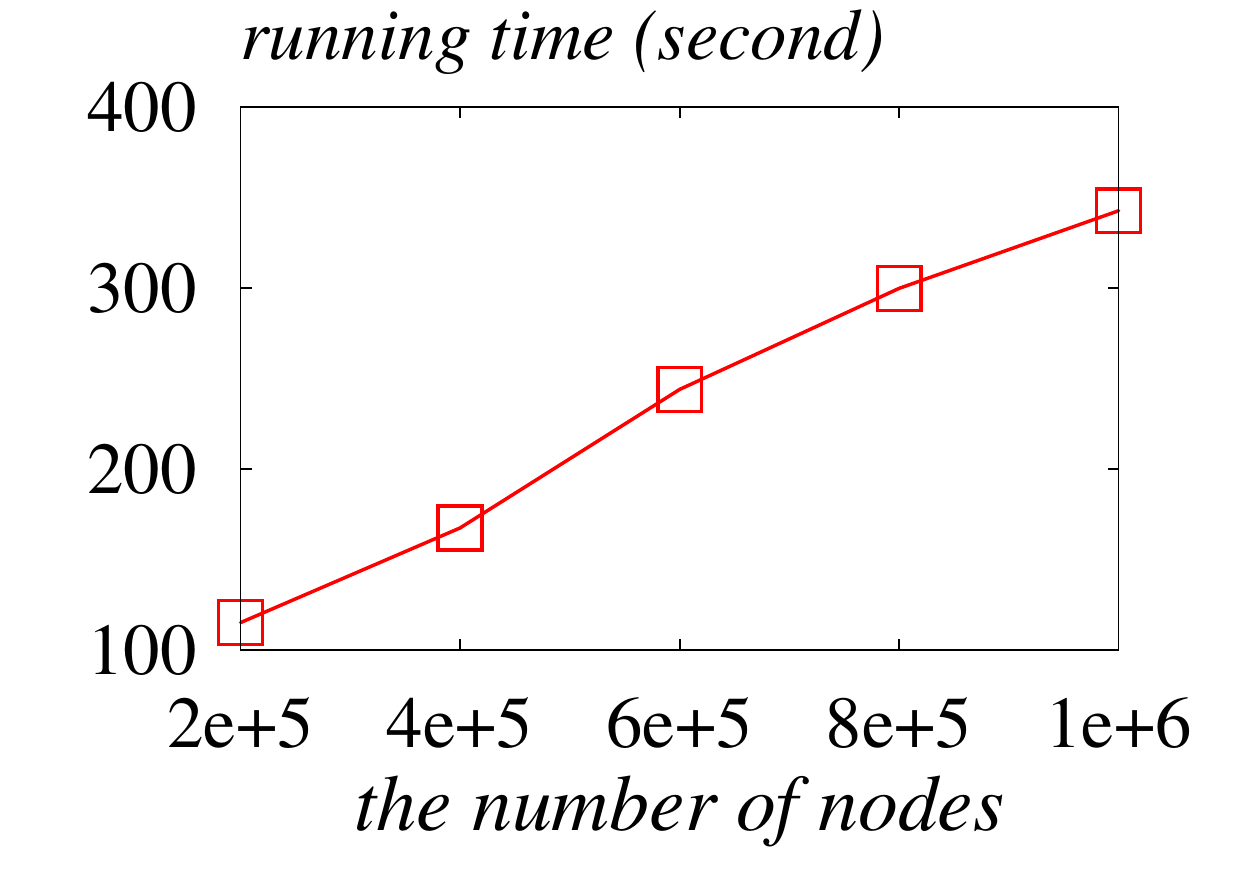}\label{fig:scale-n}} &
\hspace{-2mm}\subfloat[Varying the number of edges]{\includegraphics[width=0.5\linewidth]{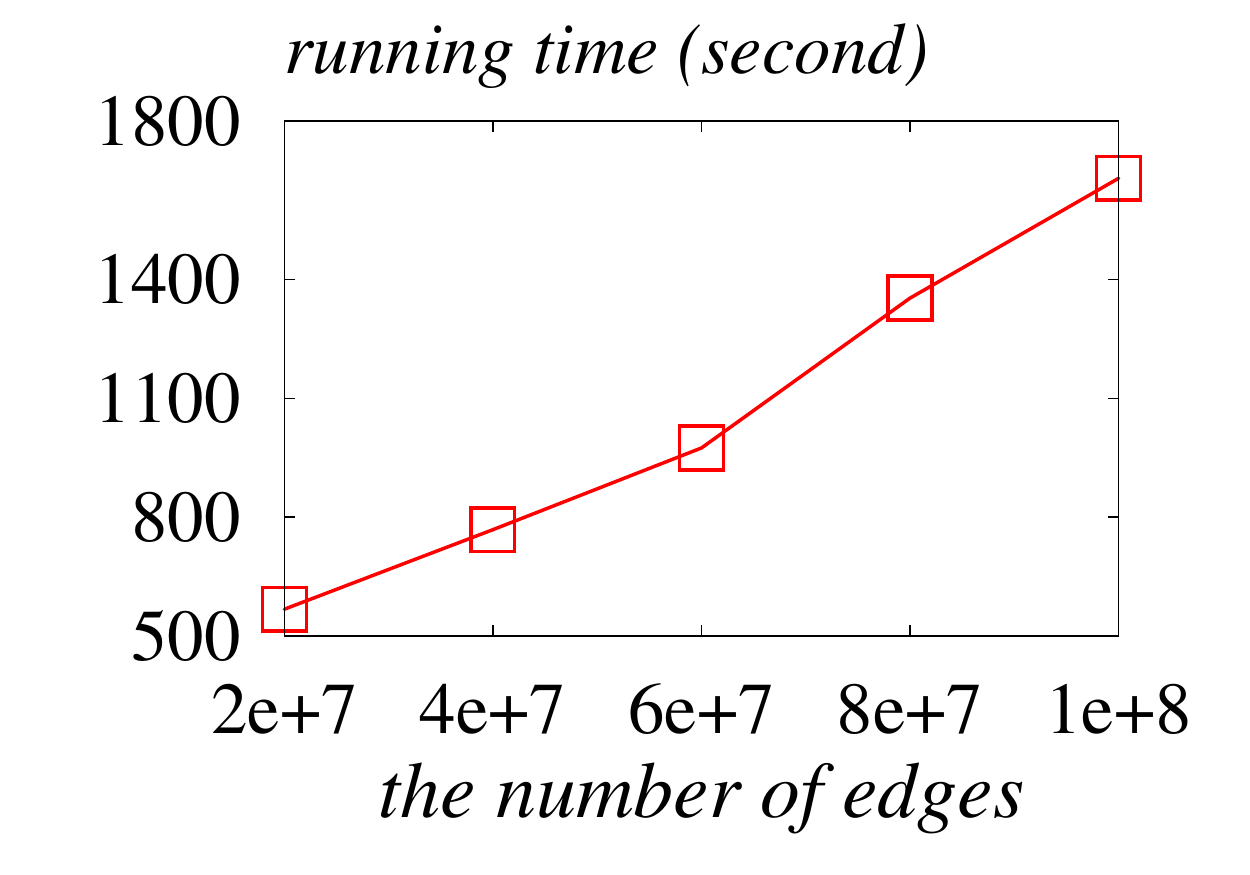}\label{fig:scale-m}}
\\[-1mm]
\end{tabular}
\caption{Scalability tests.} \label{fig:scale}
\end{small}
\vspace{-2mm}
\end{figure}

\header
{\bf Scalability Tests.} 
In this set of experiments, we verify the scalability of \nrp. Following prior work \cite{AROPE18,RandNE18}, we use synthetic
graphs of different sizes generated by the
Erdos Renyi random graph model \cite{erdHos2013spectral}. We run \nrp on these synthetic
graphs with default parameter settings described in Section \ref{sec:exp-set}. We record the running time when fixing the number of nodes
(as $10^6$) or fixing the number of edges (as $10^7$) while varying the other, {\it i.e.}, the number of edges in $\{2\times10^7,4\times10^7,6\times10^7,8\times10^7,1\times10^8\}$ and the number of nodes in $\{2\times10^5,4\times10^5,6\times10^5,8\times10^5,1\times10^6\}$, respectively.  Figure \ref{fig:scale-n} and \ref{fig:scale-m} plot the running time of \nrp when varying the number of nodes and the number of edges, respectively. It can be observed that the running time grows linearly with the number of nodes and the number of edges, respectively, confirming the time complexity of \nrp as well as verying the scalability of \nrp.

\begin{figure*}[!t]
\centering
\captionsetup[subfloat]{captionskip=-0.5mm}
\begin{small}
\begin{tabular}{cccc}
\multicolumn{4}{c}{\hspace{-3mm} \includegraphics[height=3mm]{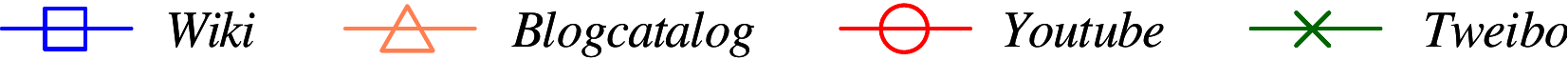}}\vspace{0mm}  \\
\hspace{-5mm}\subfloat[Varying $\ell_1$]{\includegraphics[width=0.26\linewidth]{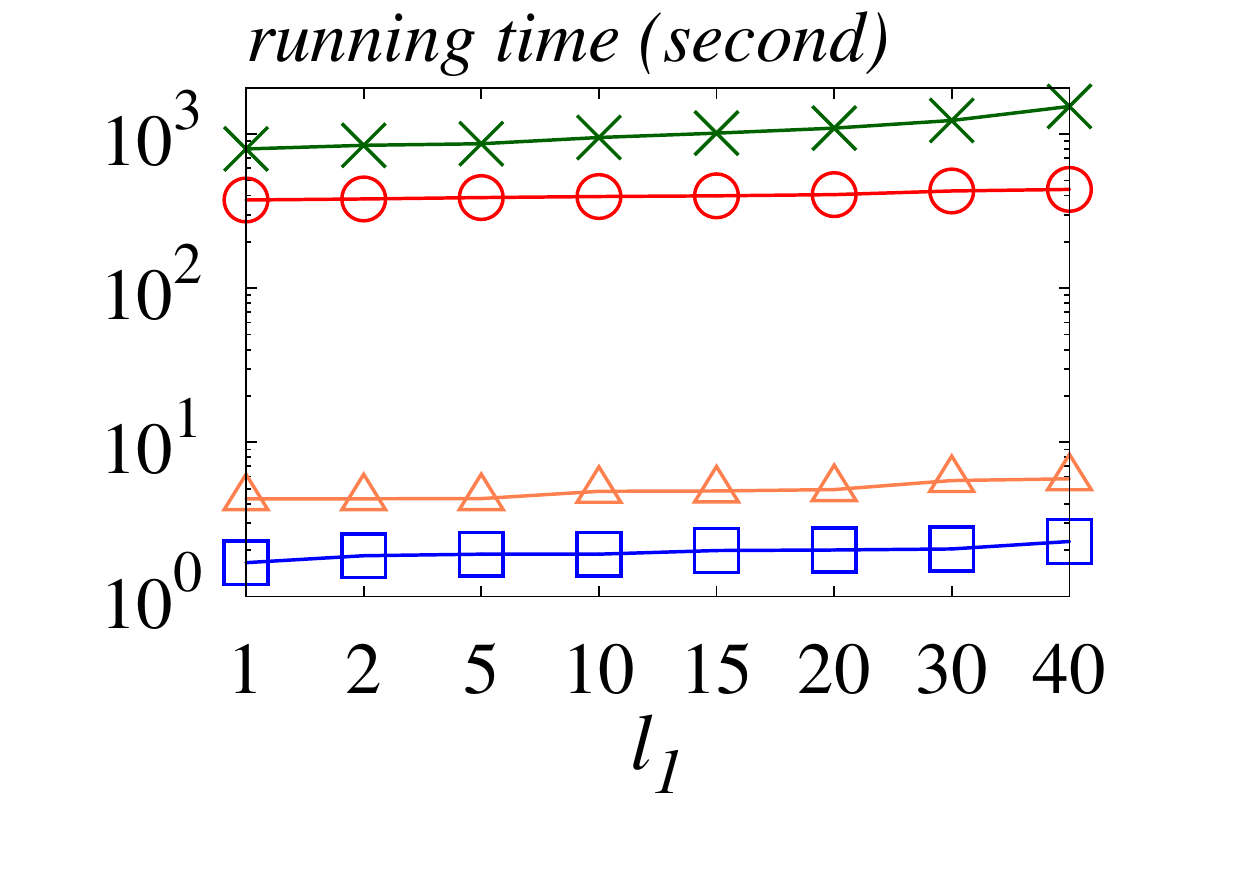}\label{fig:time-ell1}} &
\hspace{-4mm}\subfloat[Varying $\ell_2$]{\includegraphics[width=0.26\linewidth]{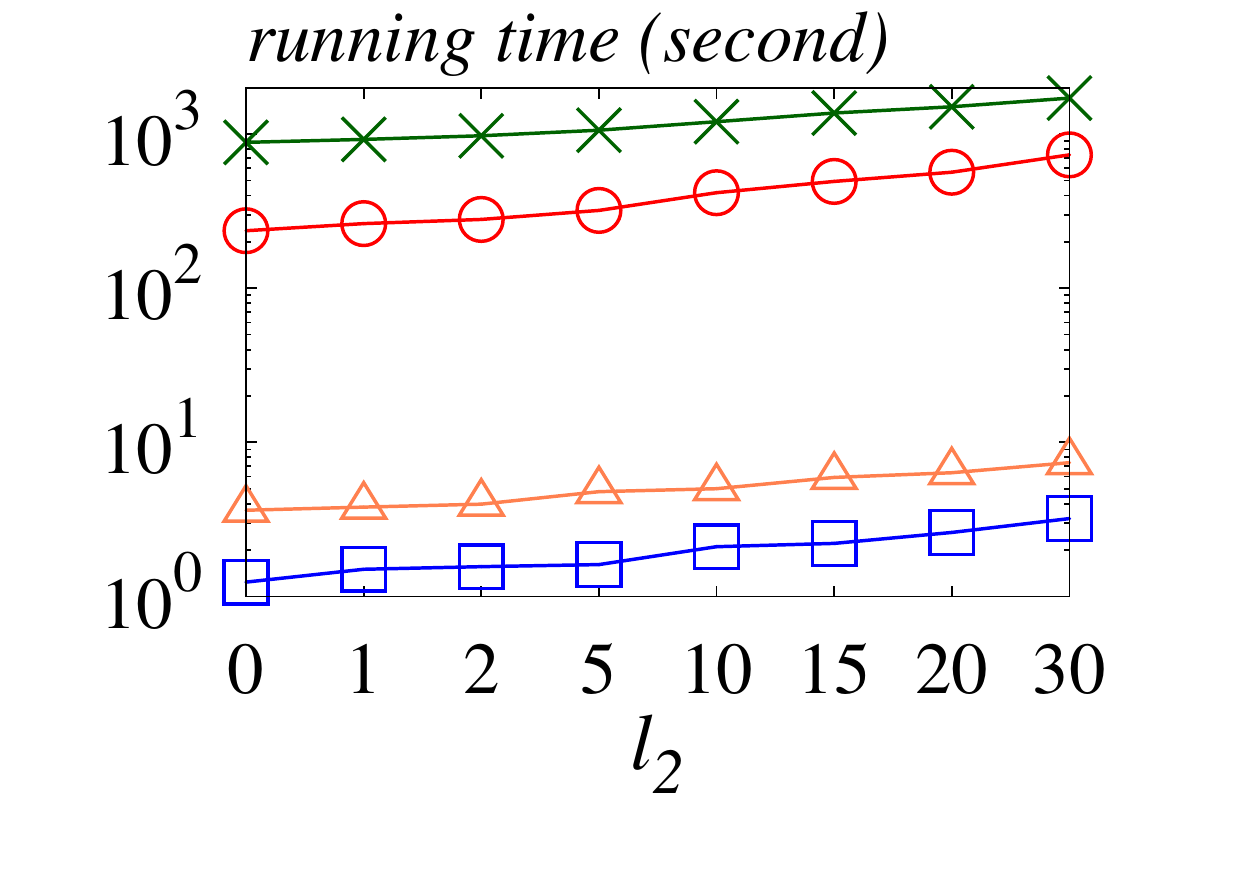}\label{fig:time-ell2}} &
\hspace{-4mm}\subfloat[Varying $\alpha$]{\includegraphics[width=0.26\linewidth]{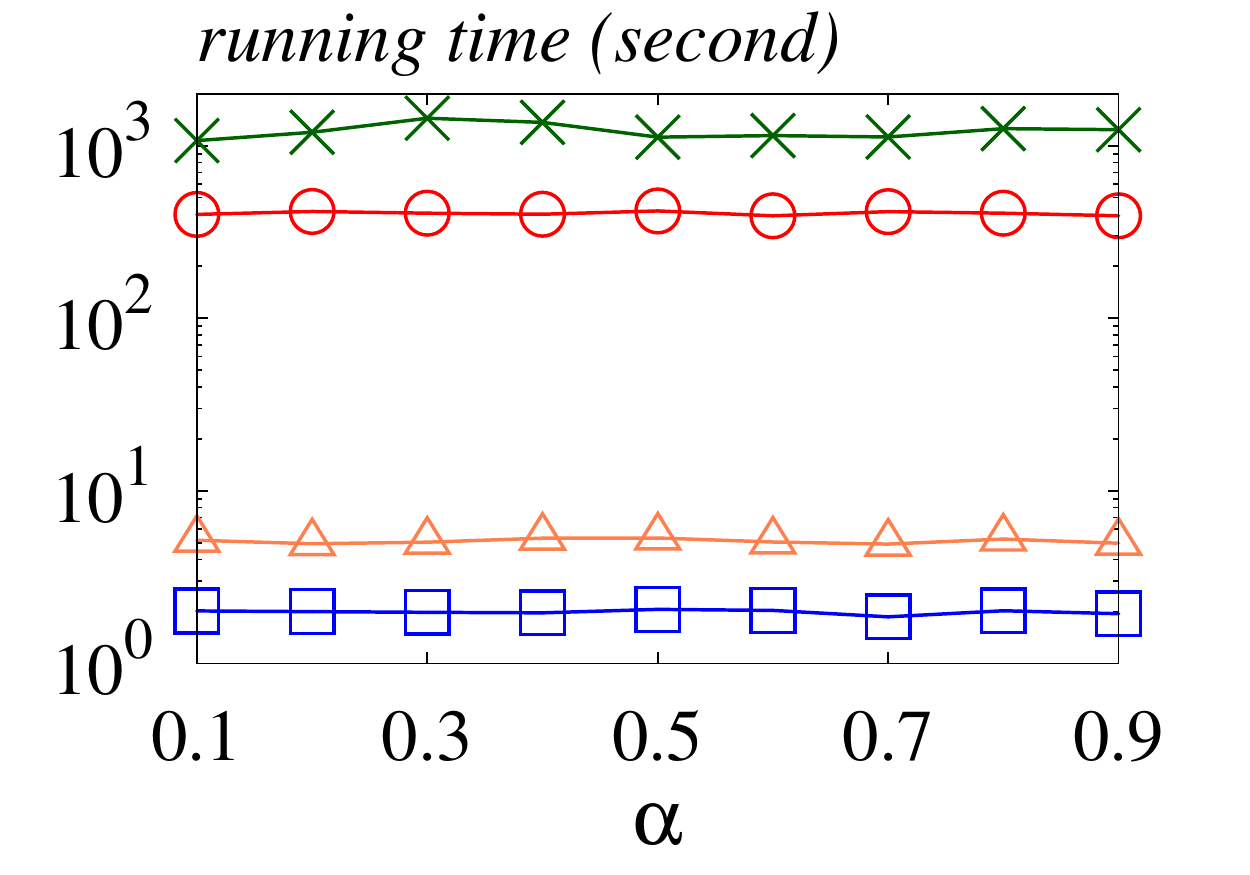}\label{fig:time-alpha}} &
\hspace{-4mm}\subfloat[Varying $\epsilon$]{\includegraphics[width=0.26\linewidth]{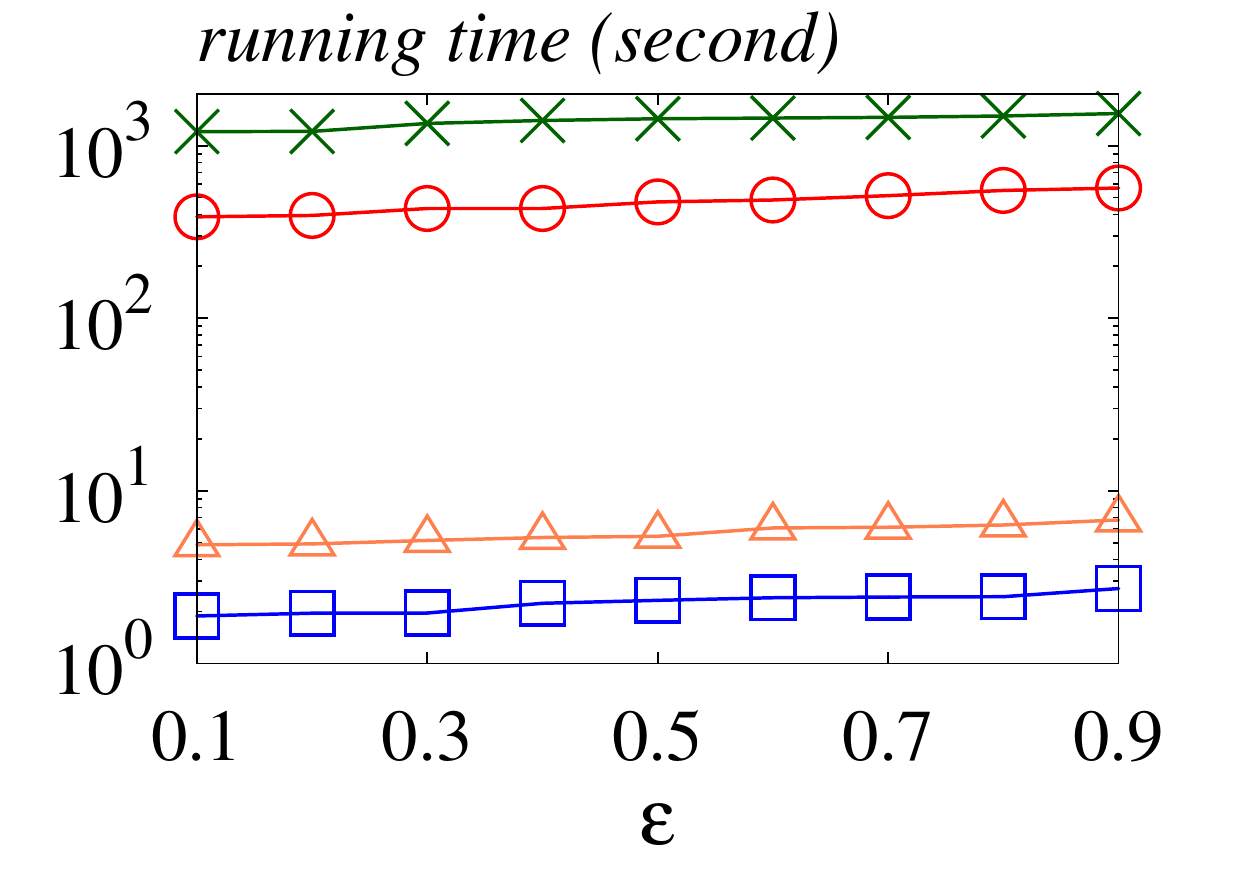}\label{fig:time-eps}}
\\[0mm]
\end{tabular}
\caption{Running time with varying parameters (best viewed in color).} \label{fig:time-para}
\end{small}
\vspace{-1mm}
\end{figure*}

\header
{\bf Running Time with Varying Parameters.}
Figure \ref{fig:time-ell1}-\ref{fig:time-eps} depict the results when varying $\ell_1$, $\ell_2$, $\alpha$ and $\epsilon$ on {\em Wiki}, {\em Blogcatalog}, {\em Youtube} and {\em Tweibo}, respectively. We can observe that the running time of \nrp grows when we increase the values of $\ell_1$, $\ell_2$ and $\epsilon$ but remain almost stable when increases $\alpha$, which accords with the time complexity of \nrp, {\it i.e.}, $O\left(\left(\frac{\log{n}}{\epsilon} + \ell_1\right) m \kp + \frac{\log{n}}{\epsilon} n{\kp}^2+\ell_2 n k^{\prime 2} \right)$. Especially, Figre \ref{fig:time-ell2} shows that $\ell_2$ has greater impact on the running time compared with other parameters. 

\bibliographystyle{abbrv}
\bibliography{main}

\end{document}